\documentstyle[aps,preprint,twoside]{revtex}
\draft
\def\real{{\cal\rm R}{\rm e} \ }                                              
\def\imag{{\cal \rm I}{\rm m} \ }                                             
\def\BRA{\left\langle}
\def\KET{\right\rangle}
\def\kubosp#1#2{\langle #1  ;  #2 \rangle}
\def\dcorr#1#2{\left( #1 \parallel  #2 \right)}
\def\dcorrz#1#2{\left( #1 \parallel    #2 \right)\LBK z\RBK}

\def\LK{\left(}
\def\RK{\right)}
\def\LBK{\left\lbrack}
\def\RBK{\right\rbrack}
\def\LB{\left\lbrace}
\def\RB{\right\rbrace}
\def\Tr{{\rm Tr}\,}

\def\a{\alpha}                                                                 \def\b{\beta}
\def\w{\omega}
\def\h{\hbar}
        
\def\E{{\bf E}}
\def\A{{\bf A}}
\def\B{{\bf B}}
\def\S{{\bf S}}
\def\p{{\bf{p}}}
\def\r{{\bf{r}}}
\def\j{{\bf j}}
\def\rp{{\bf{r}}\, '}
\def\m{{\bf m}}
\def\k{{\bf{k}}}
\def\kp{\bf{k}'}
\def\EF{E_{\rm F}}
\def\GM{G^{-}}
\def\GP{G^{+}}
\def\INTF{\int d E\, f(E)}
\def\Int{\int\limits}
\def\ty{{\rm t}}
\begin{document}

\title{Statistics and Scaling in  Disordered  Mesoscopic Electron Systems}
\author{Martin Janssen}
\address{Institut f\"ur Theoretische Physik, Universit\"at zu K\"oln, 
Z\"ulpicher Strasse 77, 50937 K\"oln, Germany}
\date{March 21, 1997}
\maketitle

\begin{abstract}
\baselineskip=0.75\baselineskip
This review is intended to give a pedagogical and unified view on the
 subject of the statistics and scaling of physical quantities in
 disordered electron systems at very low temperatures.  Quantum
 coherence at low temperatures and randomness of microscopic details
 can cause large fluctuations of physical quantities.  In such
 mesoscopic systems a localization-delocalization transition can occur
 which forms a critical phenomenon. Accordingly, a one-parameter
 scaling theory was formulated stressing the role of conductance as
 the (one-parameter) scaling variable.  The localized and delocalized
 phases are separated by a critical point determined by a critical
 value of conductance.  However, the notion of an order parameter was
 not fully clarified in this theory.  The one-parameter scaling theory
 has been questioned once it was noticed that physical quantities are
 broadly distributed and that average values are not characteristic
 for the distributions.  Based on presently available analytical and
 numerical results we focus here on the description of the total
 distribution functions and their flow with increasing system size.
 Still, one-parameter scaling theory does work in terms of typical
 values of the local density of states and of the conductance which
 serve as order parameter and scaling variable of the
 localization-delocalization transition, respectively.  Below a
 certain length scale, $\xi_c$, related to the value of the typical
 conductance, local quantities are multifractally distributed.  This
 multifractal behavior becomes universal on approaching the
 localization-delocalization transition with $\xi_c$ playing the role
 of a correlation length.
\end{abstract}

\newpage

\tableofcontents

\newpage
\baselineskip=0.7\baselineskip

\section{Introduction}\label{int}
This review is about mesoscopic conductors, statistics of related
physical quantities and their scaling behavior under change of the
size of the conductor. A few words about the terms used here
shall tell the reader what to  expect.

Mesoscopic systems  are intermediate between microscopic and macroscopic 
systems. The term goes back to van Kampen and was widely used after a
work on ``Physics of Mesoscopic Systems''  by  Imry
\cite{Imr86}.
Mesoscopic conductors  contain many (e.g.~$10^{19}$) elementary objects like
electrons and atoms. On the other hand their properties cannot be
obtained by using the {\em  thermodynamic limit} (volume and particle
number going to infinity with fixed particle density). The
thermodynamic limit is a convenient mathematical device in solid state
theory if one is interested in material constants.
In particular, in macroscopic conductors the conductivity, $\sigma$,
is such a material constant and the conductance, $G$, is then given by 
Ohm's law, $G=\sigma W/L$, where $W$ is the cross section of the
conductor and $L$ its length. This Ohmic behavior
is to be expected once the system size (denoted by one length $L$)
is much larger than characteristic length scales (see
Fig.~1). 
These are (i) the kinematic scale set by the de Broglie wavelength
$\lambda$
or, for low temperatures, by the {\em Fermi wavelength} $\lambda_F$,
(ii) the {\em elastic mean free path} $l_e$, which is the average distance
an electron travels before its initial momentum relaxes and (iii) the
{\em phase
coherence length}
 $L_\phi$, which is the average distance an individual electron
travels before its  initial and final phases become incoherent.
Once  the phase coherence length $L_\phi$ becomes {\em  larger}
 than the system size the conductor can no longer be described by
material constants. Microscopic details of the conductor
will then influence even global physical quantities like conductance
such that measurements  yield  {\em  finger prints}  of the mesoscopic
conductor.
These finger prints are due to quantum mechanical interference
effects.

Since mesoscopic conductors contain many electrons  one has to
use a quantum statistical description for their thermodynamic and
transport
quantities. However, this is not the only aspect of statistics
in mesoscopic conductors. Apart from the electronic degrees of freedom
the conductor contains other degrees of freedom which are responsible
for resistivity to applied voltages. In the simplest
case
these are static fields caused by impurities, vacancies and dislocations 
in an otherwise ideal crystal lattice. Also static inhomogeneous
electric fields caused by surrounding media are responsible for
resistivity. One refers to these static fields as {\em  static disorder
potentials}. For different realizations of  a static disorder
potential
the global conductance of an otherwise identical mesoscopic conductor will
differ -- sometimes significantly. 
Therefore, one considers ensembles of disorder potentials
characterized by a few parameters, e.g. by the average mean free path
on short scales. The task of the theory is then to determine
the {\em  probability distribution} of physical quantities like conductance
for such ensembles. Thus, on top of the quantum statistical treatment
for an individual disorder potential the statistical properties
of a whole ensemble of disorder potentials is addressed in mesoscopic
physics.

Typically, the phase coherence length is of the order of a few microns
for metals and semi-conductors  below  liquid-helium temperatures.
This is the reason that technological advances, starting in the 80's,
were needed in order  to study mesoscopic conductors systematically.
We will discuss some of the fundamental experiments in section
\ref{mes}.

Although it is true that  technological advances were needed to bring 
mesoscopic physics to the attention of a broader community in solid
state physics, a great deal of the physics had been discussed much
earlier. For example, Anderson had introduced the idea of {\em localization}
already in 1958 \cite{And58}. By localization
it is meant that strong disorder can trap electrons by quantum
interference to a finite region such that the conductor actually  
behaves as an insulator. Also, Landauer had presented a formula in
1957 \cite{Lan57}
that
describes the conductance of a phase coherent conductor in terms of
pure quantum mechanical transmission amplitudes, and a similar
important
approach to the conductance of mesoscopic systems  goes back to works
in 1972
by Edwards and Thouless \cite{Edw72}.

The works by Edwards and Thouless \cite{Edw72} and  by Wegner \cite{Weg76}
form  the starting
point of the {\em scaling theory} for mesoscopic conductors. Later in 1979
it
was formulated in a simple and predictive way by Abrahams, Anderson,
Licciardello and Ramakrishnan \cite{Abr79}.
The scaling theory for disordered mesoscopic conductors is based on
the assumption that the transport properties of systems probed at
scales much larger than the mean free path should become insensitive
to the microscopic origin of the disorder. The scaling theory
 states that once we know the conductance for a given system size we can obtain the conductance for an even larger system from a universal flow equation.
Now, since
we have already pointed out that the conductance is a random variable
which depends in a complicated manner on the particular disorder
realization
this assumption needs further explanation. As has been stressed
by Anderson {\em et al.} \cite{And80} and later by Shapiro
\cite{Sha83,Sha87} 
it can be correct only in a
probabilistic sense. The whole probability
distribution of the conductance approaches a universal distribution
function 
which depends only on very few
characteristic system parameters.  

The scaling theory leads to a classification of mesoscopic electron
systems
into three classes: depending on the initial values of the
characteristic
system parameters (called scaling variables)
for a  system size $L_0$ the system will flow 
under increase of system size to an insulating state (localized
phase),
or to an conducting state (delocalized phase), or it stays scale
independent at a {\em  critical state} . Thus a disorder induced
transition  from a localized to a delocalized phase
can occur which resembles a critical phenomenon. This transition is frequently
denoted as 
Anderson transition or  metal-insulator transition (MIT). To
distinguish this disorder induced MIT from other MITs occurring
in solid state physics
we  adopt  
the notion of  {\em  localization-delocalization} (LD) transition.
Accordingly, the critical point  is referred to as the LD transition point. 

The central aim of this article  is to provide a unique picture about
the LD  transition in mesoscopic disordered
electron systems. To achieve this aim we will introduce
the basic physical quantities, scales and theoretical models
and rely on simple calculations avoiding involved technical details.
We stress general concepts and methods such as quantum composition
rules for conductors, Fokker Planck
equations, and scaling laws for distribution functions.
Important results that are based on technically more elaborate methods
will only be quoted without detailed description. Nevertheless,
we hope that the present article  will help to find a way through the original
literature.

A number of reviews and books exist on related subjects discussed in
this article. The following short list shall help  for further reading.
The review by Y. Imry \cite{Imr86} is an introduction to mesoscopic
physics.  A review on the subject of mesoscopic transport that contains
also information about the technological aspect of devices was written
by Beenakker and van Houten \cite{Bee91}. The book by Datta
\cite{Dat95} on ``Electronic Transport in Mesoscopic 
Physics'' covers many subjects that have proved to be relevant in the
field. As review on   localization and scaling
theory  for  mesoscopic systems we recommend the ones by Lee and Ramakrishnan 
\cite{Lee85} and by Kramer and MacKinnon \cite{Kra93}.
The book by Al'tshuler and
Lee \cite{Alt91}
on ``Mesoscopic Phenomena in Solids'' contains review articles by 
 inventors
in  the field. 

The present article is organized as follows. In Sect.~\ref{mes} fundamental
experiments will be discussed. The electronic Aharonov-Bohm-Effect
makes the definition of mesoscopic conductors more precise. Universal
conductance fluctuations demonstrate the {\em  finger-print}  character
of mesoscopic conductance measurements and experiments on weak and
strong localization point to the necessity of a critical phenomenon
description of the LD  transitions.
Basic physical quantities are the linear response quantities such as
conductance. These and the corresponding relevant energy and length scales are
the subject of Sect.~\ref{bas}. To discuss the statistical problem
inherent in the problem of mesoscopic conductors we introduce
Hamiltonian
models in Sect.~\ref{mod}. There we will see that the Green's function
contains
the essential information. Only the asymptotic behavior of the Green's
function
is needed for describing global conductances resulting in a scattering
theoretical formulation of mesoscopic conductors (Sect.~\ref{sca}).
 This point of view
has been particular useful in describing strong localization
in so-called quasi-one-dimensional systems. However, the scattering
theoretical modeling is not restricted to this case but can be
extended to more general geometries by considering networks of
scattering matrices. After the basic quantities and tools have been
presented we discuss the statistical properties of mesoscopic
conductors in the idealized localized and delocalized phase,
respectively (Sect.~\ref{staide}).
To prepare for the statistics in more realistic systems we introduce
the concept of the scaling theory in Sect.~\ref{con}.
The flow of the relevant scaling variables is controlled by so-called
$\beta$-functions from which a critical exponent for the
LD  transitions can be obtained.
It describes how the correlation length of this critical phenomenon
diverges
on approaching the LD  transition point.
Based on a pedagogical model, introduced by Shapiro (cf.~\cite{Sha87}),
we emphasize  that typical values of the conductance
 distribution  can be good
candidates  for scaling variables. We pay special attention to the
statistics
of local quantities (like the local density of states) that show
power law scaling in non-localized phases. We come to the conclusion
that, at the LD transition, a spectrum of power law exponents (called
multifractal spectrum)  is needed
in order to describe the distribution. The more realistic models are
the subject of Sect.~\ref{starea}. The onset of broad distributions
characterizing global and local quantities at large, but finite,
values of typical conductance leaves open questions about their
interpretation. To clarify some of them the conductance distribution
in an idealized quasi-one-dimensional geometry is studied.
The idealization is due to the assumption of absence of transversal
diffusion
in such systems.
The calculations  give useful insight into the physics of universal conductance
fluctuations
and lead to the interpretation that the onset of broad tails in the
conductance
distribution is intimately connected to the critical regime of the
LD  transition. Such regime, as well as long
tails in the conductance distribution, is absent in the
idealized quasi-one-dimensional conductor. Giving up the idealization,
one can study the LD  transition in
quasi-one-dimensional systems by the so-called finite size scaling
method. Due to the complexity of  calculations for non-ideal
quasi-one-dimensional systems the finite size scaling  method  is often  
  restricted to numerical calculations.
However, such calculations are very successful and provide 
a lot of insight into the physics as well as quantitative results for
universal properties. 
The investigation of statistics and scaling of mesoscopic disordered
electron systems close to the LD transition in Sect.~\ref{statra}
 reveals the multifractal
properties
of electronic states and opens the possibility to consider the local
density of states as an appropriate order parameter field for the
LD  transition. The typical value, as given by
the geometric mean, is the global order parameter.
Similarly, we collect arguments that allow to view the typical
conductance
as the {\em  only} relevant scaling variable.
Our conclusions are summarized in Sect.~\ref{sum}
and a brief account of linear response theory is given in the appendix
Sect.~\ref{bri}.
  
\section{Mesoscopic Experiments}\label{mes}
In this section we will discuss some fundamental mesoscopic
experiments without paying much attention to historical order and 
completeness. Rather, we wish to give motivation to the following 
sections which are based on a theoretical physicists point of view.
 
The  electronic Aharonov-Bohm-effect  can be used for   a definition
 of mesoscopic
conductors. It is our starting
point of investigations. It points to the necessity for a quantum mechanical
description of transport. The need to describe physical
quantities
in terms of distribution functions rather than in terms of material
constants became evident after the discovery of universal conductance
fluctuations. That  one conductor can behave both, as a perfect
conductor and as a perfect insulator, when  changing system parameters
is the insight coming from experiments on weak and strong
localization. Weak localization refers to a situation where the system is 
metallic, but the conductivity is reduced as compared to its classical value.
This behavior is interpreted as a precursor of the strong localization 
phenomenon where the system is insulating 
due to spatial localization of electronic states.
 Weak localization effects have been found in various
materials. The  experimental verification of strong localization is most
clear-cut in the quantum Hall effect, and first experimental
investigations of fluctuations at the LD
transition point have been performed.
\subsection{Electronic  Aharonov-Bohm-Effect}\label{subele}
A ring-shaped conductor shown schematically in Fig.~2
with wire thickness of $\approx 40$~nm and a diameter of $\approx
820$~nm
 that was etched out of a high quality polycrystalline  gold film
of thickness $\approx 40$~nm was used to perform the first realization
of the most fundamental experiment in mesoscopic physics, the
electronic Aharonov-Bohm-Effect.
At temperatures less than $1$~K the resistance oscillates as a
function of the magnetic flux enclosed by the ring \cite{Was86}.
Figure~3 shows the result for the resistance oscillations 
found in a ring-shaped conductor fabricated in high mobility
GaAs-Al GaAs heterostructures performed by Timp {\em et al.} \cite{Tim88}.
Here the diameter of the ring is $2 \mu$~m. The period in the
magnetic field ($\approx 5$~mT) corresponds to a periodicity
in the magnetic flux enclosed by the ring  of one flux quantum $h/e$
where $-e$ is the elementary charge of an electron and $h$ is Planck's
constant. 
The qualitative explanation of the oscillations  relies on the 
concept that electrons in a magnetic field $\B = {\rm curl} \, \A$,
where $\A$ is the vector potential, pick up a quantum mechanical phase,
\begin{equation}\label{AB1}
	\varphi= \frac{e}{\h}\int\limits_{C} \A(\r)\cdot  d\r \, ,
\end{equation}
when they move along a classical path $C$ that connects two points in
space. Now, for an electron  coming from the current source the path
splits up into the upper and lower branch of the ring and the phase
difference $\varphi_u -\varphi_l$ between both paths is given by 
the total magnetic flux $\phi$ enclosed by the ring
\begin{equation}\label{AB2}
	\varphi_u -\varphi_l = \frac{e}{\h}\oint\A(\r) d\r=
	\frac{e}{\h}\phi\, . 
\end{equation}
In terms of the the flux quantum $\phi_0:=h/e$ the phase difference
is  $2\pi \phi/\phi_0$. Taking for granted that the resistance
is  sensitive to quantum interference of electronic waves  the periodicity
of the resistance is given by $\phi_0$.
The electronic Aharonov-Bohm-Effect demonstrates
that the electrons are to be treated quantum mechanically along the
ring
and that their initial phase is not randomized before leaving the ring.
This phenomenon is called phase coherence and the Aharonov-Bohm-Effect
gives rise to an operative definition of the phase coherence length
$L_\phi$ (see also~Sect.~\ref{subrel}).
On increasing  the temperature  or the diameter of the
ring  the phase coherence gets lost and the Aharanov-Bohm effect
disappears. The system size at which the Aharanov-Bohm effect
(gradually) disappears quantifies  the temperature dependent phase
coherence
length
$L_\phi$. 
The disappearance  of the Aharanov-Bohm effect
 is caused by an increasing number of
inelastic scattering events which destroy  the phase coherence.
The average distance of inelastic scattering events can be identified
with $L_\phi$.

\subsection{Universal Conductance Fluctuations}\label{subuni}
Measurements of conductance in mesoscopic systems that have been
performed since the  80's,  showed
sample specific, reproducible, statistical fluctuations when
system parameters  like Fermi energy,  magnetic  field or disorder
configurations were changed \cite{Was92}.
In Figure~4 the results of a recent experiment
are shown. Gold-nanowires of very high purity with 
 cross-section $30^2$~nm$^2$ and lengths
between
$400$ and $1000$~nm  were used. The conductance $G$ was measured as a
function of applied magnetic field below $16$~T in steps of  $1$~mT
at $60$~mK. 
Although the average conductance varies from $\approx 3000$~$e^2/h$
(for $L=400$~nm) to $\approx 1400 e^2/h$ (for $L=1000$~nm),
the variance of the conductance stayed constant and was $\approx 0.1
(e^2/h)^2$. 
This phenomenon is referred to as {\em universal conductance fluctuations}
(UCF). In a mesoscopic conductor with high conductance (in atomic
units $e^2/h$) the conductance fluctuations are  reproducible 
{\em  finger-prints}  of the conductor. However, the variance turns out to
be approximately a constant
\begin{equation}\label{UCF1}
	{\rm var} (G) = {\rm const.} \sim {\cal O} (1) \times (e^2/h)^2\, .
\end{equation}
Two aspects of UCF are interesting: reproducibility and universality.
That the fluctuations are
 reproducible
  can be understood again by a semiclassical
description of conducting electrons.
Consider electrons moving along classical paths between elastic
collisions, but having a quantum
mechanical phase attached. Since the phase information is not lost
over distances of the phase coherence length (mesoscopic regime) 
a sample specific interference pattern arises. The interference
pattern will be altered e.g. by applying a magnetic field.
This gives rise to reproducible conductance fluctuations.
The fact that the order of magnitude
of these fluctuations turns out to be independent of the average
conductance
and does only depend weakly on dimensionality and symmetry properties
of the conductor is, however,
 not easy to understand without having a transport
theory
of mesoscopic conductors. A striking consequence of the phenomenon is 
that mesoscopic conductors can not be characterized by an average conductivity
being a material constant, even if the system size $L<L_\phi$ becomes very
large.
Conductivity $\sigma$ is related to conductance by Ohm's law. For a
cubic
geometry the conductance behaves as
$G = \sigma L^{d-2}$. For UCF  the relative fluctuations of
conductance are $\delta G/G \propto L^{2-d}$ which is in contrast to
the classical behavior $\delta G/G \propto L^{-d/2} $. Thus, even in 3D
metallic systems the relative fluctuations are much stronger than
$1/\sqrt{\rm volume}$. 
\subsection{Weak And Strong Localization}\label{subwea}
Early measurements of conductivity of amorphous Silicon \cite{Bey74}
showed that, at low enough temperatures, the temperature behavior 
of conductivity follows Mott's $T^{-1/4}$-law,
\begin{equation}\label{Loc1}
	\sigma(T) =\sigma_0 \exp \LK -(T_0/T)^{1/(d+1)}\RK \, \;\;\;
	{\rm for}\;\; 
	d=3\, ,
\end{equation}
over two orders of magnitude with some constants $T_0$ and $\sigma_0$
(see Fig.~5).
This 
 behavior of the conductivity could be understood when assuming
that the transport is mediated by phonon assisted hopping processes
between localized states \cite{Mot79}.
 Localized states are spatially
localized to 
some finite volume  
within the conductor.   The finite volumes are characterized by a
 localization radius $\xi_r$. The phonon assisted hopping is possible because
localized states  can be energetically 
close to each other if the corresponding localization centers are
separated by distances much larger than the localization radius.  
The  observation of Mott's law 
made it clear that electronic states can indeed be 
localized
by disorder which has been pointed out by Anderson \cite{And58}
much earlier. The effect is referred to as {\em strong localization}
since the localization radius is smaller than system size.

A further experimental hint  to  localization came from
 weakly disordered metallic films at low temperatures.
They showed a logarithmic increase of resistance when the temperature
is decreased (see Fig.~6). 
Although the system behaves metallic with large conductance
the conductance was lower than its classical value.
The effect was interpreted
as being due to quantum interference corrections to classical
transport and is referred to as {\em weak localization}.
In weak localization the localization length is larger than system
size but the effect is interpreted as a precursor of strong
localization, i.e.~under further increase of system size (while still
being mesoscopic, i.e~$L<L_\phi$) the system would show strong
localization. This is, of course, in practice very hard to realize
when localization lengths are larger than microns.

The weak localization effect (predicted by Gor'kov {\em et al.} \cite{Gor79})
can be explained by the phenomenon
of {\em enhanced backscattering} due to quantum interference.
 It relies on the  fact
that the quantum
coherent
probability of the sum of two probability amplitudes $|A+ A|^2= 4|A|^2$ is 
 two times larger than the incoherent sum  of the corresponding
probabilities, $|A|^2+|A|^2=2|A|^2$.
Consider the chance for an electron traveling in a phase coherent
conductor (see Fig.~1)
to return back to some   point within the conductor. In
Fig.~7
a possible path is drawn for such a path (straight line).
 If the motion is time reversal symmetric also the time reversed path
(broken line)
is  possible. The amplitudes of both paths add coherently in a semiclassical
calculation of the return
probability. Thus, the quantum mechanical return probability
is twice the classical value. Without explicitly calculating
the conductivity it is obvious from this observation that the conductivity
must be reduced as compared to its classical value. 

 The coherence is limited by the
phase coherence length $L_\phi$  which varies as some inverse power of
temperature, $L_\phi(T)\propto T^{-p}$
For system sizes $L$ {\em  larger} than the phase coherence length $L_\phi(T)$
the logarithmic corrections to the classical conductivity could be
interpreted as being due to the quantity $-\ln L_\phi/l_e$ where $l_e$
is the average elastic mean free path.

Magnetic fields destroy the time reversal symmetry of
the electronic motion and thus the weak localization effect
vanishes (see upper set of Fig.~8)
A very interesting effect is due to spin-orbit scattering processes
that are time reversal symmetric. In this case the time reversed 
path, however,
gives rise to destructive interference in the return probability.
As a result 
the sign of the conductance corrections is reversed 
(see
lower set of Fig.~8) and
the enhancement of conductivity above its classical value is denoted as 
{\em  weak anti-localization} (first predicted by Hikami {\em et
al.} \cite{Hik80}). 

Weak localization experiments
are an important tool e.g. to determine the phase coherence length
$L_\phi$ and the spin-orbit scattering length.
For more details we refer to the  review by
Bergmann \cite{Ber84}.

To reach experimentally the regime of strong localization
where the localization radius is less than the system size
$L<L_\phi$ 
 one has to
use strongly disordered systems or work in energy regions with very
low density of states.
Localization-delocalization (LD) transitions have been observed in 
several experiments (for reviews see \cite{Paa91,Sar95}). Figure~9
shows the finding of an experiment by Stupp {\em et al.} \cite{Stu94} 
indicating that the
conductivity
vanishes on approaching a critical charge carrier concentration.
A major problem in comparing theoretical models with the experiments
comes from the fact that in the low density of states regimes Coulomb
interaction between electrons becomes essential, but most theoretical
models are based on an independent electron picture.

The most clear-cut experiment for the occurrence of
LD 
transitions is the quantum Hall effect discovered by von Klitzing
\cite{Kli80}. The quantum Hall effect that occurs in 
two-dimensional electron gases in the presence of strong perpendicular
magnetic fields is characterized by a
step-function like behavior of the Hall conductivity $\sigma_H$
as a function of the so called filling factor (a dimensionless
quantity proportional to carrier concentration and  inverse magnetic field)
and by a vanishing dissipative conductivity $\sigma$ in the Hall
plateau regimes (see Fig.~10; 
Note, that for non-vanishing  Hall conductivity
a vanishing dissipative conductivity causes also a vanishing
longitudinal
resistivity.). The peaks in the dissipative conductivity as a function
of the filling factor have a clear interpretation in terms of
LD  transitions. However, no metallic phase
exists.
All states, except for those at {\em  critical fillings} where
the localization radius is larger than system size, are strongly
localized.
The systems showing the quantum Hall effect (quantum Hall systems)
are well suited to study the properties of critical states.
Recently, in an experiment by Cobden and Kogan \cite{Cob96}
it was possible to extract the whole conductance distribution
at criticality for a  truly mesoscopic quantum Hall system.
The main finding was a conductance 
distribution which is independent of system size
(within the mesoscopic regime),
almost uniform between $0$ and $e^2/h$. This means in
particular
that the fluctuations are of the same order as the average value.
Thus, fluctuations in the conductance can become much stronger than
one could expect from classical transport theory where the relative
fluctuations scale like the square root of the inverse volume.

\section{Basic Physical Quantities}\label{bas}
This section prepares for the discussion of transport theory in mesoscopic
electron systems. 
We 
introduce the basic physical quantities of disordered
mesoscopic systems for which the phase coherence length is
larger than the  system size. 
They are introduced here on a phenomenological level as response quantities to
applied fields. The modeling of mesoscopic systems and the tools
 to calculate these quantities is the subject of the following 
Sect.~\ref{mod} (and the appendix~\ref{bri}).
The basic physical quantities  
 are  the thermodynamic density of states and the transport
quantities resistance, conductance and
diffusion. Related physical scales are thermodynamic and kinematic scales
like level spacing, Fermi energy, Fermi wavelength, magnetic length
 and transport
scales like Thouless energy, elastic mean free path
  and the localization length. 
\subsection{Linear Response Quantities}\label{sublin}
The thermodynamic density of
states 
\begin{equation}
	\rho(\mu) = \frac{\partial n}{\partial \mu}\label{2.2}
\end{equation}
determines  the change of the global  particle density $n$ 
with respect to
a change in the chemical potential $\mu$.
The local thermodynamic density of states $\rho(\r;\mu)$
is defined similarly when  replacing $n$ by the local particle density
$n(\r)$.

In charge transport one considers a stationary though non-equilibrium
situation.
Global transport quantities are the resistances in a multi-probe setup
\begin{equation}
	{\cal R}_{kl,mn}=\frac{U_m-U_n}{I}\, ,\;\;\; I=I_k=-I_l\, ,\;\;\;
	I_{i\not=\LB k,l\RB}=0 
				\label{2.3}
\end{equation}
where probes are characterized by electro-chemical potentials
$U_i$. The current $I$  is driven from the current source
(probe $k$)
to the current sink contact (probe $l$) (cf.~Fig.~11). 

The  resistances ${\cal R}_{kl,mn}$ are determined by 
 the 
 conductance coefficients $G_{kn}$ \cite{Buet86}
\begin{equation}
I_k = \sum_n G_{kn}(U_k-U_n)\, .
				\label{2.4}
\end{equation}
With the condition of  total current conservation  $\sum_k I_k =0$ the
conductance coefficients $G_{kn}$  fulfill the relations 
\footnote{
Equations (\ref{2.3},\ref{2.4}) can also be considered for time
dependent transport 
 with the quantities $U_m$, $I_k$ being the Fourier components for
frequency $\omega$. In that case ${\cal R}_{kl,mn}$ and $G_{km}$ will
depend
on $\omega$, too. The total current conservation and Eq.(\ref{2.5})
will still be valid as long as electric fields which are generated 
when applying voltages to the system (including the coherent probe
region)
 have their sources and sinks
within the system (see e.g.~\cite{Buet95}).
}
\begin{equation}
	\sum_k G_{kn}=\sum_n G_{kn}=0\, .
				\label{2.5}
\end{equation}

In a two-probe geometry the transport is determined by the two-probe
 conductance
$G$ (Ohm's law)
\begin{equation}
		G=I/U =R^{-1}
	  			\label{2.6}
\end{equation}
where $U=U_1-U_2$ and $G=G_{12}$ and $R={\cal R}_{12,12}$. 

A full description of the local charge transport is possible with the help
of  the
{\em conductivity tensor}, $\sigma_{\mu \nu}(\r ,\rp;\w)$, 
\begin{equation}
	{j}_{\mu}(\r )= \int d^dr'\, \sum_\nu \sigma_{\mu \nu}(\r ,\rp;\w)
	E_{\nu} (\rp )	\label{2.7}
\end{equation}
relating local electric field, $\E(\r)$, to the local current density,
$\j(\r)$. Both can here be  considered as 
the Fourier components for frequency $\omega$
 of the corresponding time-dependent quantities ($\sim e^{-i\w t})$.

Local currents can be also  generated by charge density gradients
under conditions of local equilibrium, i.e. the chemical potential $\mu$
is a function of position.
The relevant transport coefficient in this case is the 
{\em diffusion function},
$D_{\mu\nu}(\r,\rp;\w)$,
\begin{equation}
	j_\mu(\r)= -\sum_\nu \int d^dr'\,
	D_{\mu\nu}(\r,\rp;\w) \frac{\partial}{\partial
	r'_\nu}
 	q(\rp)
	\label{2.10}
\end{equation}
 This name is
justified
since the continuity equation between current and charge density, $q(\r)$,
 ${\rm div}\; \j (\r) = i\omega q(\r)$ tells that  $q(\r)$ obeys a non-local
diffusion equation 
\begin{equation}
	-i\w q(\r)= \sum_{\mu\nu} \int d^dr'\,
	\frac{\partial}{\partial r_\mu}
	 D_{\mu\nu} (\r,\rp;\w) \frac{\partial}{\partial r'_\nu} q(\rp)
		\label{2.11}
\end{equation}
Under the conditions of local equilibrium the charge density can be related
to the gradient of the chemical potential via the  thermodynamic
charge response, $\Pi(\r,\rp)= e^2 \delta n(\r)/ \delta \mu(\rp)$,
\begin{equation}
	\frac{\partial}{\partial
	r_\mu} q(\r) = (-e)^{-1} 
	\int d^dr'\, \Pi(\r,\rp) \frac{\partial}{\partial
	r'_\mu} \mu(\rp) \, .
	\label{2.12}
\end{equation}
Since the linear response quantities conductivity and diffusion 
belong to the same system and describe currents in response to local
electric and chemical  potential gradients, respectively, they can be
related by identifying the currents and driving fields. This constitutes
a generalization of the important {\em Einstein relation} between conductivity
and diffusivity,   
\begin{equation}
	\sigma_{\mu\nu}(\r,\rp;\omega) = \int d^dr'' 
	D_{\mu\nu}(\r,{\bf{r}}\, '';\w) \Pi({\bf{r}}\, '',\rp) \
			\label{2.13}
\end{equation}

Equation~(\ref{2.7}) reduces to Ohm's law
for spatially averaged currents and electric fields
\begin{equation}
		j_\mu = \sum_\nu \sigma_{\mu\nu}(\w) E_\nu
			\label{2.14}
\end{equation}
and for the {\em 
global  conductivity tensor}, $\sigma_{\mu\nu}$, Eq.~(\ref{2.12})
reduces to the  global
Einstein relation between conductivity and   diffusion 
\begin{equation}
		\sigma_{\mu\nu}(\w)=
	 e^2 D_{\mu\nu}(\w) \frac{\partial n}{\partial \mu}	\, .
	\label{2.15}
\end{equation}

In the following we will concentrate on the d.c. limit of transport
($\w\to 0$) and on a longitudinal (dissipative) component
 of the conductivity tensor, say $\sigma :=\sigma_{xx}$.
In the case  where off-diagonal components of conductivity	
vanish, e.g. in the absence of magnetic field, the two-probe
conductance $G$
of a system with box geometry (characterized by length $L$ and cross
section
$L_t^{d-1}$, see Fig.~12)
is related to the  conductivity $\sigma$ by
\begin{equation}
		G=\sigma L_t^{d-1}/L		\, .\label{2.16}
\end{equation}
Equation~(\ref{2.16}) is also referred to as Ohm's law.
Whenever off-diagonal components of conductivity are present
Eq.~(\ref{2.16})
refers to the dissipative conductance which can be measured in an
appropriate four-probe setup.
We stress that the global conductivity is treated here as a   spatial average
over the conductivity tensor, $\sigma=\int d^dr \int d^dr'
\sigma(\r,\rp)/(LL_t^{d-1})$, and may depend on $L,L_t$.

\subsection{Relevant Scales}\label{subrel}
The Einstein relation, Eq.~(\ref{2.15}), and Ohm's law,
Eq.~(\ref{2.16})
allow for a very interesting interpretation of the conductance
measured in atomic units $e^2/h$. Defining a diffusion time $t_D$
by $t_D:=L^2/D$ and a quantum mechanically related energy scale
\begin{equation}
		E_{\rm Th}:=h/t_D	\, ,	\label{2.17}
\end{equation}
called Thouless energy, and the thermodynamic level spacing,
$\Delta(\mu)$,
by
\begin{equation}
		(LL_t^{d-1})\frac{\partial n}{\partial \mu}=:
		(\Delta(\mu))^{-1}
			\label{2.18}
\end{equation}
we obtain the Thouless formula \cite{Thou74} 
for the dissipative conductance, $G=(e^2/h)g$,
\begin{equation}
			g= E_{\rm Th}/\Delta \, .	\label{2.19}
\end{equation}
This means that the dimensionless conductance, $g$, can be expressed
as the 
ratio of two characteristic energy scales of the problem: a transport
related energy scale $E_{\rm Th}$ and a thermodynamically defined
 energy
scale
$\Delta$. As we will see later this  expression will help a lot in
classifying
mesoscopic electron systems.

So far  we have not
introduced any particular
 dynamical properties  of the electron system. In a linear response
treatment as outlined in the appendix~Sect.~\ref{bri} thermodynamic as
well as dynamic response quantities 
can be calculated by appropriate correlation
functions. 
In a first attempt we make a phenomenological ansatz to describe the
global dissipative conductivity $\sigma$  thereby making contact with standard
macroscopic transport theory.

In the absence of magnetic fields and in the presence of static
disorder
one can make the plausible assumption that an electron starting with
velocity ${\bf v}$ will decelerate 
owing to scattering by impurities. This process can be modeled in the
time dependent velocity correlator (see Eq.~(\ref{4.4},\ref{4.7b})
\begin{equation}
	\kubosp{v_\nu}{v_\mu(t)} = v_0^2 e^{-t/\tau}\delta_{\mu\nu}
	\,\label{Dru1}
\end{equation}
where $\tau$ is a phenomenological  momentum relaxation
time  and $v_0^2$ the static velocity
auto-correlator 
for a given direction and serves also as a measure of
the kinetic energy of the initial state. With the help of the Kubo
relation, Eq.~(\ref{4.10}),
 this correlator is found to be
\begin{equation}
	v_0^2=\frac{N}{\beta m} \label{Dru2}
\end{equation}
($N$ being the number of electrons, $\b$ the inverse temperature
$1/k_BT$ and $m$ being the electrons mass).
The  global conductivity is proportional to the Laplace transform of 
the velocity correlator (see Eq.~(\ref{4.14b})) which yields the Drude formula
for the d.c. conductivity
\begin{equation}
	\sigma_{xx}=\frac{e^2 n\tau}{m}\, ,\;\;\; \sigma_{yx}=0\, .\label{Dru3}
\end{equation}
For
non-interacting electrons, at $T=0$, Eq.~(\ref{Dru3}) can be interpreted as
\begin{equation}
	\sigma_{xx}(E_{\rm F})= e^2
n(E_{\rm F})\tau(E_{\rm F})/m 
=e^2\rho(E_{\rm F})\frac{v_F^2\tau(E_{\rm F})}{d} \label{Dru4} 
\end{equation}
where $E_{\rm F}$ is the Fermi energy, $v_F=\sqrt{(2/m)E_{\rm F}}$ 
 the Fermi velocity  and
$\rho(E_{\rm F}) $ the density of states.

In a finite system of volume $L^d$
the density of states defines the  level spacing
\begin{equation}
		\Delta = (\rho L^d)^{-1}\label{Dru4a}
\end{equation}
which is the smallest quantum kinematic energy scale of the 
system.

The Fermi velocity sets a quantum kinematic length scale, the {\em  Fermi
wavelength}
\begin{equation}
	 \lambda_F=\frac{h}{v_F m}\label{Dru4b}
\end{equation}
The momentum relaxation time is equivalent to an elastic
 {\em  mean free path}, $l_e$,
that describes a typical distance between elastic collisions 
\begin{equation}
	l_e=v_F\tau\,.\label{Dru5}
\end{equation}
The corresponding diffusion constant follows from the Einstein
 relation, Eq.(\ref{2.15}),
\begin{equation}
		D=\frac{v_Fl_e}{d} \, .\label{Dru6}
\end{equation} 
Assuming that in a large but finite system the diffusion constant
is indeed a constant the Thouless energy, Eq.~(\ref{2.17}), and the
conductance, Eq.~(\ref{2.18}),
 can be determined as
\begin{equation}
		E_{\rm Th}=\frac{hD}{L^2} \,, \;\; g= \frac{S_d}{d}\LK\frac{
		L}{\lambda_F}\RK^{d-2} \frac{ l_e }{\lambda_F } 
		 .\label{Dru7}
\end{equation}
Here $S_d$ is a geometrical number, the surface area 
 of  the unit sphere in $d$ dimensions, e.g. $2\pi$ in 2D.
The situation with $l_e\gg \lambda_F$ corresponds to large $g$ and
 will be denoted as weak scattering case. 

Within macroscopic transport theory the transport is determined by
material constants, e.g. the diffusion constant, and the basic task
is to calculate them from microscopic models. The situation changes when
the system becomes smaller than  
the phase coherence length $L_{\phi}$.

Consider  a given conducting system with fixed geometry and at fixed
temperature. Imagine that we can pierce a hole in the conductor and put a
magnetic flux through it.
As quantum mechanics tells, such flux can only influence the phase
of the electrons wave functions. In case that one observes
periodic oscillations in the resistances  when changing the flux over
several
flux quanta, $h/e$, (electronic Aharonov-Bohm effect, see Sect.\ref{subele})
the electronic system is  
mesoscopic, i.e.
the electronic wave functions are phase coherent over the entire
conducting
system. As soon as the charge carrying wave packets experience
scattering processes with dynamical degrees of freedom, e.g. phonons, that
can dissipate energy from the electronic system the phase coherence
will be destroyed  and the electronic Aharonov-Bohm effect gets lost.
It is worth noticing that elastic scattering cannot
destroy the phase coherence. 
Thus, the criterion for an electron system to be mesoscopic is set by
the {\em  phase coherence length} $L_\phi$ being larger than the system
size
$L$,
\begin{equation}
		L_\phi \geq  L \, .		\label{2.20}
\end{equation}
 Here $L_\phi$ can be considered as a typical distance before the
charge carrying states experience an inelastic scattering event.
It is evident that $L_\phi$ will usually decrease with increasing
temperature.
In practice one has to go below liquid Helium temperatures and use 
micro-structured
electron systems to reach the mesoscopic regime. 

The main consequence of the phase coherence in mesoscopic conductors
is the fact that quantum interference  plays a dominant role for
physical
quantities to be observed. As we have already seen in Sec.~\ref{mes}
it can cause
two important effects: unexpectedly large  fluctuations and localization.

The first effect can be qualitatively
described as follows. A wave packet entering
the
mesoscopic conductor will be scattered by some elastic
collisions. After a while it will form a complicated interference
pattern.
If one changes an external parameter only
slightly, e.g. magnetic field,
one
has to be aware of drastic effects on the measured physical quantity
since   the interference pattern changes. A similar effect can be
caused by a slight modification in the arrangement of elastic
scatterers.
At least, one cannot expect that  a measured
physical quantity is close
to  an average value obtained   from a large number of
 measurements under slightly  different conditions.
Instead, the quantities will show  fluctuations which can become
unexpectedly large. However, such fluctuations are reproducible.
The magnitude of mesoscopic  fluctuation effects will depend on system
specific  parameters. This will be the topic of Secs.~\ref{con},\ref{starea},\ref{statra}.
What can be learned by this reasoning is that  a characterization of
a mesoscopic sample is incomplete unless the whole distribution function of
the  physical quantity of interest is obtained. It is no longer
possible to describe  it by material constants (see
Sect.~\ref{subuni}).
  This insight forms
 the motivation to ask
for the statistical properties of mesoscopic electron systems.

The second effect, localization, was 
pointed out by Anderson in a by now famous paper \cite{And58}.
He demonstrated  in a model calculation that the conductance of a phase
coherent disordered
(mesoscopic) system can decrease exponentially with increasing system
size. Denoting $L$ as the linear system size, the conductance can
scale as
\begin{equation}
			g(L)\propto \exp (-2L/\xi)	\label{2.21}
\end{equation}
where $\xi$ is called the {\em  localization length}\footnote{The 
factor of $2$ is due to conventions which will be
discussed later.}.
The origin of this phenomenon lies in the multiple interference
of the electrons wave packet after several elastic collisions
that can lead to destructive interference such that the wave packet 
is bound to a finite region, the localization volume,
 with exponentially decreasing amplitude outside of this region. 
For system sizes that are several times larger than the localization length
the conductor actually behaves as an insulator.
The effect becomes even more interesting since it is possible that one system
can change from the insulating, localized, phase to a conducting, delocalized,
phase when parameters are changed continuously. Such parameters can be the
degree of disorder, electron density or 
external fields like pressure, electric or magnetic field
(see Sect.~\ref{subwea}). Thus,
mesoscopic systems can undergo a disorder induced metal-insulator
transition which we called  {localization-delocalization
transition} (LD transition). 
\section{Modeling of Mesoscopic Systems}\label{mod}
In this section we introduce model systems appropriate
for disordered mesoscopic systems and describe how to
obtain relevant physical quantities.
 Since we do not want to describe
coupling
of electrons to dynamic degrees of freedom that can dissipate energy
from the electronic system we ignore from the very beginning
electron-phonon
interactions and restrict ourself to zero temperature, unless
otherwise stated. In addition, we will  treat electrons mainly
as independent additive systems characterized by effective mass, only.
This is a drastic restriction and it is in general {\em  not}
justified to incorporate electron-electron interactions into such
effective
one-particle models. However, with  one-particle models one can already
 gain some insight into the mechanism of LD transitions and
fluctuation effects. The interplay of disorder and electron-electron
interaction forms a challenging research field which has still to be
developed. Furthermore,
internal degrees of freedom of the electrons will be ignored to a
large extent, leading to models of non-interacting spin-less fermions in the
presence of static disorder potentials.

\subsection{Hamiltonian Modeling}\label{subham}
Independent electrons of effective mass $m$ moving in a
random one-particle potential $V(\r)$ are described by
the Hamiltonian
\begin{equation}
		H=\frac{(\p+e\A)^2}{2m}+V(\r)		\label{3.1}
\end{equation}
where  the kinetic momentum $\p+e\A$ 
may include
 magnetic fields $\B={\rm curl}~\A$.  Randomness
of the disorder potential  has to be specified by a distribution
function. It is often assumed that 
the essential parameters of the distribution are contained in the
first and second cumulant, i.e.~a Gaussian
distribution, fixed by the 
 mean value $\BRA V(\r)\KET$ and a finite range correlation function
$\BRA V(\r)V(\rp)\KET$, will be general enough to cover the essential physics.
It is reasonable to assume that neither specific  points, nor specific
directions are preferred by the whole ensemble of potentials. Then one
can characterize the ensemble by two  parameters, 
a potential correlation strength, $V_0$, 
and a potential correlation length, $l_V$,
\begin{eqnarray}
		\BRA V(\r)\KET	&\equiv& 0 	\label{3.2a}\\
		\BRA V(\r)V(\rp)\KET &=& V_0^2\exp
		(-|\r-\rp|^2/l_V^2)\label{3.2b} 
\end{eqnarray}
In case that $l_V$ can be considered to be smaller than other relevant
microscopic scales, it can be set to zero, leading to a Gaussian white-noise
potential 
\begin{equation}
		\BRA V(\r)V(\rp)\KET =\gamma^2
		\delta(\r-\rp)		\label{3.3}
\end{equation}
where $\gamma^2$ corresponds to $(\sqrt{\pi}l_V)^dV_0^2$.
Within the Gaussian white-noise potential model the disorder is characterized
by one parameter, $\gamma^2$, of dimension $[{\rm Energy}^2\, {\rm Volume}]$.
We will see later, Eq.~(\ref{3.26}),
 how this parameter corresponds to an  elastic 
 mean free path, $l_e$. 

A homogeneous magnetic field
 pointing in a fixed direction, say $\B=(0,0,B)$, introduces a quantum
 mechanical length scale, the {\em  magnetic length $l_B$} defined by the
 size
of an area which is penetrated
by one flux quantum $h/e$
\begin{equation}
		l_B=\sqrt{\hbar/(eB)} \, .		\label{3.4}
\end{equation}
The magnetic length comes along with chirality due to the  axial vector
character of magnetic field.
As long as the magnetic length is larger than the actual system size,
the effect of the magnetic field on the classical paths of electrons
can be neglected and it will only influence  the phases of wave functions.
 As soon as the magnetic length is of the order of the mean free path
$l_B\sim l_e$ the dynamics is, to a large extent,
 determined by the magnetic field
and we refer to this situation as strong magnetic field case.

A simple modeling to account for possible spin-orbit scattering
events
that can flip the electron's spin is given by the addition of 
\begin{equation}
			\S\cdot({\rm grad}\, U_{\rm SO} \times \p)
	\label{3.5}
\end{equation}
to the Hamiltonian in Eq.~(\ref{3.1}) \cite{Efe83,Fasdis92}.
 Here $\S$ is the spin operator
and
$U_{\rm SO}$ denotes an
average spin-flip potential the gradient of which is taken to be a
constant into all directions
\begin{equation}
		{\rm grad}\, U_{\rm SO}=(\sigma_U,\sigma_U,\sigma_U)\,
		. \label{3.6}
\end{equation}
Depending on the kinematic properties of the electron's state
the  inverse of $\sigma_U$ gives rise to a distance, the
{\em  spin-orbit
scattering length}
$l_{SO}$,
after which the electrons  spin relaxes on average. 

So far, we have considered a Hamiltonian modeling in a continuous real
space representation. The randomness is characterized completely by a
diagonal potential, Eqs.~(\ref{3.2a},\ref{3.2b},\ref{3.3}). It is
instructive
to consider the same model in a discrete matrix representation and to
ask for the resulting statistical properties. 
In a finite cubic system (with volume $L^d$) with periodic boundary
conditions
we can use the momentum representation, $\p \left.\mid\k\KET = \hbar \k
\left.\mid\k\KET$,
 which diagonalizes the kinetic
energy 
in Eq.~(\ref{3.1})($\B=0$)
\begin{equation}
	H_{\k\kp}= \frac{(\hbar\k)^2}{2m}\delta_{\k\kp}
			+ \tilde{V}(\k-\kp) 			\label{3.7}
\end{equation}
where $\tilde{V}(\k)$ is the Fourier transform of $V(\r)$.
The Gaussian white noise results in the correlation
\begin{equation}
			\BRA\tilde{V}(\k_1-\k_2)\tilde{V}(\k_3-\k_4)\KET=
		L^{-d}\gamma^2\delta_{\k_1-\k_2,\k_3-\k_4}  \,.        	\label{3.8}
\end{equation}
Thus, the random matrix $H_{\k\kp}$ has a diagonal deterministic part
and a random  part the average value of which vanishes, but
with strong statistical correlations among those elements which
fulfill momentum conservation. What can be seen here is that
the statistical properties of a representing matrix for the
Hamiltonian do, in general, depend on the chosen basis in Hilbert space.

The matrix $H_{\k\kp}$ can be taken to be finite dimensional if one
restricts the wave lengths to be larger than a convenient microscopic
scale, $a$. Then the matrix dimension is $\sim (L/a)^d$ and the system can
be studied on a computer. 

A direct modeling of disordered mesoscopic systems 
by a finite dimensional Hamiltonian matrix is given by
a tight-binding version
of the Gaussian white-noise potential and was introduced in the work
of  Anderson \cite{And58}. This Anderson model is defined on a cubic
lattice
with lattice constant $a$ 
\begin{equation}
	H=\sum_{\m} \varepsilon_{\m} \left| \m \KET\BRA \m
	\right|
	+ \sum_{\BRA \m , \m'\KET } t_{\m ,\m'} \left| \m \KET\BRA
	\m' \right| 
	\, ,
	\label{3.9}
\end{equation}
where $\left|\m\KET$ denote tight-binding states situated at
lattice point $\m$ and $\BRA \m , \m'\KET $ means nearest neighbors only.
The site energies $\varepsilon_{\m}$ correspond to the (random)
potential energy and the (deterministic)
hopping matrix elements $t_{\m ,\m'}$ to the
kinetic energy. 
Again, the
 matrix dimension is $\sim (L/a)^d$ for a cubic system of linear size
$L$. However, the matrix is sparse since the nearest neighbor
condition
leaves  most elements  vanishing.

In the absence of magnetic field the  hopping strength
is
taken to be equal for all nearest neighbors (with
coordination number $Z$) and set to $1$. This
defines
the kinetic energy band to be of width $2Z$.

A homogeneous  magnetic field $\B=(0,0,B)$ can be 
 included in the kinetic energy
by the Peierls substitution \cite{Pei33}. In the Landau gauge,
$\A=(0,-Bx,0)$, it reads in 2D
\begin{equation}
	t_{\m ,\m'}(B)  =  t_{\m ,\m'}(0)
 	\LB e^{\pm 2\pi i\alpha
 	m_x}\delta_{m'_x,m_x}\delta_{m'_y,m_y\pm 1}\RB\, .
	\label{3.10}
\end{equation}
Here $0< \alpha < 1$ is the number of flux quanta $h/e$ per unit cell.
In contrast to continuum models a feature of commensurability appears.
Requiring commensurability between the lattice
constant and the magnetic length restricts $\alpha$ to take rational
values only 
\begin{equation}
	\alpha=p/q\, .
\label{3.11}
\end{equation}
Furthermore, in this model the magnetic length is restricted to be
larger
than the lattice constant since the effect of magnetic field is
assumed to be weak on the size of a unit lattice cell, i.e.  only its
influence
on phases is taken into account by the Peierls substitution.

The statistical properties of the model are fixed once a
distribution
of the independent diagonal energies is chosen. A convenient choice
is  a box distribution on the interval $[-V_0,V_0]$
with correlation
\begin{equation}
		\BRA \varepsilon_\m\varepsilon_{\m'} \KET = \frac{V_0^2}{3}
			\delta_{\m\m'}
			\label{3.12}
\end{equation}
The parameter $V_0^2/3$ can be identified with $\gamma^2/(\sqrt{\pi}a)^d$
($\gamma$ taken from
the white-noise
model of Eq.~(\ref{3.3})) and determines the mean
free path $l_e$ (see Eq.~(\ref{3.26})).
We mention that the inclusion of spin-orbit scattering into the 
Anderson model is also possible \cite{Eva87,And89,Fas91} and gives rise
to a spin-orbit scattering length $l_{SO}$. 

Equations~(\ref{3.1},\ref{3.9})  correspond to the most prominent models of
disordered mesoscopic electron systems. In the following we want to
discuss, how one can use these models to calculate some of the basic physical
quantities. 
\subsection{Physical Quantities by Green's Function}\label{subphy}
As outlined in the appendix, for non-interacting
 Fermion models we can concentrate
on zero temperature (see Eq.~(\ref{4.22})) and only need the resolvent of the
Hamiltonian
\begin{equation}
		G^{\pm}(E)=(E^{\pm}  -H)^{-1}\, ,\;\;E^{\pm}=E\pm
			i\epsilon\, ,\;\;  \epsilon \to +0
			\label{3.13}
\end{equation} 
to calculate the basic physical quantities (Eqs.~(\ref{4.19} --
\ref{4.27})). Choosing  a convenient basis one has to perform
traces involving $G^{\pm}(E)$.
The knowledge of $G^{\pm}(E)$ is equivalent to finding the
eigenvalues $\varepsilon_\a$ and eigenvectors $\left.\mid \psi_\a\KET$
of $H$
since 
\begin{equation}
		G^{\pm}(E)=\sum_\a
	\frac{\left.\mid \psi_\a\KET \BRA \psi_\a \right.\mid}
	{E^{\pm}  -\varepsilon_\a }
			\label{3.14}
\end{equation} 

The dynamic content of the resolvent is expressed by its relation to
the time evolution operator $U(t)=e^{-iHt/\h}$
\begin{equation}
		i\h G^{\pm}(E)=\int\limits_0^{\pm \infty} dt\,
e^{(iE\mp\epsilon )t/\h}U(t)\, .\label{3.15} 
\end{equation}
With this it can be concluded that the long time average of the
time dependent probability, $P(\r,\rp;t)$,
to find an electron  at point $\rp$  
that was created at point $\r$ 
 is related to the Green's function,
$G^{\pm}(\r,\rp;E)=\BRA \rp \mid G^{\pm}(E)\mid \r \KET$, by
\begin{equation}
	\overline{P(\r,\rp;t)}=\lim_{\epsilon \to +0}
	\frac{\epsilon}{2\pi}
	\int dE\, \mid \GP(\r,\rp;E)\mid^2 \, .\label{3.16}
\end{equation}
That leads to the important interpretation that 
$(\epsilon/2\pi) \mid \GP(\r,\rp;E)\mid^2$
 defines the long time transition probability from $\r$
to $\rp$ for a given energy $E$.

The local density of states (LDOS) at $T=0$, 
$\rho(\r;\EF)=\BRA \r\mid  \delta (\EF -H)\mid \r\KET$ can be calculated by
(see Eq.~(\ref{4.17}))
\begin{equation}
	\rho(\r;\EF)= -\pi^{-1} \imag \GP (\r,\r;\EF)=\frac{\epsilon}{\pi}
 \int d^dr'\mid \GP(\r,\rp;\EF)\mid^2 =\sum_\a \mid \psi_\a(\r) \mid^2
 \delta(\EF-\varepsilon_\a)  \label{3.17} 
\end{equation}
and the global (averaged over volume $L^d$) density of states (DOS) by
\begin{equation}
	\rho(\EF)= L^{-d} \sum_\a \delta 
	(\EF-\varepsilon_\a) \,.\label{3.18}
\end{equation}
For a finite closed system the DOS consists of isolated
$\delta$ peaks  which have to be smeared out by attaching a finite
width of the scale of the level 
spacing   
to the peaks   or by averaging over the  ensemble of disorder realizations. 
The average level spacing $\Delta$ at $\EF$ can thus be obtained by
\begin{equation}
	\Delta= -\pi \BRA \imag \Tr \GP\KET^{-1}\label{3.19}
\end{equation}

Furthermore, it is of conceptual importance to mention that the
Green's function $G(\r,\rp;E)$ can be represented by a path integral, since
\begin{equation}
		\BRA \rp\left.\mid U(t)\mid\right.\r\KET
		=\int d[\r(t)] \exp (i/\h)S[\r(t)] \, ,\label{3.19b}
\end{equation} 
where the integration runs over all paths $\r(t)$ connecting initial ($\r$)
and final ($\rp$)  points and $S[\r(t)]$ denotes the classical action
associated to the Hamiltonian of the electron. The stationary
path  of $S[\r(t)]$ yields the solution of the
 classical equation of motion.

As a first instructive example we  calculate the ensemble average
of the Green's function in the continuous model without magnetic field,
 using the momentum representation, Eqs.~(\ref{3.7},\ref{3.8}).
 Since we cannot solve the Schr\"odinger equation for arbitrary
disorder
potential we rely on a perturbative analysis, 
\begin{equation}
	\GP  = {\GP}^{0}\sum_{n=0}^{\infty} (V{\GP}^{0})^{n}
	\label{3.20}\, ,
\end{equation}
where $ \BRA \kp \right| {G^{\pm}}^{0}(E)\left| \k\KET = (E^{\pm}-(\h
\k)^2/2m)^{-1}\delta_{\k\kp}$ is the Green's function of the
kinetic energy term. Due to the property Eq.~(\ref{3.8})
the ensemble averaged Green's function is diagonal in momentum
representation, too. It can be written in the form
\begin{equation}
	\BRA \k \right| G^{\pm}(E)\left|\k\KET=\LK E^{\pm}
	-(\h\k)^2/2m -\Sigma^{\pm}(\k)\RK^{-1}\label{3.21}
\end{equation}
where $\Sigma^{\pm}(\k)$ is called the self-energy the imaginary
part of which defines an inverse time scale which can be interpreted
as an elastic mean free time $\tau$
\begin{equation}
	\frac{\h}{2\tau}=\imag \Sigma^{+}(\k)\, .\label{3.22}
\end{equation}
We mention that, in general, this $\tau$ must not be
identical to the average momentum relaxation time due to random collisions.
 
The self-energy can now be calculated order by order in perturbation
theory
(for details cf. \cite{Abr88}).
For weak scattering, i.e. for $l_e\gg \lambda_F$,  it turns out to be 
sufficient to take the first non-vanishing contribution which yields
\begin{equation}
	\Sigma^{+}(\k)= L^{-d}{\gamma^2} \sum_\k \BRA \k \right|
	{\GP}^{0} \left|\k\KET \label{3.23}
\end{equation}
and improve it  by replacing ${\GP}^{0}$ with the full resolvent
$\GP$ in Eq.~(\ref{3.23}) making the approximation for the self energy
self-consistent. It leads to
\begin{equation}
		\tau=\frac{\h}{2\pi \gamma^2 \nu}\label{3.24}
\end{equation}
where $\nu$ is the average DOS which, for Fermi energies far
off
the kinetic energy  band edges, is not changed as compared to
the  DOS  without disorder, i.e.
\begin{equation}
	\nu (\EF) = 
	\frac{mS_d}{(2\pi)^d\h^2}
	\left(\frac{2\pi}{\lambda_F}\right)^{d-2}\label{3.25}
\end{equation}
Consequently, the elastic mean free path is related to  $\gamma$ and
$\lambda_F$ by
\begin{equation}
	l_e=
	v_F\tau=\frac{h^2\h^2\lambda_F^{d-3}}{m^2\gamma^2 S_d}\propto
	\frac{\EF^2\lambda_F^d}{\gamma^2}\lambda_F \, .\label{3.26}
\end{equation}

Calculating the disorder average of the conductivity requires already
the average over a product of two Green's functions (called
two-particle Green's function) which is already a non-trivial task.
Starting from the Kubo formula, Eq.~(\ref{4.23}), for the velocity
correlator, using the momentum representation 
and replacing the matrix elements of the resolvent by their average
values
leads back to the Drude formula with the momentum relaxation time
given by Eq.~(\ref{3.24}). However, this procedure ignores completely
that the average of two Green's functions does not decouple
into a product of averages. In the context of the  path integral
representation for the Green's functions, Eq.~(\ref{3.19b}), 
 it becomes obvious that
 this method cannot cope
with quantum interference effects. Thus, the mean free time as given
by Eq.~(\ref{3.24}) corresponds to a Drude like conductivity when
quantum
interference is suppressed.

The symmetry properties of the disorder ensemble yield a translational
and 
rotational invariant two-particle Green's function	
\begin{equation}
	\BRA|G^{+}(\r ,\rp ,E)|^2\KET =: K (|\r -\rp |, E)\, .\label{3.27}
\end{equation}
With this function the longitudinal conductivity, as given by 
Eq.~(\ref{4.27}), is
\begin{equation}
	\sigma = \frac{2e^2{\epsilon}^2}{h}
	\int d^dr\,   x^2  K(|\r |,\EF )\, .\label{3.28}
\end{equation}
The  diffusion constant
$D(E)$ is defined
 by the long time limit of the square displacement with respect to the
probability distribution $\epsilon K(r,E)$ as
\begin{equation}
	\rho(E)D(E) t(\epsilon) := 
	\epsilon \int d^dr\,
 x^2 K(|\r |,E)\label{3.29}
\end{equation}
where 
$t(\epsilon)=h /2\epsilon$
is a growing time scale as $\epsilon$ is sent to $+0$.
Thus,
Eq.~(\ref{3.28}) demonstrates the validity of the Einstein relation.

Furthermore, it allows for a first  definition of a localization length
by means of the Green's function, i.e. if $\epsilon K(|\r |,E)$ stays
finite
for arbitrary finite $\r$, but falls off as
\begin{equation}
	\epsilon K(|\r |,E)	\propto \exp(-2|\r|/\xi_0)\label{3.30}
\end{equation}	
the average conductance of a large volume $L^d$ will show the
localization
phenomenon introduced in Eq.~(\ref{2.21}) with $\xi_0$ defining a 
localization
length. Note, that this length is defined with respect to the average
of $|G^{+}|^2$ and must not be identical to
just $1/q$  of  the exponential fall off of  averages of 
powers $|G^{+}|^{2q}$. Nevertheless, the factor of $2$ in
Eq.~(\ref{3.30})
follows the convention to associate the localization length
 with the modulus of the
Green's function.

\subsection{The Statistical Problem}\label{subthe}
We are now in the position to formulate the statistical problem
of calculating physical quantities in mesoscopic electron systems.
The system is described by an ensemble of  Hamiltonian matrices
in a certain matrix representation. The relevant physical quantities
can be obtained from  the Green's function $\GP(\r,\rp;E)$. 
The average value will only determine
the average DOS and the short distance mean free path, however not
the global transport behavior. To account for the latter, the averaged
two-particle Green's function is needed  in order to see localization
effects. Fluctuation effects in global transport  cannot
be obtained on the basis of averaged two-particle Green's functions;
at least four-particle Green's functions have to be considered.
We see that this situation calls for a general approach to the
distribution functions of $\GP(\r,\rp;E)$.
The field theoretic approaches to mesoscopic systems
commonly known as non-linear
sigma-models
are constructed to fulfill this purpose\cite{Weg79,Efe83}.
 We will not enter this
subject because of its technical complexity, but will here focus on
some general considerations. 
   
Owing to Eq.~(\ref{3.14}) the statistics of $\GP(\r,\rp;E)$
is contained in the joint probability distribution
of eigenvalues, $\varepsilon_\a$, and  eigenvectors, $\psi_\a$,
\begin{equation}
	{\cal P} \LK \varepsilon_{\a_1}, \psi_{\a_1},\varepsilon_{\a_2},
	\psi_{\a_2},\varepsilon_{\a_3}, \psi_{\a_3}, \ldots \RK \, .
	\label{3.31}  
\end{equation}
Of course, it seems hopeless to find a general method to determine
this function for any given ensemble of Hamiltonian matrices.

To get some feeling about the nature of the statistical problem
we consider the
 Hamiltonian  in a finite basis 
 $\LB\left| i\KET\RB_{i=1,\ldots,N}$ as hermitean $N\times N$ matrix
$H_{ik}=\BRA i\mid H\mid k\KET$.
The diagonalizing unitary matrix $U\in {\cal U}(N)$
with matrix elements $U_{k\a}$ fulfills
\begin{equation}
		\sum_{ik} U^{\dagger}_{\b i}H_{ik} U_{k\a}=\varepsilon_\a
		\delta_{\a\b} \, .\label{3.32}
\end{equation}
It is related to the amplitude of an eigenstate $\psi_\a$ in the
$\LB\left| k\KET\RB$ basis by
\begin{equation}
		\psi_\a (k):=\BRA k | \psi_\a\KET= U_{k\a} 			\label{3.33}
\end{equation}
We can think of the randomness of $H$ as being controlled
by a large number of independent parameters, e.g. strength and
position of point scatterers in space, or  the on-site
energies in the Anderson model. Each realization of the Hamiltonian
represents one point in this high dimensional parameter space. We can
ask what happens for a  certain stochastic process $H(s)$ 
in this parameter space
where $s$ denotes a fictitious time and each element $H(s)$ is weighted
by the corresponding  
probability. The original $H$ corresponds to
some arbitrary point, say $H(s_0)$.
We can increase $s$ in small steps $\delta s$ which leads to
\begin{equation}
		H(s+\delta s)= H(s) + \delta H (s,\delta s)
  			\label{3.34}
\end{equation}
The corresponding unitary matrix $U(s)$ behaves then as
\begin{equation}
		U(s+\delta s) = U(s) \tilde{U}(s,\delta s)
			\label{3.35}
\end{equation}
where $\tilde{U}(s,\delta s)$ denotes the unitary matrix that diagonalizes the
Hamiltonian $H(s+\delta s)$ with respect to the eigenvector basis of
$H(s)$.
Thus, $U(s)$ evolves in a {\em  multiplicative} manner.
One could think of treating $\tilde{U}(s,\delta s)$ by
perturbation theory with respect to $\delta s$ to derive closed
evolution equations for the probability distribution of $U_{k\a}(s)$ and the
corresponding set of eigenvalues $\varepsilon_\a(s)$. This
has, so far, not been undertaken. However, recently Chalker et
al. \cite{Cha96} 
adopted a related picture to investigate the eigenvalue statistics
 separately by relying on simplifying approximations. 
Here, we only wanted to stress, that the problem of the 
statistics of eigenvectors can be viewed 
as the problem of a random multiplicative process for  unitary matrices.

As to the problem of the statistics of eigenvectors one can easily
imagine two extreme situations.
In the first situation the probability distribution of  $U$ is peaked
at a 
single fixed matrix.
This matrix singles out a certain basis of eigenstates. 
 We  choose  this matrix as the unit matrix, 
such that the  Hamiltonian is   diagonal in the
initial basis
\begin{equation}
		H_{ik}=\varepsilon_i \delta_{ik}
			\, .\label{3.36}
\end{equation}
The
corresponding eigenstates are localized at certain `sites',
i.e.
\begin{equation}
		\psi_\a(k)=\delta_{\a k}			\label{3.37}
\end{equation}
The second extreme situation corresponds to an {\em  isotropic}
distribution
for the unitary matrix. By this we mean that the probability density,
${\cal P}(U)$, 
to find a certain unitary matrix $U$ within the  
volume element,
$d[{\cal U}(N)]$,  
is equal for all elements $U\in{\cal
U}(N)$. The  volume element  itself stays
invariant under the action of group transformations (invariant
measure)
$d[{\cal U}(N)]$, i.e.
it does not single out any particular element. 
The corresponding eigenstates are also isotropically distributed among
all possible eigenstates, no basis is preferred. 

Of course, both extreme situations are not generic ones and we have to
see for which values of physical scales they may appear.
\section{Scattering Matrix modeling}\label{sca}
For mesoscopic conductors with several probes the d.c. conductance coefficients
$G_{km}$ are determined by asymptotic current correlators, Eq.~(\ref{4.15}),
and hence for independent Fermions at $T=0$ by asymptotic Green's functions,
Eq.~(\ref{4.26}). In a geometry where probes
are represented as infinite leads attached to the conductor
the asymptotic Green's functions  determine the matrix of transmission
 amplitudes,$t_{km}^{\a\b}$,  to scatter from channel $\a$ in lead
$k$ to channel $\b$ in lead $m$ and the B\"uttiker formula relates
them to the conductance, Eq.~(\ref{4.27b}), 
\begin{equation}
	G_{k\not=m}(\w =0,T=0)=
	 \frac{e^2}{h}\; \Tr \LB {t}_{km} t^{\dagger}_{km}
	\RB (\EF)\, .\label{3.38}
\end{equation}
This 
formula tells that the dimensionless
conductance, especially in a two-probe geometry,
is the number of effective transmitting modes, 
\begin{equation}
	g= \sum_{\a} T_{\a} \, ,\;\; T_\a = \sum_\b \mid
	t_{\a\b}\mid^2\, ,\;\; 0\leq T_\a \leq 1
	\label{3.39}
\end{equation}
and it is thus of similar conceptual
importance as the Thouless formula, Eq.~(\ref{2.19}).

The transmission matrix can be calculated by asymptotic Green's
functions. These have to be determined under the requirement of
attaching the leads to the conductor.
For this purpose it is advantageous to divide the Hilbert space
into two parts by the 
 projection operator ${ P}$ onto the Hilbert space
of the conductor and its counterpart ${ Q}=1 -{ P}$ that
projects onto the leads. Denoting  projected operators $A$ by
$PAP=:A_{PP},PAQ=:A_{PQ}$, etc.
 the general algebraic solution of the problem to calculate the
projected
Green's function reads
\begin{equation}
	G^{+}_{PP}(E)=(E^{+}- \tilde{H}_{PP}(E))^{-1}
	\label{3.41}
\end{equation}
where the effective (in general energy dependent and non-hermitean), 
Hamiltonian $\tilde{H}_{PP}$
 is defined via the couplings and the Green's
function
of the leads $\tilde{G}^{+}_{QQ}(E):= (E^{+}-H_{QQ})^{-1}$
\begin{equation}
	\tilde{H}_{PP}(E)= H_{PP}+ H_{PQ}\tilde{G}^{+}_{QQ}(E)H_{QP}\, . 			\label{3.42}
\end{equation}
Since
Eq.~(\ref{3.41}) is defined for a finite system and the infinite clean leads
Green's function can be determined analytically for not too
complicated geometries the method of effective Hamiltonian 
 also helps in doing the calculations numerically.  The Green's
function evaluated at the surface between leads and conductor will then 
determine the conductance coefficients.

The B\"uttiker formula points out that the mesoscopic conductor
 has very much in common with optical wave guides where transmission
 probabilities  between incoming and outgoing wave modes are the
 central physical quantity of interest.   It is thus tempting to model
mesoscopic conductors  by  scattering matrices in analogy to  
  optical wave guides.

\subsection{Quasi-One-Dimensional Conductor}\label{subqua}

As a guiding example let us consider a  box shaped system (see
 Fig.~12 with two semi-infinite
 leads attached serving as particle reservoirs (contacts)).
 The cross section $L_t^{d-1}$
is kept fixed while the 
the length $L$ is treated variable. 
Such systems are denoted as quasi-one-dimensional (quasi-1D) conductors.
The  leads are  characterized by discrete branches of energy dispersion
(see~Fig.~13)  due to the finite extension  in transverse
direction. For a given value of the Fermi energy there exist 
$N_c$ quantum states of incoming and outgoing waves (channels).
As a rough estimate one may think of $N_c$ being the number of
lattice points one can put on the cross section of the conductor 
with lattice spacing of half a Fermi wavelength, $N_c\approx (2
 L_t/\lambda_F)^{d-1}$.

A direct consequence of the B\"uttiker formula is the integer
quantization of dimensionless 
conductance in ideal ballistic conductors where no scattering occurs,
or the mean free path $l_e$ is much larger than $L_t,L$.
Then $T_\a=1$ and the conductance displays quantized plateaus as a
function of the Fermi energy (or particle density). Jumps from one
plateau
to another occur whenever a new channel is occupied, i.e. the Fermi
energy
crosses a new branch of the leads energy dispersion. In contrast to
classical physics the conductance of an ideal conductor 
 is not infinite but bounded by the finite number of quantum
channels. The classical behavior can only be recovered for vanishing
Fermi wavelength or, equivalently, for infinite number of channels.
The finite resistance of ideal ballistic conductors can be attributed
to the contacts where the equilibration of the electrons take place.
Therefore,
one may think of a decomposition of the resistance  into
a {\em  contact resistance}, $R_{c}=(e/h^2)1/N_c$, and an 
{\em  intrinsic resistance},
$\hat{R}$, such that $R=R_c+\hat{R}$.
This yields for the corresponding  intrinsic conductance
$
	\hat{g}= \sum_\a T_\a/(1-N_c^{-1}\sum_\a T_\a) 
$,
a result that was 
obtained by Langreth and Abrahams \cite{Lan81}
generalizing the one-channel version of  Landauer's pioneering works
\cite{Lan57,Lan70}.
However, one should keep in mind
that the notion of intrinsic conductance is a theoretical construction
and not a quantity to be measured in a two-probe
experiment. Such experiment corresponds to the conductance as
 described by the
B\"uttiker formula. 
 
In general, the quasi-1D  conductor can  be described by a $2N_c\times 2N_c$
scattering matrix $S$ 
\begin{equation}
		S = \left( \begin{array}{cc} r & t' \\ t & r'  \end{array}
		\right)\label{3.43}
\end{equation}
 connecting incoming and outgoing channels left and right of a scatterer, $t$,
$r$, $t'$ and $r'$ being $N_c\times N_c$ matrices of transmission and
reflection coefficients for scattering from left to right and vice
versa, respectively (see Fig.~14). Here $t$ is precisely the
matrix of transmission amplitudes that appear in the B\"uttiker
formula.
Due to probability-flux conservation  $S$ is unitary,
\begin{equation}
	S^{\dagger}= S^{-1}\, .\label{3.43b}
\end{equation}

To model a long quasi-1D  system one can now add several conductors
in series.
However,
 the total  $S$-matrix   is not multiplicative. Instead of
\begin{equation} \label{3.44}
	\left( \begin{array}{c} O \\ O' \end{array} \right) = 	S
\left( \begin{array}{c} I \\ I' \end{array} \right)
\end{equation}
we are seeking a {\em  transfer matrix} 
${M}$ with the property ${M}_{1+2}={M}_2{M}_1$, i.e.
\begin{equation} \label{3.45}
	\left( \begin{array}{c} O' \\ I' \end{array} \right) = 	{M}
\left( \begin{array}{c} I \\ O \end{array} \right)\, .
\end{equation}
A straightforward  calculation yields
\begin{equation} \label{3.46}
	{M} = \left( \begin{array}{cc} \LK t^{\dagger}\RK^{-1} & r't'^{-1} \\
-t'^{-1}r & t'^{-1} \end{array} \right)\, .
\end{equation}
Making use of probability-flux conservation 
\begin{equation}\label{3.46b}
	M\Sigma_z M^\dagger =
	\Sigma_z\, ,\;\; \Sigma_z=\left( \begin{array}{cc} 1 & 0 \\ 0 & -1  
	\end{array}
		\right)
\end{equation}
where $1$ stands for the $N_c\times N_c$ unit matrix
  we can  rewrite the dimensionless
 two-probe conductance as
\begin{equation}\label{3.47}
	g =
 	\Tr\frac{2}{{M}{M}^\dagger+({M}{M}^\dagger)^{-1}+2} \, .
\end{equation}

A modeling of a mesoscopic quasi-1D 
conductor can now be based directly on the
scattering matrix or, equivalently, on  the transfer matrix.
This can be done by fixing the statistical properties of 
the $S$-matrix corresponding to a small strip of length $\delta L$,
denoted as strip $S$-matrix $S(\delta L)$ and compose the whole
conductor  by putting statistically independent strip $S$-matrices in series. The total $S$-matrix follows then from the
multiplication of the corresponding strip transfer matrices $M(\delta
L)$. The assumption of statistical independence is justified if the
strip length $\delta L$ is larger than the microscopic potential
correlation
length $l_V$ introduced in Sect.~\ref{subrel}. 
Furthermore, this modeling allows for a simple description of the
 mean free path corresponding to the strip $S$-matrix. As long as 
the corresponding 
reflection probabilities $\mid r_{\a\b}\mid^2=:R_{\a\b}$ are
small compared to $1$ the mean free path $l_e$ is large compared to
$\delta L$ and can be defined as follows
\begin{equation}
		\frac{\delta L}{l_e} :=  N_c^{-2} \sum_{\a\b}\BRA
 	R_{\a\b}(\delta L)\KET\, .
 	\label{3.48}
\end{equation} 
The  $S$-matrix fulfills the requirement of unitarity. To model
systems with specific symmetry properties one can impose further
symmetry constraints on $S$. The most important case is that of time
reversal symmetry. This symmetry can be broken by  magnetic fields.
To model systems with magnetic field one faces the problem that the
magnetic length (see Eq.~(\ref{3.4})) 
as a chiral length scale has to be incorporated into
the model. It is not obvious how to do that in the scheme described
here. However, weak magnetic fields only break time reversal
symmetry (the magnetic length being larger than system
size) and  can be simply incorporated in the symmetry properties of $S$.
For example,
time reversal operates by interchanging incoming and outgoing
channels and complex conjugation of amplitudes. In case of time
inversion symmetry the $S$-matrix and transfer matrix $M$ fulfill 
\begin{equation}
		S=S^{\rm T} \, ,\;\; M^{*}=\Sigma_x M
		\Sigma_x\, , 
	\;\; \Sigma_x =\left( \begin{array}{cc} 0
		& 1 \\ 1 & 0  
	\end{array}
		\right) \, .\label{3.49}
\end{equation} 
Internal degrees of freedom, such as spin, can also be incorporated by
taken the corresponding time reversal transformation  into
account. 

So far, the modeling rests on a microscopic length scale, the mean
free path $l_e$, and symmetry properties.
The statistical problem is thus defined by (i) fixing the distribution
of the strip $S$-matrix $S(\delta L)$ and (ii) by applying the
composition law for the corresponding transfer matrices
\begin{equation}
		M(L+\delta L)= M(\delta L) M(L)	\, .\label{3.50}
\end{equation} 
This defines a stochastic multiplicative {\em  process} where the {\em  time
variable} is the system length. This resembles Eq.~(\ref{3.35}).
(iii) The statistical properties of the increment  
\begin{equation}
	\delta M(L,\delta L)=M(L+\delta L) - M(L) \label{3.50a}
\end{equation}
 are known by construction. Following the steps (i)~--~(iii) it
 is possible to construct a stochastic
differential equation (Fokker-Planck equation) for the distribution
function
${\cal P}(M;L)$ (see e.g.~\cite{Mell91}). 
Still, a general solution of this
equation is presently not available and we will not discuss it in
detail.
A related approach, based on simplifying assumptions, will be
discussed in Sect.~\ref{substaqua}.  

The matrices $tt^{\dagger}$ and $MM^{\dagger}$ occurring in the
conductance
formulas, Eqs.~(\ref{3.38}, \ref{3.47}), are hermitean and, thus, can
be diagonalized. The positive eigenvalues are denoted as $0\leq {\cal T}_i
\leq 1$ (for $tt^{\dagger}$) and as $0\leq e^{\nu_i} < \infty$ (for
$MM^{\dagger}$). The eigenvalues of $MM^{\dagger}$ appear in inverse
pairs and we can restrict to those with $\nu_i \geq 0$.
The conductance reads in the corresponding eigenvalue representation
\begin{equation}
		g=\sum_i {\cal T}_i = \sum_i \frac{2}{1+\cosh \nu_i}
	\label{3.51}
\end{equation} 
Here the eigenvalues ${\cal T}_i$ have to be distinguished from the
total
probability to be scattered from a channel $\a$ into forward
direction, $T_\a$, that occurs in Eq.~(\ref{3.39}).
A related {\em  polar
parameterization} (cf.~\cite{Sto91})
of the transfer matrix $M$ in terms of {\em  radial coordinates} $0\leq
\lambda_i<\infty $ and four 
unitary
$N_c\times N_c$ matrices $u^{i}$  is also used frequently 
\begin{equation}
	M=\left(\begin{array}{cc}
	u^1 & 0\\
	0 & u^3
	\end{array}\right)
	\left(\begin{array}{cc}
	\sqrt{1+\lambda} & \sqrt{\lambda}\\
	\sqrt{\lambda} &
\sqrt{1+\lambda}\end{array}\right)
	\left(\begin{array}{cc} 
	u^2 & 0\\
	0 & u^4
	\end{array}\right)\, .
	\label{3.52}
\end{equation} 
Here $\lambda={\rm diag} (\lambda_1,  \ldots, \lambda_{N_c})$
and
\begin{equation}
	{\cal T}_i = (1+\lambda_i)^{-1} \, .\label{3.52a}
\end{equation}

It worth noticing that in these $S$-matrix models the conductance
itself appears as a linear statistics \footnote{By linear statistics
one denotes quantities that are sums of a specific function $f$ of the
random parameters, $\sum_i f(\lambda_i)$.} of eigenvalues of a certain
random matrix. However, one should keep in mind that the statistical
properties of these eigenvalues are, in general, not independent of
those of the associated eigenvectors.

\subsection{Localization}\label{subloc}
Before we introduce generalizations of the $S$-matrix modeling we will
show that the present one is  well suited to discuss the problem
of localization in quasi-1D conductors.

An  instructive example is the 1D conductor for which $N_c=1$.
The composition law Eq.~(\ref{3.50}) tells that the composition law
for the transmission reads (see Fig.~15)
\begin{equation}
		t_{12} = \frac{t_1t_2}{1-r'_1r_2}	\, .\label{3.53}
\end{equation} 
Exploiting unitarity of $S$ yields for the transmission probability
$T=|t|^2=1-R$, $R=|r|^2=|r'|^2$,
\begin{equation}
		T_{12}=\frac{T_1T_2}{1-2\cos \phi \sqrt{R_1R_2} +
			R_1R_2}
	\label{3.54}
\end{equation} 
where $\phi$ is the sum of phases of $r'_1$ and $r_2$.
Based on Eq.~(\ref{3.54}) one can derive an evolution equation for the
probability distribution of $T$ with increasing length $L$. We will
postpone this to Sect.~\ref{subscadis}. Here we will take advantage of the fact
that an average of $\ln T_{12}$ over a uniform distribution of phases
$\phi$
yields the simple expression \cite{And80}
\begin{equation}
		\BRA\ln T_{12} \KET_{\phi}= \ln T_1 + \ln T_2	\, .
\label{3.55}
\end{equation} 
Thus, the phase-average of the logarithm of the transmission turns out
to be additive. Iterating this procedure tells that the logarithm of
the transmission will be distributed in a Gaussian way according to
the central limit theorem for  independent additive random numbers.
This motivates to call $ T_{\rm t} :=\exp \BRA \ln T \KET $ the {\em  typical}
transmission,  and to write down a {\em  quantum} series composition law
for conductance $g=T$ in 1D conductors   
\begin{equation}
		g_{\rm t}(L_1+L_2)=g_{\rm t}(L_1)g_{\rm t}(L_2)
		\label{3.56}
\end{equation}
This composition law is in striking contrast to the macroscopic composition
law for incoherent conductors, $G^{-1}(L_1+L_2)= G^{-1}(L_1)+G^{-1}(L_2)$.
Equation~(\ref{3.56}) immediately tells that the conductance will
exponentially decrease with increasing system length $L$, which means
localization. 
To determine the localization length $\xi$ we calculate
\begin{equation}
		\frac{d g_{\rm t}(L)}{dL}= g_{\rm t}(L) \frac{g_{\rm t}
		(\delta L) -1}  {\delta
		L}\, , \;\; \delta L\to 0
		\, .\label{3.57}
\end{equation}
Thus, $g_{\rm t}(L)=g_{\rm t}(L_0)\exp(-2(L-L_0)/\xi)$ with $\xi/2= \delta L/(1-T_{\rm t}(\delta
L))$, and by Eq.~(\ref{3.48}) 
\begin{equation}
		\xi = 2 l_e \, .\label{3.58}
\end{equation}
This result demonstrates that in one-dimensional conductors all states
are localized. Furthermore, the localization already occurs on the
scale of the mean free path. This is a crucial mesoscopic 
effect as it is a consequence of the multiplicative quantum series composition,
Eq.~(\ref{3.56}). 

For arbitrary channel numbers $N_c$ the above 1D result does not
apply.
However, one can make an educated guess what might happen
to the localization length, if localization occurs at all.

To this end, we notice that the conductance in 1D for length $L$ much
smaller than $l_e$ is given by $g( L)= 1- L/l_e$ in
accordance with our definition of $l_e$.
Since the result $g=1$ corresponds to ideal transmission of the wave
guide
an Ohmic series decomposition into {\em  contact resistance} $1$ and 
{\em  intrinsic
resistance}, $ L/l_e$,  is meaningful.
Assuming that for $N_c >1$ and $L\ll \xi(N_c)$ Ohm's parallel
composition law is applicable yields the {\em  classical conductance}
\begin{equation}
		g_{ cl}(L)= \frac{N_c l_e}{L} \label{3.59}
\end{equation}
and the natural limit for this law to hold, $g(L\approx \xi)\approx 1$,
 yields an estimate of $\xi(N_c)$
\begin{equation}
		\xi(N_c)\approx N_c l_e    \, .\label{3.60}
\end{equation}

We are now going to discuss rigid results that confirm the
localization effect for arbitrary $N_c$. In Sect.~\ref{substaqua} 
we will see that our
estimate of the quasi-1D
localization length is in accordance with known results.

A crucial point in the 1D calculation was the multiplicative
composition law for typical conductance. The  transfer matrix obeys
also a multiplicative composition law and the question arises what can
be said about the statistical properties of a large product of random
transfer matrices. 
Recall that the product of independently distributed positive random numbers
$X(i)$,
\begin{equation}
	{\cal X}(N)=\prod_{i=1}^N X(i) = \exp \LBK \sum_{i=1}^N \ln X(i)\RBK
        \, , \label{3.61}
\end{equation}
gives rise to a central limit theorem for its logarithm:
For large $N$, the distribution of $\ln {\cal X}(N)$ is well
approximated by a Gaussian with mean value $N\BRA \ln X\KET$ and
variance
\begin{equation}
		\sigma_N^2 =N\LBK \BRA (\ln X)^2 \KET - \BRA \ln X
		\KET^2 \RBK \, .\label{3.62}
\end{equation} 
For the original variable ${\cal X}(N)$ the distribution is 
called
{\em  log-normal distribution} and reads
\begin{equation}
		{\cal P}({\cal X},N) d {\cal X}
	= \frac{1}{\sqrt{2\pi}\sigma_N}\exp \LBK
	-\frac{1}{2\sigma_N^2} \LK \ln {\cal X} - N \BRA \ln  X\KET\RK^2
	\RBK d \ln {\cal X}
	\label{3.63}
\end{equation}
Furthermore, one can conclude that for almost all realizations
 the geometric mean asymptotically
coincides with the typical value
\begin{equation}
		\lim_{N\to \infty} \LK {\cal X}(N)\RK^{1/N} =\exp\LK
		\BRA \ln X \KET \RK =: X_{\rm t} \, .\label{3.64}
\end{equation}

There
 exist some extensions of the law of large numbers
to the case of products of independent random transfer matrices (see
\cite{Cri93}).
A version of the
 theorem of Oseledec \cite{Osel68}  guarantees the existence (with
 probability $1$) of
eigenvalues of the diagonalizable
limiting matrix \cite{Tutu68,Virst70} 
\begin{equation} \label{3.65}
	{\cal M} := \lim_{N\to\infty}(M^\dagger(N)M(N))^{\frac{1}{2N}}
\end{equation}
($N=L/L_0$, $L_0$ some fixed reference length) of the form
\begin{equation}\label{3.66}
 	{\cal M} =
	(e^{\gamma^{}_{N_c}},\ldots,e^{\gamma^{}_1},
e^{-\gamma^{}_1},\ldots,e^{-\gamma^{}_{N_c}}).
\end{equation}
Motivated by the analogy between transfer
matrices and time evolution matrices considered in chaotic dynamics
the quantities $\gamma^{}_1 < \ldots  < \gamma^{}_{N_c}$
are 
called {\em 
Lyapunov exponents}. 

The Lyapunov exponents $\gamma_i$
are asymptotic non-random numbers and are the
so-called self averaging limit of the random numbers $\nu_i(N)$, the
exponents of eigenvalues of $M^{\dagger}(N)M(N)$,  divided
by $2N$
\begin{equation}\label{3.67}
	\gamma^{}_i = \lim_{N\to\infty} \frac{\nu_i(N)}{2N}
\end{equation}
A system with channel number $N_c$ has thus $N_c$ characteristic length scales
\begin{equation}\label{3.68}
	\xi_{i}= \frac{L_0}{\gamma_i}=\lim_{L\to\infty}
	\frac{2L}{\nu_i(L)}
\end{equation}

According to the conductance formula Eq.~(\ref{3.51}) the
 inverse of the smallest Lyapunov exponent 
 yields  the relevant localization length
\begin{equation}\label{3.69}
	\xi = \frac{L_0}{\gamma_1}\, .
\end{equation}

A theorem by F\"urstenberg \cite{Fuer63} 
states that the largest Lyapunov exponent
 is always positive. This proves localization in strictly 1D, but
 is
meaningless for $N_c >1$. Virster \cite{Virst70} has later proofed that
$\gamma^{}_1$ is strictly positive, as long as $N_c$ is
finite. Thus, 
in the quasi-1D limit $L\to\infty$ the localization length $\xi$ is
finite.

We postpone  the discussion of the statistics of
$\nu_i(L)$ to Sect.~\ref{substaqua} and proceed by introducing more general
$S$-matrix models.
\subsection{Networks Of Scatterers}\label{subnet}
So far, the modeling of mesoscopic systems by scattering matrices was
restricted to the quasi-1D case. 
More general $d$-dimensional
mesoscopic systems can be described by a 
model introduced by Shapiro
\cite{Sha82,Edre89}. Instead of taking a strip $S$-matrix one can
consider
a regular network (lattice spacing $a$) consisting of $N={\rm
volume}/a^d$ 
sites 
and $N_b$ bonds (see
Figs.~16, 17).
 Each site represents a scatterer and is described by a
unitary
$2d\times 2d$ scattering matrix $S$. The bonds represent free
propagation between the blocks. Each bond carries two waves
propagating in opposite directions. Each $S$-matrix  transforms $2d$
incoming amplitudes into $2d$ outgoing amplitudes. To each bond
amplitude a random phase is added which simulates an irregular lattice
of scatterers which are distributed between distance $a$ and
$a+\lambda_F$ where $\lambda_F$ is the wavelength of scattered waves.

By specifying appropriate boundary conditions one can
describe multi-probe
conductors within this network model. Consequently, physical
quantities like conductance coefficients and even channel dependent
 transmission amplitudes can be studied with less effort than
in systems described by a Hamiltonian.

Interesting is the fact that the $S$-matrix modeling does also allow
to
determine 
 energy eigenstates and eigenvalues. This comes along with the
following
interpretation: mapping incoming to outgoing states by scattering
matrices usually refers to a process consuming infinite time in order
to guarantee energy conservation. However, if one allows for a certain
uncertainty in the energy, a finite time is enough to follow the
scattering process. Therefore, one can take the point of view that the
scattering matrices at each site describe a unitary time evolution of
states in one unit of time. By this interpretation the finite energy
uncertainty is treated as  negligible. Accordingly we define the unitary
matrix of time evolution in a unit of time by \cite{Edre89,Fert88,Kles95}
\begin{equation}
	\sum_k U_{ik}\psi(k;t)= \psi(i;t+1)\, .\label{3.70}
\end{equation}
 Here  $\LB \psi(k;t)\RB$ forms, for each time $t$,
a set of $2N_b$ complex bond amplitudes. Normalizing them, $\sum_i^{2N}
|\psi(i)|^2=1$, makes them a proper choice for local wavefunctions.
The unitary $2N_b\times 2N_b$ matrix $U$ 
is uniquely determined by the network of
scattering matrices since it summarizes nothing but the scattering
conditions at the sites.
Energy eigenstates $\psi$ are stationary solutions of the unitary
operator
$U$
\begin{equation}
	U\psi =\psi \label{3.71}
\end{equation}
and thus correspond to the eigenstates of $U$ with eigenvalue $1$.
The energy of the underlying electron system  enters here only parametrically
through the scattering amplitudes. For simplicity, let us assume that
the scattering strengths are equal at all nodes
(abbreviated by one symbol $s$) and randomness enters
only through the bond phases. Then $U$ will have an eigenstate to
eigenvalue $1$ only for a discrete set of scattering strengths, which
correspond to a discrete set of energy eigenvalues of the underlying
electron
system. Instead of looking for the discrete set of 
 energies one can also fix the energy and ask for the more
general
eigenvalue problem 
\begin{equation}
     U(E)\psi_\a(E)= e^{i\phi_\a(E)} \psi_\a(E) \,
     .\label{3.72} 
\end{equation}
$\psi_\a(E)$ are, for fixed $E$, a set of $2N_b$   
eigenstates at energy $E$ of slightly
modified
disorder configurations, each of the modifications being an  overall shift in
random phases on the links.  
The eigenphases $\phi_\a(E)$ can be considered as an energy 
excitation spectrum at
energy $E$.  
 Thus, the network models allow  to investigate energy eigenvector
statistics and, via the phases $\phi_\a(E)$, energy eigenvalue
statistics.

The network models have a further advantage as compared to Hamiltonian
models concerning numerical calculations. In case that only a finite
resolution 
of energy scales is required one does not need to fully
diagonalize
the unitary operator. Instead, one  iterates the time evolution step by step
until
a certain time scale is reached, the inverse of which yields the required
energy resolution.

Finally, we like to point out that within the network models there are
 possibilities not only to include global symmetries but also
characteristic length scales. For example by allowing for
 internal degrees of freedom (e.g.~spin) the number of amplitudes on
each bond can be increased and scattering strengths can be introduced
that take the change of internal degrees of freedom into account. 
The inclusion
of strong magnetic fields with their characteristic chiral character
is also possible.  No commensurability features
like in the Anderson model with Peierls substitution do occur. Especially, one 
model of this category has become widely known as the
Chalker-Coddington model \cite{ChaCod88}
 in 2D which only allows  for scattering to the
left and right (see Figs.~18, 19).

\section{Statistics in Idealized Situations}\label{staide}
In this section we consider idealized mesoscopic systems and ask for
the statistics of energy eigenvalues and eigenvectors. The 
idealization is due to the assumption that the statistical properties
of both can be studied separately. We distinguish  between
the localized situation imitated by a diagonal Hamiltonian and the
delocalized
situation imitated by a standard random hermitean matrix having independent
Gaussian distributed  random entries.

We will review   the energy level statistics in these systems
and   the statistics of the local box-probability $P(L_b)$ for
 an electron that is in an eigenstate $\psi(\r)$ to be found 
in a certain box of volume $L_b^d$,
\begin{equation}
  	P(L_b):=\int\limits_{\rm box} d^dr \mid \psi(\r)\mid^2 \, .\label{5.1}
\end{equation}
The complete information about the (energy) level statistics of a
$N\times N$ Hamiltonian is contained in the joint probability density
\begin{equation}
		{\cal P}  \LK\varepsilon_1, \ldots,\varepsilon_N\RK\, .
\label{5.2}
\end{equation}
To simplify notations we will assume throughout that any reordering of
levels 
in ${\cal P}$ does not change the probability. 
The simplest statistical quantity to be extracted is the
average
level density
\begin{equation}
		\nu(E) := \BRA \rho(E)\KET =N^{-1}\BRA \sum_\a \delta
		(E-\varepsilon_\a)\KET = \int d\varepsilon_2 \ldots
		d\varepsilon_N \, {\cal P}  \LK E,\varepsilon_2,
		\ldots,\varepsilon_N\RK \label{5.3}
\end{equation}
defining the average level spacing $\Delta (E) =  \LK N \nu(E) \RK^{-1}$.
The two-level correlation function is
\begin{equation}
		R(s)= \Delta^2 \BRA \sum_\a \delta
		(E-\varepsilon_\a) \sum_\b \delta
		(E'-\varepsilon_\b)\KET - 1 \, ,\;\;
		s:=\frac{E-E'}{\Delta}\, .
 \label{5.4}
\end{equation}
For simplicity we assumed that it is
translational invariant within a band of
states with $\Delta$  being independent of energy. By writing
the energy separation in units of average level spacing ($s$) one
focuses  on those properties that are independent of the chosen
energy units.
Quantities which contain information about $n$-level correlation
functions
are the $n$-level spacing distribution $Q(n,s)$ which yields the
probability to find exactly $n$ levels in an energy interval of width
$s$
(the center of which can be arbitrary).
 The probability to find
exactly no level inside the interval defines the so-called
level spacing distribution $P(s)=d^2 Q(0,s)/ds^2$ which yields the
probability density
to find a level spacing $s$ with respect to  $ds$.
The average number
$
	\BRA n (s)\KET 
$
 of levels
in an interval $s$ and its variance 
$
	\Sigma^2:= \BRA n^2(s)\KET - \BRA
	n(s)\KET^2
$ are also of interest.
 The derivative of $\Sigma^2$ with respect to $\BRA n
\KET$
defines (for large $\BRA n\KET$) a {\em  level compressibility}
\begin{equation}
	\chi:= \lim\limits_{\BRA n\KET \to \infty} \frac{d \Sigma^2}{d
	\BRA n \KET} \, .\label{chichakrav}
\end{equation}
\subsection{Localized Phase}\label{sublocpha}
We consider $N$ orthogonal energy 
states $\psi_\a\, ,\; \a=1,2,\ldots ,N$ 
each of which is  localized to one of $N$
sites of a regular lattice (lattice spacing $a$) in an arbitrary dimension
$d$, leading to  a {\em  vanishing localization length}.
 The corresponding energy eigenvalues are independent random
variables taken from a distribution
\begin{equation}
	p(\varepsilon)= \frac{1}{\sqrt{2\pi}\varepsilon_0}\exp \LK
	-\frac{\varepsilon^2}{2\varepsilon_0^2}\RK \, .\label{5.5}
\end{equation}
The corresponding random Hamiltonian is diagonal in site representation
\begin{equation}
	H_{ik}=\varepsilon_i \delta_{ik} \label{5.6}
\end{equation}
with eigenstates 
\begin{equation}
	\psi_\a(i)= \delta_{\a i} \;\;  {\rm for} \;\; \varepsilon_\a =
	\varepsilon_i \label{5.7}
\end{equation}
This situation was already  addressed in Sect.~\ref{subthe}
as one of two extreme cases for the eigenvector statistics.
As compared to the Anderson tight-binding Hamiltonian,
Eq.~(\ref{3.9}),
the Hamiltonian in Eq.~(\ref{5.6}) ignores the kinetic energy completely.
The energy level statistics is contained in the joint probability
distribution
\begin{equation}
	{\cal P}  \LK\varepsilon_1, \ldots,\varepsilon_N\RK =\prod_{\a=1}^N
	p(\varepsilon_\a)\, .\label{5.8}
\end{equation}
Consequently, the average level density is 
\begin{equation}
	\nu(E)=  p(E) \label{5.9}
\end{equation}
and the average level spacing within the band ($|E|\ll \varepsilon_0 $)
is $\Delta=1/(Np_0)$, $p_0:= (\sqrt{2\pi}\varepsilon_0)^{-1}$.
The two-level correlation function $R(s)$ reflects the fact that there
is only auto-correlation between levels (up to $1/N$ corrections).
\begin{equation}
		R(s)= \delta (s) - N^{-1}\label{5.10}
\end{equation}
The level spacing distribution follows the Poisson law 
\begin{equation}
		P(s)=\exp \LK -s\RK \label{5.11}
\end{equation}
(that is why the ensemble is called {\em  Poisson ensemble}) and the
 characteristic
 relation
between number variance and average number of levels can be obtained
\begin{equation}
		\Sigma^2 = \BRA n \KET \, \label{5.12}
\end{equation}
indicating that  the levels are  compressible with compressibility  $1$.

The probability to find an electron in a box of size $a$  in
the eigenstate $\psi_\a$ is
\begin{equation}
	P_\a (a; i) = |\psi_\a(i)|^2 = \delta_{\a i}\, .\label{5.13}
\end{equation}  
Increasing  box sizes to $L_b>a$  does not change the result as long as the
boxes are chosen to be non-overlapping and centered at the original
point. Independently of the box size the probability is either zero,
if the state is not located in the box or $1$ in the opposite case.

For a given energy $E$ we consider the microcanonical average
of $P_\a$
\begin{equation}\label{5.13b}
	P(E,L_b;i):= \frac{\sum_\a \delta
(E-\varepsilon_\a)P_\a(L_b;i)}{\sum_\b \delta (E-\varepsilon_\b)}
\end{equation}
where a smearing-out of $\delta$-functions  is understood. The smearing width
$\Gamma(E)$
has to be adjusted such that isolated $\delta$-peaks are broadened, but
 not over many neighboring levels. A convenient choice is to take $\Gamma(E)$
of the order of the actual level spacing at $E$ in a given realization.
The denominator in Eq.~(\ref{5.13b}) is then $1/\Gamma(E)$.
Since the eigenvalues are independently distributed,
the centers of the
corresponding eigenstates are  randomly distributed over the lattice
points. 
Therefore, in almost any case  one will find zero box-probability
at a certain  site $i$, unless the energy $E$ coincides (within the
scale of the smearing $\Gamma(E)$) with eigenvalue $\varepsilon_i$, 
in which case the
box-probability is $1$. The average box-probability is
finite and given by
$\BRA P(E) \KET_{L_b}= (1/N)(L_b/a)^d= \BRA P\KET_{L_b}$.

Thus, the probability distribution ${\cal P} (P; L_b)$  of the
box-probability
is obtained as 
\begin{equation}
		{\cal P} (P; L_b) dP = \LBK \delta(P) \LK 1-\BRA P
		 \KET_{L_b}\RK + \delta (P-1)\BRA P \KET_{L_b}\RBK dP 
		 \, .\label{5.14}
\end{equation}
The average value is not characteristic for this distribution, since
the main weight is given to $P=0$. However, e.g. the geometric mean  
\begin{equation}
	P_{\rm t}(L_b) :=\exp \BRA \ln P \KET_{L_b} = 0\label{5.15}
\end{equation}
reflects a {\em  typical} box-probability. Alternatively, one may
consider the median of the distribution (that value up to which half
of the total
 weight is used up) which vanishes also and could be termed
{\em  typical} as well. 

The LDOS  $\rho(E,i)=\sum_\a
\delta(E-\varepsilon_\a)|\psi_\a(i)|^2$  with a smearing-out of
$\delta$-functions on the scale of $\Gamma(E)$  yields for
the LDOS in a box, $\rho(E,L_b;i)$
\begin{equation}
	\rho(E,L_b;i) := \Gamma(E)^{-1}P(E,L_b;i) \, .\label{5.16}
\end{equation}
However, we must mention that this notion of LDOS in a box differs 
from that used by other authors (see e.g.~\cite{EfePri93}). If one considers
the LDOS with a parametric Lorentzian smearing-out of $\delta$-functions,
\begin{equation}
	\rho(E;\r)= \sum_\a \frac{\Gamma}{2\pi}\frac{1}{\LK
	E-\varepsilon_\a\RK^2 + \Gamma^2/4}|\psi_\a(\r)|^2 \, ,\label{5.27}
\end{equation}
where $\Gamma/\Delta $ can vary from 
$0$ to $\infty$ one can investigate parametrically the
 interplay of level and eigenvector
statistics.

 To simplify the discussion we will be often  more restrictive and call
\begin{equation}\label{5.26b}
	\tilde{\rho}(E,L_b;i) := \Delta^{-1}P(E,L_b;i) \,
\end{equation}
the {\em  reduced} LDOS in a box. This definition
concentrates on the eigenvector statistics.
Consequently, the  distribution function reads
\begin{equation}
	{\cal P} (\tilde{\rho};  L_b) d{\tilde{\rho}} 
	= \LBK \delta({\tilde{\rho}}) \LK 1-\Delta\BRA {\tilde{\rho}} 
	\KET_{L_b}\RK
	+ \delta ({\tilde{\rho}}-\Delta^{-1})\Delta\BRA {\tilde{\rho}}	
\KET_{L_b} \RBK d{\tilde{\rho}} 
		 \, .\label{5.17}
\end{equation}

In summary, ideal localized systems (with vanishing localization
 length $\xi$)
 are characterized by a
compressible spectrum of uncorrelated energies and a LDOS in a box that
typically vanishes, although the average DOS is finite.  
\subsection{Delocalized Phase}\label{subdel}
For the idealized delocalized phase we make the assumption that
eigenstates are distributed isotropically within the space of all possible
 eigenstates for a $N\times N$ Hamiltonian
matrix, $H_{ik}$.
 In other words the diagonalizing unitary matrices $U_{k\a}$,
Eq.~(\ref{3.32}), are uniformly distributed with respect to the invariant
measure $d[{\cal U}(N)]$. This situation was already  addressed in 
Sect.~\ref{subthe}
as the second extreme case for the eigenvector statistics.

Therefore, we  assume the  distribution  of the  Hamiltonian matrix to
be unitarily invariant. One of the simplest possible choices is
the so-called Gaussian unitary ensemble (GUE) introduced by Wigner
\cite{Wig51}
when studying the level statistics of complex nuclei,
\begin{equation}
	{\cal P}\LK\LB \real H_{ik}, \imag H_{ik}\RB\RK d \LBK H\RBK =
	{\tilde{C}}_N
	\exp \LK - \frac{N}{2 {\cal E}_0^2} 
	\Tr H^2 \RK d\LBK H\RBK \, .\label{5.17a}
\end{equation}
Here $\tilde{C}_N$ is a normalization constant, ${\cal E}_0$ some
arbitrary energy scale  and 
the volume element is defined in terms of the independent matrix
elements of $H$ as
\begin{equation}
	d \LBK H\RBK = \prod_1^N dH_{ii}
	\prod_{i<k}^N d \LK\real H_{ik}\RK  d\LK \imag H_{ik}\RK \, .
	 \label{5.17b}
\end{equation}
Since $\Tr H^2 = 2 \sum\limits_{i<k} \LBK \LK \real H_{ik}\RK^2 + \LK \imag
H_{ik}\RK^2\RBK
+ \sum_i H_{ii}^2 $  the GUE describes a random matrix with all its
elements
uncorrelated. Each of the elements vanishes on average and its
absolute value fluctuates, $\BRA |H_{ik}|^2 \KET ={\cal E}_0^2/N$.
If we think of the model in site-representation and compare it to
the Anderson model we see that by Eq.~(\ref{5.17a})
 hopping terms are dominant and
hops
from one site to another are equally probable independent of the
spatial distance. This means that we can associate a
{\em  vanishing  time scale} $t_D$  to travel through the
system.  

Introducing eigenvalues $\varepsilon_\a$ and eigenvectors
$\psi_\a(k)=U_{k\a}$
one can transform the probability  ${\cal P}(\LB \real H_{ik}, \imag
H_{ik}\RB) d \LBK H\RBK$ to these variables at the expense of
introducing
a Jacobian between the set of variables. Fortunately, the Jacobian can
be calculated, the integration over $d[{\cal U}(N)]$ being a trivial
normalization (since $\Tr H^2$ is unitarily invariant), 
  and one obtains for the joint probability of
eigenvalues
\begin{equation}
	{\cal P}\LK \varepsilon_1, \ldots ,\varepsilon_N\RK  =
	C_N \prod_{\a < \b} \LK \varepsilon_\a -\varepsilon_\b\RK^2 
	\exp \LK - \sum_\a^N \frac{N}{2{\cal E}_0^2}\varepsilon_\a^2 \RK
	\, . \label{5.18}
\end{equation}
The factors in front of the exponential are due to the Jacobian and
ensure
that the probability to find two levels close to each other vanishes; a
phenomenon which is denoted as {\em  level repulsion}. 
The factors can be rewritten as $\exp\LK 2 \sum\limits_{\a <\b} \ln |
\varepsilon_\a -\varepsilon_\b |\RK$ such that the joint probability
describes a classical Gibbs ensemble,
\begin{equation}
	{\cal P}\LK\LB \varepsilon_\a\RB\RK
	= C_N 
	\exp\LBK -\beta {\cal H}
	\LK\LB \varepsilon_\a \RB\RK\RBK \, , \label{5.19a}
\end{equation}
of a gas of {\em  particles} with coordinates $\varepsilon_\a$ and a
Hamiltonian
\begin{equation}
	{\cal H}\LK\LB \varepsilon_\a 
	\RB\RK = \frac{1}{2}\sum_{\a\not=\b}
	U(\varepsilon_\a,\varepsilon_\b)
	 +\sum_\a
	V(\varepsilon_\a)\, \label{5.19b}
\end{equation}
that contains a logarithmic two-body interaction $U(x,y)=- \ln  |x-y|$
and a one-body (confining) potential $V(x)= Nx^2/ (2\b {\cal E}_0^2)$.
The {\em  inverse temperature} is $\b=2$ for the GUE. 
A related  ensemble of real
symmetric matrices, reflecting time inversion symmetry, denoted as
Gaussian orthogonal ensemble (GOE), gives rise to the same Gibbs
ensemble, the only change being $\beta=1$
 ($\beta=4$ corresponds to spin systems with time
reversal symmetry, for more details and review see~\cite{Meht91}).
The interpretation of the joint probability distribution of levels as a
Gibbs-ensemble has led to an important mean-field approach to level
statistics
\cite{Dys72} which we will use later in the context of quasi-1D
conductance (cf.~Sect.~\ref{substaqua}).
In the standard random matrix  of GUE and GOE a number of
 results are well known:
\begin{enumerate}
\item	
The average DOS is (for $N \gg 1$) given by Wigner's {\em  semi-circle
law},
\begin{equation}
	\nu(E)= \frac{1}{\pi{\cal E}_0}\sqrt{1-\LK
	\frac{E}{2{\cal E}_0}\RK^2 }\,, 
	\label{5.20}
\end{equation}
with average level spacing ($E\approx 0$) $\Delta= \pi{\cal E}_0/N$.
\item 
The two-level correlation function shows level-repulsion for
 $s\ll 1$ and decays $\sim -1/s^2$ for  $s\gg 1$. In the GUE it
reads ($s\not=0$, $N\to \infty$)
\begin{equation}
	R(s)=- \LK \frac{\sin \pi s}{\pi s}\RK^2\,
	.\label{5.21}
\end{equation}
\item 
	The level spacing distribution is very well approximated by
	Wigner's surmise
\begin{equation}
		P(s)= A_\b\, s^\b \exp \LK - B_\b\, s^2\RK \label{5.22}
\end{equation}
where e.g. $A_2=\pi/2, B_2=4/\pi$. The characteristic small $s$
behavior
$\sim s^\b$ reflects the power of level-repulsion characteristic of
the symmetry of the Hamiltonian. 
\item
The level number variance $\Sigma^2$ depends only logarithmically on
the average level number, 
\begin{equation}
	\Sigma^2 \propto \ln \BRA n\KET \, ,\label{5.23}
\end{equation}
which yields a vanishing level compressibility 
($\sim \BRA n\KET^{-1}\to 0$).
\end{enumerate}

Eigenvectors $\psi_\a(k)=U_{k\a}$  are isotropically
distributed 
 uncorrelated Gaussian 
variables up to $1/N$ corrections. More precisely,
in the GUE the eigenvector components are complex with uncorrelated Gaussian
distributed real and imaginary parts. The average of each is zero
while the variance is, due  to normalization, given as
\begin{equation}
   \BRA \LK \real\psi_\a(k)\RK^2 \KET = \BRA \LK \imag\psi_\a(k)\RK^2
   \KET = \frac{1}{2N}\, .
	\label{5.24}
\end{equation}
For the GOE the components are real with vanishing average and variance
$\BRA |\psi_a(k)|^2\KET = N^{-1}$.

This fixes the distribution of the probability $P(E;a)$  
to find an electron in
a box  of size $a$ ($a$ lattice spacing).  In the GUE one finds
the Rayleigh distribution
\begin{equation}
	{\cal P} (P; a) = N \exp \LK -PN\RK \, , \label{5.25}
\end{equation}
and in the GOE the Porter-Thomas distribution
\begin{equation}
	{\cal P}(P;a) = \LK\frac{N}{2\pi P}\RK^{1/2} \exp\LK - PN/2\RK \, .
	 \label{5.26}
\end{equation}
Both distributions follow immediately once an intensity of waves $\psi$ is
considered
which are large sums of either complex  or real numbers. This leads,
by the law of large numbers, to Gaussian distributions for $\psi$. 
Both distributions can be generalized to arbitrary box size $L_b$ by replacing
$N^{-1}$ with $\BRA P \KET_{L_b}  =  N^{-1} (L_b/a)^d$.
Characteristic of these distributions is the exponential tail and the
fact
that no parameter besides $\BRA P \KET_{L_b} $ occurs. Still, the
average
value is a typical value for the distribution; e.g. the median differs
from the average value by only a factor of $\ln 2$.

The statistics for  the reduced LDOS in a box,
${\tilde{\rho}}(E;L_b)=\Delta^{-1}P(E;L_b)$, is then straightforward. 

In summary, ideal delocalized systems (with vanishing time scale
$t_D$)
are characterized by an incompressible spectrum of correlated energies
and a LDOS in a box the distribution of which develops 
 an exponential  tail. Still,  the average value is typical. 
\section{Concept of Scaling}\label{con}
The level and eigenvector statistics
of the idealized systems show a remarkable feature: No physical scale
besides the kinematic scales, level spacing $\Delta$, total volume
$Na^d$ and box volume $L_b^d$ enters the  statistical quantities.
The parameters of the models, $\varepsilon_0$ (for the localized
phase), 
and ${\cal E}_0$ only define the energy unit for $\Delta$.
No transport scale enters the models. This is consistent
with the following interpretation: In the  localized phase
the  only relevant scale is the localization length, which was treated
as zero.
In the idealized delocalized phase the only relevant scale is the time
to cross the system, or equivalently the corresponding Thouless energy
$E_{\rm Th}=h/t_D$ which was treated as infinite. 
We may thus expect that the statistics described in the foregoing
section correspond to asymptotic  situations for which the dimensionless
conductance is either zero (localized phase) or infinite (delocalized)
phase. 

The question then arises how  the statistical properties do change for
finite conductance. Of particular importance are  the
statistical properties when approaching the LD transition. Which
physical 
scales determine the
statistical properties?  
A related question concerns the change of the statistical properties
with increasing system size. 

\subsection{Relevant Scales And $\beta$-Function}\label{subrelbet}
Looking at the conductance formulas by Thouless and B\"uttiker,
Eqs.~(\ref{2.19},\ref{3.39}), it is tempting to assume that the
conductance itself is the (perhaps only) relevant quantity to distinguish
localized and  delocalized phases and to characterize the LD transition.  
\begin{enumerate}
\item
The localized phase with 
\begin{equation}
	g\ll 1 \label{6.1} 
\end{equation}
corresponds to {\em  only closed}
transport modes (channels) (B\"uttiker formula with $T_\a\ll 1$) and
the existence 
of a finite localization length $\xi \ll L$, $g\sim \exp(-2L/\xi)$.
\item
The delocalized phase with
\begin{equation}
	g\gg 1 \label{6.2} 
\end{equation}
corresponds to many {\em  open channels} (B\"uttiker formula with
$N_c\approx g$ modes of $T_\a\approx 1$) and a Thouless energy much
larger than the level spacing, $E_{\rm Th}\gg \Delta$.
\item
The LD transition is expected to occur when scales match, i.e. 
\begin{equation}
	E_{\rm Th}\sim {\cal O} \LK \Delta\RK  \, ,\;\; g\sim {\cal
	O}\LK 1\RK \,.\label{6.3} 
\end{equation}
According to the B\"uttiker formula only a number of ${\cal O}(1)$
modes are open and the localization length diverges.
This divergence turns out to be algebraically 
on approaching the transition point,
\begin{equation}
	\xi \propto \tau^{-\nu} \label{6.4} 
\end{equation}
where $\tau$ measures the deviation of a tuning parameter (e.g. Fermi
energy)
from its {\em critical value} at the LD transition. The exponent
$\nu$ is called {\em critical exponent} of the localization length. 
\end{enumerate}

By this reasoning a relevant length scale $\xi_c$ can be defined via the
conductance: $\xi_c$ is that fictitious system size for which 
\begin{equation}
	g(\xi_c) \sim {\cal O}\LK 1 \RK\, .\label{6.5}
\end{equation}
 In the  localized phase $\xi_c$ can therefore be
identified with the localization length, while in the delocalized
phase
$E_{\rm Th}(\xi_c)\sim {\cal O}\LK \Delta (\xi_c)\RK$. 
 
As our considerations call for a description of the
LD transition as a critical phenomenon 
 we briefly review some aspects of the latter.

A thermodynamic system is described in terms of the coordinates of 
its equilibrium states manifold $\cal M$ (e.g. volume $V$, particle
number $N$ and temperature $T$). Any thermodynamic
  quantity of interest can be
calculated from  a thermodynamic potential
(e.g. the free energy ${\cal F}(V,T,N)$) which is an analytic function of
these coordinates (e.g. the specific heat
$c_V=\partial^2{\cal F}/{\partial T^2}$).
Generally, one can change the parameterization of $\cal M$ (e.g. 
volume $V$ $\to$
pressure $p$) to describe the same equilibrium state -- at least
locally.
In the thermodynamic limit where $N$ and $V$ tend to infinity
simultaneously  leaving  the density 
$n=N/V$ constant,  it may happen that the thermodynamic potential is no
longer an analytic function of its coordinates. In such a situation
the state of the system is not uniquely described by only one set of
coordinates manifesting in phases.
When crossing the coexistence curve the thermodynamic potential
jumps. Such a process  is
called a first-order phase transition. The jump is quantitatively
described by the value of the latent heat. Assume that
the coexistence curve ends
in some point of the $(p,T)$ diagram which is called the critical
point.
Here the latent heat vanishes, the thermodynamic potential
is continuous and the difference of densities between the 
 phases vanishes. Yet, at the critical point, the thermodynamic
potential is not analytic since its second derivatives are
discontinuous. It  is then helpful 
 to look for  a quantity
which serves as an  {\em  order parameter}
 of the transition.  It should be finite in one phase and vanishes in
the other phase. A more restrictive requirement is that an order parameter
   should correspond to a local quantity $\varphi(\r)$ the average
value of which $m=\BRA \varphi(\r) \KET$
 vanishes  with some power
law at the critical point (e.g. the difference of
the corresponding densities in the  transition),
\begin{eqnarray}
	m & = & 0 \;\;\, ,\;\;\; T < T_c \nonumber\\ 
	m & \propto & \tau^{\beta}  \, ,\;\;\; T > T_c \, .\label{6.6}
\end{eqnarray}
 The exponent $\beta$ is called the { critical
exponent} of
the order parameter and $\tau=|(T-T_c)/T_c|$. 
It is a necessary condition for such  power law behavior
that the phase transition is not of first order since 
a power law 
is the paradigm of a physical law with no  scales
 involved. In a first order transition, however, the jump in the
thermodynamic potential sets a physical energy scale.
A defining characteristic  of a critical phenomenon  is 
that there is also no
 relevant length scale at the critical point. Thus, the
{\em  correlation length}
 $\xi_c$ of the order parameter, $m=\BRA \varphi({\bf r})\KET$, defined by
the statistical properties of the local order parameter 
field
\begin{equation}\label{6.7}
	\chi(r):=\BRA \varphi(0)\varphi(\r)\KET \propto 
	\exp\LK-r/\xi_c\RK\, ,
	\;\; r:=|\r|\, ,
\end{equation}
has to diverge at $T_c$
with a power law with some critical exponent $\nu$, 
\begin{equation}\label{6.8}
	\xi_c \propto \tau^{-\nu} \, .
\end{equation}

The crucial assumption in any scaling approach to critical phenomena
is that the critical exponents $ \b$,  $\nu$ have their origin in
the divergence of a single relevant length scale, the correlation length
$\xi_c$\index{correlation length}.
 At the critical point where $\xi_c$  diverges the
correlation
 function obeys a power law
\begin{equation}\label{6.9}
	\chi(r)\propto r^{-\tilde{\eta}}\, .
\end{equation}
Using $\xi_c$  as a cutoff when calculating the global susceptibility
\hbox{$\chi =\int d^{d}r\,\chi(r)$} 
we find a {\em  scaling
relation}
\begin{equation}\label{6.10}
	{\tilde{\eta}}=\frac{2\beta}{\nu} \, .
\end{equation}

Let us now return to finite systems of linear size $L$
and assume that both the distance $r$ 
and the finite system size $L$  are within the  regime without 
length scales  and   that  the  
function
\hbox{$C(r,L):= \BRA \varphi(0)\varphi({\r})\KET_L$}
obeys  a power  law with respect to both lengths
\begin{equation}
	C(r,L)\propto r^{-\eta}L^{-y}\, .\label{6.11}
\end{equation}
The  new exponent $y$ describes the system length dependence and
$\eta\not=\tilde{\eta}$ describes the distance dependence. To consider 
 this case turns out to be relevant in the LD transition problem 
as will be discussed
 in more detail in Sect.~\ref{submul}. A similar reasoning as above yields now 
\begin{equation}
	 y+\eta ={\tilde{\eta}}= \frac{2\beta}{\nu} \, .\label{6.12}
\end{equation}
Notice that  the quantity $\eta$ is not determined by
$\beta $ and $\nu$ alone but requires the additional
knowledge of the unusual exponent $y$.

Far away from the critical regime 
there may still exist a {\em  scaling variable}
 $\Lambda(L)$ which replaces the role of $T$ in the following sense.
The state of the system is described for a finite size $L$ already by
the quantity $\Lambda(L)$ and the flow with increasing system size $L$
is determined by a single function of this variable, the so called
$\beta$-function
\begin{equation}\label{6.13}
	\beta\LK\ln\Lambda\RK:= \frac{d \ln\Lambda (L)}{d \ln
	L}
\end{equation}
which should be a  function of $\ln\Lambda$ alone. Provided the
$\beta$-function
is smooth, the flow is determined by the solution of the differential
equation,Eq.~(\ref{6.13}),
frequently  called a {\em  renormalization group equation}.  The
regime where $\beta(\ln \Lambda)$ 
can be linearized around a fixed point $\Lambda^{\ast}$ of the
flow ($\beta \LK\ln\Lambda^{\ast}\RK =0$), 
\begin{equation}\label{6.15}
	\beta\LK \ln \Lambda\RK = \beta'\LK \ln \Lambda
	-\ln\Lambda^{\ast}\RK\, ,
\end{equation}
with $\beta'$ being the slope of the $\beta$-function at
$\Lambda^{\ast}$, is called the critical regime.
It is in this regime where power law scaling can be expected. 
If we
start with a system of size $L=L_0$, $\Lambda=\Lambda_0$ chosen
close to $\Lambda^{\ast}$,   and turn on the renormalization flow until the
system reaches a size $\xi_c$ determined by the range of validity  of
Eq.~(\ref{6.15}) (which is here by definition the { correlation
length})  we get
\begin{equation}\label{6.16}
	\LK\frac{\xi_c}{L_0}\RK^{\beta'} =
	\left(\frac{\ln\Lambda-\ln\Lambda^\ast}{\ln\Lambda_0-\ln\Lambda^\ast}
	\right)	 		\, .
\end{equation}
Assuming that the critical regime is narrow, we can expand
$\Lambda(\xi_c)$ and $\Lambda_0$ around $\Lambda^{\ast}$ and get for the
correlation length
\begin{equation}\label{6.17}
	\xi_c =
	L_0
	\left(\frac{\Lambda-\Lambda^\ast}{\Lambda_0-
	\Lambda^\ast}\right)^{1/\beta'}  \, .
\end{equation}
The width of the critical regime  
\hbox{$\Delta\Lambda=
\Lambda_0
-\Lambda^{*}$} is  determined by $\tau$ such that
\begin{equation}\label{6.18}
	\xi_c \propto \LK\Delta\Lambda\RK^{-\nu}\propto  \tau^{-\nu}
	\, .
\end{equation}
Thus, the exponent $\nu$ is given by the inverse slope of the
$\beta$-function at the fixed point  
\begin{equation}\label{6.19}
	\nu=1/\beta'(\Lambda^{*})\, .
\end{equation}
The behavior of the scaling variable close to the transition is therefore
of the form 
\begin{equation}
	\Lambda (L;\tau) = \Lambda^\ast + A \tau L^{1/\nu} + {\cal
	O}\LK \tau^2\RK  \, .\label{6.19b}
\end{equation} 
The coefficient $A$ can be brought to the form $A \tau L^{1/\nu}=
\tilde{A} \LK L/\xi_c\RK^{1/\nu}$ where $\tilde{A}$ is of order
$\Lambda^\ast$.
Note, that, in general, 
 the coefficient $\tilde{A}$ can differ on both sides of the
transition, while the critical exponent $\nu$ is the same.

It is crucial that any other scaling variable $\tilde{\Lambda}$ which
is a smooth function of the previous one leads to the same value of
$\nu$,
although the overall form of its $\beta$-function can differ.
This reflects the fact that critical exponents are universal numbers
characterizing a whole class of systems. Such systems may differ by
microscopic details, however the critical behavior is characterized
by universal critical exponents of a few relevant quantities.

In  Sect.~\ref{subphy}
we have learned that 
knowing  the  Green's function 
is, in principle, enough to describe the transport properties
including the LD transition. A natural guess for the order parameter
field
is the LDOS since it typically vanishes in the localized phase and is
finite in the delocalized phase. Consequently, the average DOS
 (which is determined by the one-particle Green's
 function) appears as the natural (formal) order parameter.
Since 
 the average DOS  does not show
the LD transition but is a smooth function of the 
energy (which, in the problem of
the LD transition, is the analog of  temperature in equilibrium phase
transitions) 
the {\em  critical exponent}  $\beta$ of this formal order parameter vanishes
\begin{equation}\label{6.20}
	\beta  =  0  \, .
\end{equation}
The LD transition is not manifested  in the  one-particle Green's
function, but in transport  related quantities such as the
two-particle Green's function. 
We like to mention that this fact
 rules out the possibility of a straightforward mean-field
theory for the LD transition.

Furthermore, as
the scaling relation Eq.~(\ref{6.10}) tells  us the correlation
exponent
of the local susceptibility (a two-particle Green's function) vanishes,
\begin{equation}\label{6.21}
	{\tilde z} =0\, .
\end{equation}
This result seems to rule out a  power law behavior with finite power
for the two
particle Green's function (say the  density correlator) at criticality.
In spite of that a  finite power has been observed \cite{ChalDan88} for
the density correlator at a LD transition.
However, it turned out that it  corresponds to
a regime where the correlation function shows a power law (with
unusual exponent $y$) also with
respect to the system size $L$.

A further unusual feature of the LD transition, which at first glance 
seems to be independent of the previous ones, was first pointed out  by
Al'tshuler {\em et al.} \cite{Alt86-89} when
reexamining
the 
phenomenological scaling theory  of Abrahams et
al. \cite{Abr79}. 

This theory picks up the ideas that we introduced in the beginning of
this section. The 
conductance $g(L)$
   of a cube with linear dimension $L$ is treated as a scaling
 variable,
i.e. $g$ obeys a differential equation in terms of 
a $\beta$-function
which, for a certain {\em  universality class}, is a unique function of
$\ln g$. The fact that one scaling variable is enough to describe the
transition is referred to as {\em one-parameter} scaling theory.

That such treatment makes sense can be seen when calculating the
$\beta$-function for extreme situations. In a good metal with a
conductivity that is a material constant, Ohm's law, Eq.~(\ref{2.16}),
yields
\begin{equation}\label{6.22}
	\b (\ln g) \equiv  d-2
\end{equation}
while in the exponential localized phase one obtains
\begin{equation}\label{6.23}
	\b (\ln g) = \ln g + A_d
\end{equation}
where $A_d$ is a constant  for each dimension $d$.
Assuming $\b(\ln g)$ to be a smooth function one may interpolate between the
asymptotic branches (see Fig.~20). From this picture one
concludes:
In 1D all states are localized, in 3D a generic LD transition takes
place and for 2D the situation depends on the sign of
weak-localization effects for large $g$. If these tend to localize,
the $\beta$-function stays negative and all states will localize in
the large $L$ limit. For weak anti-localization  (found in 2D
systems with spin-orbit interaction \cite{Hik80,Ber84})
 a LD transition should occur.
It has been observed by numerical model calculations (see
e.g.~\cite{Fas91}). 
In an
experiment by Kravchenko {\em et al.} \cite{Kra95} on a strongly interacting
2D electron system a LD transition was observed. At present it is not clear if
it can be  due to strong  spin-orbit scattering.
  The quantum
Hall effect, as mentioned in the introduction, is exceptional   in
this picture. This is due to  the absence of a metallic phase.
The corresponding $\b$-function terminates from below at the critical point.

Although the one-parameter scaling theory by Abrahams {\em et al.} \cite{Abr79}
  is
simple, predictive and qualitatively correct, it has an obvious
shortcoming as was pointed out by Al'tshuler {\em et al.} \cite{Alt86-89}.  
The
conductance of a mesoscopic system strongly depends on the individual
system properties (e.g.~on the given disorder potential). As a
result, an ensemble of different systems exhibits conductance
fluctuations which can become so strong that the corresponding
distribution cannot be characterized by the mean value alone. The
latter as well as higher moments,
 can be drastically influenced by the far tails of the
distribution. It can thus happen that the mean value of the
conductance $\BRA g \KET$ can {\em not} serve as a scaling variable.

Instead,
 an appropriate scaling approach
to the LD transition
has  to be set up in terms of 
the complete distribution function of the conductance. However,
as pointed out by Shapiro \cite{Sha87}
it may be possible to apply the renormalization group approach to
certain
parameters occurring in that  distribution function.
If an appropriate parameter of the distribution exists which is less
 influenced by the far tails and which can serve to
 define a {\em  typical} conductance, ${g_{\rm t}}$, it may be possible
to find 
 a flow equation in
 terms of a $\b$-function for this typical conductance.
In this sense the theory by Abrahams {\em et al.} can
 still be complete.
For this scenario to be realistic,
 at least in the vicinity of the transition point,
only {\em  one} length scale $\xi_c$ should diverge at criticality. 
Then the LD transition   obeys one-parameter
scaling  and the exponent $\nu$ 
of the  correlation length $\xi_c$ is  given by
the  $\beta$-function of ${g_{\rm t}}$.

It is worth mentioning that the conductivity, defined by
the conductance as $\sigma(L) = G(L)L^{2-d}$, is often referred to as
an order parameter of the transition. It vanishes  at the transition
point in $d>2$ and  $\sigma(L\to \infty,\tau)
 \propto \tau^s$ defines an critical exponent $s$.
However, this notion of conductivity gives no further information than
that already contained in  the conductance. Only for $d>2$ the conductivity
does vanish at the transition point and the exponent $s$ is trivially
related to $\nu$ by $s=(d-2)\nu$  since $G(\xi)- G^{\ast}\propto \tau
\xi^{1/\nu}$. This relation is sometimes referred to as a scaling relation
between critical exponents $s$ and $\nu$ suggesting that this replaces
the scaling relation Eq.~(\ref{6.10}) in common critical phenomena. 
To the author this interpretation is misleading. Firstly, the
conductivity is defined via the conductance, which is the scaling
variable
of the problem. For any critical phenomenon with a scaling variable
$\Lambda(L)$ 
one can define similar {\em  order parameters} by simply multiplying the
scaling variable by a negative  power of $L$. It is obvious that this
does not give further insight. Secondly,
for LD transitions occurring in $d=2$ the conductivity stays finite at the
transition, which invalidates the notion of an order parameter for $\sigma$.
  Thirdly, the conductivity is not related to a local
field of the problem. 

In Sect.\ref{statra}  we will  see that the choice of an
 appropriate order parameter
and the occurrence of broad distributions of physical quantities are intimately
interrelated.

\subsection{Scaling Of Typical Values}\label{subscatyp}
In this section we will exploit our experience with the notion of
typical conductance in 1D and make an extension to higher dimensions
following a work by Shapiro \cite{Sha87}.
 The extension relies on uncontrolled
approximations and may  thus be only of  pedagogical interest.
However, the results are surprisingly far-reaching and consistent with 
commonly accepted results about LD transitions. 

Recall that the typical transmission probability in 1D followed
a simple multiplicative composition law which  under a scale
transformation $L\to bL$ leads to the scaling law in 1D
\begin{equation}
	T_\ty (bL)= \LK T_\ty(L)\RK^b\, .\label{6.24}
\end{equation}
We would like to generalize this law to higher dimensions. The
multiplicative nature of the composition law is a consequence of the
multiplicative nature of transfer matrices when put in series. If the
cross section of the conductor is finite the transfer matrix acquires
more modes. The parallel composition law for classical conductance
treats all modes equal and so do we. 
To this end, we first have to define an appropriate conductance
for a quasi-1D system for which the parallel composition makes sense.
The notion of intrinsic resistance $\hat{R}$  and
conductance $\hat{g}=\hat{R}^{-1}$, respectively, is helpful. In 1D we have
$\hat{R}=(1-T)/T$ which can take all positive values. By
Eq.~(\ref{6.24})
the series 
composition law reads  
\begin{equation}
	\hat{R}(bL)= \LK 1+ \hat{R}(L)\RK^b -1 \, .\label{6.25}
\end{equation}
We adopt  this law  for any series composition in quasi-1D
when we put $b$ blocks of size $L^d$ in series.
That this makes sense can be seen by considering the cases with
$\hat{R}(L)\gg 1$ and $\hat{R}(L) \ll 1$, respectively.
For weak resistivity of each block, the resistances behave additive  like in
classical resistor networks. However, 
for strong resistivity of each block we recover the multiplication law
leading to strong localization which is consistent with the finding of
strong localization
in quasi-1D. Once we have put the $b$ blocks in series, forming a
quasi-1d conductor of cross section $L^{d-1}$ and length $bL$ we put 
$b^{d-1}$ of such conductors in parallel, resulting in a cubic
conductor of total volume $(bL)^d$. For this composition we assume the
classical
parallel composition law to be valid \footnote{Note, that this procedure
is not capable of neither strong magnetic field nor spin-orbit
interaction.
In both of these cases a quantum parallel composition has to be constructed.
},
\begin{equation}
	\hat{R}(bL) = b^{1-d}\LBK \LK 1+ \hat{R}(L)\RK^b -1\RBK \,
	.\label{6.26}
\end{equation}
The application of parallel composition to
arbitrary
scaling factors can become quite inadequate. However, 
 we will now restrict to very small scaling factors
$b=1+\zeta$, $\zeta =\delta L/L \ll 1$ since we are only interested in
the change  of $\hat{R}$ under a small increase of system size.
By this we can easily derive the corresponding $\beta$-function for $\hat{R}$,
\begin{equation}
	\frac{d\ln \hat{R}}{d\ln L}= (1-d) +
	\frac{1+\hat{R}}{\hat{R}}\ln \LK 1+ \hat{R}\RK \, .
	\label{6.27}
\end{equation}
For easier comparison with the scaling theory of  Abrahams {\em  et
al.}
we reformulate the result in terms of the intrinsic conductance 
\begin{equation}
	\b(\ln \hat{g})= \frac{d\ln \hat{g}}{d\ln L}= (d-1)   
	-\LK 1+\hat{g}\RK\ln \LK 1+ \hat{g}^{-1}\RK \, .
	\label{6.28}
\end{equation}
One sees immediately that this $\b$-function  has the same asymptotic
behavior as in the scaling theory by Abrahams {\em et al.}, i.e.
for $\hat{g}\gg 1$,  $\beta(\ln \hat{g}) \approx d-2 -
\hat{g}^{-1}/2$,
and for $\hat{g}\ll 1 $, $\beta(\ln \hat{g}) \approx \ln \hat{g}$.
Furthermore, 
it has no zero for $d\leq 2$ (see Fig.~21) and for
$d=2+\epsilon$ with $\epsilon\ll 1$ one recovers the results
$\hat{g}^{\ast}=1/\epsilon= \nu$ of a much more involved field
theoretic calculation \cite{Weg80} which is expected to be exact in the
limit $\epsilon \ll 1$. In addition, one finds for $d=3$ a value of
$\nu \approx 1.65$ which is close   to  known results ($\nu\approx
1.4$ \cite{Hen94,Hof93,Hof94})
of numerical
calculations for 3D models .

\subsection{Scaling Of Distribution Functions}\label{subscadis}
We are now going to address the flow of distribution functions.
As a guiding example we consider the 1D conductor. Here localization
on the scale of the mean free path $l_e$
dominates and we can expect to get an almost complete description.

The composition law for transmission, Eq.~(\ref{3.54}), translates to that
of intrinsic resistance, $\hat{R}$,
\begin{equation}
	\hat{R}_{12}= 2\hat{R}_1\hat{R}_2 + \hat{R}_1+\hat{R}_2 -
	2\cos\phi \sqrt{\hat{R}_1\hat{R}_2\LK 1+\hat{R}_1\RK\LK 1+
	\hat{R}_2\RK }\label{6.29}
\end{equation}
We can use this to derive an evolution equation for the distribution
function ${\cal P}(\hat{R};L)$. Since the procedure is of general
importance
we describe the essential steps (cf.~\cite{Ris89}):
a stochastic quantity $X(t)$ depending on time $t$ may increase in time
$\delta t$ by $\delta X (t)= X(t+\delta t)- X(t)$. If the stochastic
 process is Markovian, i.e. it is determined by transition
probabilities
from one time step to the next and is not influenced by the history of
the process the essential information is contained in the moments,
 $\BRA \LK \delta X (t)\RK^n \KET$. If, in addition, only the
first
two   moments scale linear with $\delta t$ (while all higher moments
scale with higher powers) a Fokker-Planck equation (FPE)
 results 
\begin{equation}
	\frac{\partial {\cal P}(X;t)}{\partial t}=
	 -\frac{\partial \LBK{\cal D}^1(X) {\cal P}(X;t)\RBK}{\partial X} +
	 \frac{\partial^2 \LBK {\cal D}^2(X){\cal P}(X;t)\RBK}{\partial X^2} 
	\label{6.30}
\end{equation}
where ${\cal D}^1(X)= \BRA  \delta X (t) \KET /\delta t$ and 
${\cal D}^2(X)= \BRA \LK  \delta X (t) \RK^2 \KET /2\delta t$ (for
$\delta t \to 0$) are
called
{\em  drift} and {\em  diffusion} function, respectively.

In our problem, the stochastic quantity is $\hat{R}(L)$
and by Eq.~(\ref{6.29}) we obtain
\begin{equation}
	\delta \hat{R} (L) = 2\hat{R} (L)\hat{R}(\delta L) 
	+\hat{R}(\delta L)  -
	2\cos\phi \sqrt{\hat{R}(L)\hat{R}(\delta L)\LK 1+\hat{R}(L)\RK\LK 1+
	\hat{R}(\delta L)\RK } \, .\label{6.31}
\end{equation}
The average is to be taken over random phases $\phi$ and over the
strip-resistance $\hat{R}(\delta L)$ which introduces the mean free
path
$\BRA \hat{R}(\delta L)\KET = \delta L/l_e$. This definition is often
referred to as {\em  local weak scattering condition} and is the basis of
a FPE approach to the statistics. Equation~(\ref{6.31})
shows that only the first and second moments are linear in $\delta L$
while
higher moments scale with higher powers. Thus, a 
FPE  can be derived for ${\cal P}(\hat{R},L)$ with drift and
diffusion given by
\begin{equation}
	{\cal D}^1= \frac{2\hat{R} +1}{l_e}\, ,\;\; {\cal D}^2 =
	\frac{\hat{R}^2+\hat{R}}{l_e}\, .\label{6.32}
\end{equation}
The FPE can be brought to the form \cite{Mel81}
\begin{equation}
	\frac{\partial {\cal P}(\hat{R};L)}{\partial L}=
	 \frac{1}{l_e}
	 \frac{\partial}{\partial \hat{R}}\LBK \LK \hat{R}^2
	 +\hat{R}\RK
	\frac{\partial {\cal P}(\hat{R};L)}{\partial \hat{R}} \RBK\, .
	\label{6.33}
\end{equation}
This distribution depends on only one parameter, the mean free path
$l_e$.
Two limiting cases can be considered now: (i) $\hat{R}\ll 1$ and (ii)
$\hat{R}\gg 1$.

For case (i) the solution is an exponential distribution 
\begin{equation}
	{\cal P}(\hat{R};L)= \frac{l_e}{L}\exp\LK -\hat{R}l_e/L\RK \,	
	\label{6.34}
\end{equation}
with average value $\BRA \hat{R}(L)\KET = L/l_e$. This
solution corresponds to system sizes $L\ll l_e$.

For case (ii) the solution is a log-normal distribution
\begin{equation}
	{\cal P}(\hat{R};L) d\hat{R} = \frac{1}{\sqrt{4\pi L/l_e}} \exp
	\LBK -\frac{\LK \ln \hat{R} -L/l_e \RK^2}{2(2L/l_e)} \RBK d\ln\hat{R}
	\,\label{6.35}
\end{equation}
with a typical value $\hat{R}_{\rm t}=\exp \BRA \ln \hat{R}\KET =
\exp\LK L/l_e\RK $. This solution corresponds to 
localization with $L\gg l_e$, and a typical localization length
of $\xi=2l_e$, consistent with our previous finding (see
Eq.~(\ref{3.58})).
Interesting is the fact that the second parameter of the log-normal
distribution, the log-variance $2L/l_e$, is simply related to the
log-average value
\begin{equation}
	\BRA \LK \ln \hat{R}- \BRA \ln
	\hat{R} \KET \RK^2  \KET = 2\BRA \ln \hat{R}\KET = 2L/l_e \, .
	 \label{6.36}
\end{equation}
The log-normal form of the distribution for $L\gg \xi$ is also
consistent with  the general statement, Eq.~(\ref{3.67}), that the
 localization length ( $\frac{2L}{\nu_1}$) corresponds to normally
distributed quantities that become self averaging in the limit $L\to
\infty$.

The distribution ${\cal P}(\hat{R}; L)$ 
is broad even for $L\ll \xi$ (since an exponential
tail
already introduces a strong growth of moments  $\propto
n!$)  and  causes  large fluctuations for $\hat{R}$.
The observation of broad distributions is central to mesoscopic
systems and therefore we  make some general statements about
them. Following Shapiro \cite{Sha87} we formulate the concept of
one-parameter scaling for broad distributions as follows.

A distribution of a physical quantity $X$ that depends on system size
$L$ and a set of initial parameters $\LB \alpha_n
\RB$,
${\cal P}(X;L;\LB \alpha_n
\RB)$, obeys  one-parameter scaling if (for large $L$)
it is approximately a function of only $X$ and one scale dependent
parameter
$\alpha_L$, 
\begin{equation}
	{\cal P}(X;L;\LB \alpha_n
\RB)\approx F(X;\alpha_L) \, ,\label{6.37}
\end{equation}
and the flow of the parameter $\alpha_L$ is determined by a $\b$-function
\begin{equation}
	\b(\ln \alpha_L) = \frac{d \ln \a_L}{d \ln L}\, .\label{6.38}
\end{equation}
That means that the whole set of initial parameters $\LB \alpha_n
\RB$, needed to specify the initial small-scale distribution, disappear
into the single parameter $\alpha_L\LK\LB\alpha_n\RB\RK$. The exact
distribution {\em  does} contain all the information about the initial
set of parameters; however, the information can be accumulated in the
very far
tails of the distribution. By {\em  very far 
tail} of a distribution we denote that
range
of values $X < X_1$, $X>X_2$ 
that is not statistically significant to some small but
fixed 
probability $\delta$ 
\begin{equation}
	\int\limits_{X_1}^{X_2} dX {\cal P}(X;L) = 1-\delta \, .\label{6.39}
\end{equation}
By Eq.~(\ref{6.37}) it is meant that the approximation for the main part
of the distribution is exact up to corrections of order $\delta$.
For sufficiently large $L$ it  can be taken to be arbitrarily small.
Still, the moments of the distribution might be dominated by the
very  far tails of the exact
distribution. Therefore, one cannot expect that in general the parameter
$\alpha_L$ is simply related to moments of the distribution.

As a consequence of one-parameter scaling there should exist a universal
limiting distribution at a critical  point
\begin{equation}
	\lim\limits_{L\to\infty} {\cal P} (X;L)= {\cal P}^\ast (X)
	\label{limitdis}  
\end{equation}
which is  fixed by the critical value
$\alpha^\ast=\lim\limits_{L\to\infty} \alpha_L$ of the scaling variable.

A distribution is said to be broad if its moments $\BRA X^n\KET$ grow 
like  $\exp f(n)$ with $f(n)$ increasing  stronger than linear with $n$,
or if its moments might even diverge.
This notion refers to the fact that the corresponding distribution
of the sum of $N$ independent realizations of the  variable $X$ will
still
deviate substantially from a Gaussian, even for very large $N$. 
According to  this definition of {\em  broadness} a  weak case of a broad
distribution
is realized by the
exponential distribution ($X\geq 0$)
\begin{equation}
	{\cal P}(X) = \frac{1}{\BRA X\KET }\exp \LK - X/\BRA X\KET \RK
	\label{6.40}
\end{equation}
for which the moments grow as $\BRA X^n\KET \propto n! \sim \exp (n\ln
n)$.
The prototype of a broad distribution is the log-normal distribution
\begin{equation}
  {\cal P}(X)dX 
	= \frac{1}{\sqrt{2\pi}\sigma}\exp \LBK
	-\frac{1}{2\sigma^2} \LK \ln X - \BRA \ln  X\KET\RK^2
	\RBK d \ln  X \label{6.41}
\end{equation}
the moments of which grow as $\BRA X^n\KET \propto \exp( \sigma^2
n^2/2)$.
It represents the generic distribution for random multiplicative
variables. 

After these general remarks we would like to 
 calculate the conductance distribution in higher dimensions
including the LD transition. However, this  is difficult.
An interesting attack has been undertaken by Shapiro \cite{Sha87}.
He used the classical 
parallel composition law  to derive Fokker-Planck equations
for the conductance distribution in higher dimensions. However, the
 uncontrolled approximations involved overestimate fluctuations. 
Analytical approaches are usually restricted to dimension
 $d=2+\epsilon$
where the critical conductance is of order $\epsilon^{-1}$ and
perturbation theory together with renormalization techniques can be
 used.
We do not enter this technically involved subject and we will briefly
review
 the main conclusions in Sects.~\ref{starea},\ref{statra}. 

Here, we will turn  to local quantities and address the question 
of power law scaling relevant to the delocalized phase and the LD
transition.

In the delocalized phase wavefunctions are extended all over the
system. The correlation length $\xi_c$ in this phase can be estimated
from
\begin{equation}
	g(\xi_c)= {\cal O} (1)\, .\label{6.42}
\end{equation}  
Taking the metallic limit, Eq.~(\ref{Dru7}), one finds that ($d>2$)
\begin{equation}
	\xi_c \sim \LK \frac{\lambda_F}{l_e}\RK^{\frac{1}{d-2}}l_e\, 
	\label{6.43} 
\end{equation}
is a truly microscopic scale. Thus, 
 one can expect a power law scaling of moments of the box
probability
$P(E,L_b)$ for $\xi_c \ll L_b\ll L$. In the critical regime of the LD
 transition the correlation length $\xi_c$ is larger than system size
 L  and one can thus also expect  a power law scaling of moments of the box
probability
$P(E,L_b)$ for $ l \ll L_b\ll L \ll \xi_c$ (where $l$ stands for
 microscopic scales). 
Therefore, we will now consider distribution functions the moments of
 which
display power law scaling.
To simplify notations we consider a random variable $X\in [0,1]$ and a scale
 $\lambda \in [0,1]$ with distribution ${\cal P}(X;\lambda)$.
We assume that for sufficiently small $\lambda \ll 1$ the moments
 obey
power law scaling 
\begin{equation}
	 \BRA X^q \KET_{\lambda} = \int\limits_0^1 dX \, {\cal
	 P}(X;\lambda) X^q = c_q \lambda^{{\tilde{\tau}}(q)} \, \label{6.45}
\end{equation}
reflecting that scales in the system are separated.
Here $c_q$ and ${\tilde{\tau}}(q)$ are treated as functions of arbitrary
real values $q$.

The power law scaling condition is very restrictive and allows for a
number of conclusions:
\begin{enumerate}
\item 
Normalization requires
\begin{equation}
	c_0=1 \, ,\;\; {\tilde{\tau}}(0)=0 \, .\label{6.46}
\end{equation}
\item
Since ${\tilde{\tau}}(q)$ is independent of $\lambda$ one can calculate its
derivatives and send $\lambda \to 0$. By this one can conclude two
strong inequalities 
\begin{equation}
	{\tilde{\tau}}'(q) \geq 0\, , \;\; {\tilde{\tau}}''(q) \leq 0 \,\label{6.47}
\end{equation}
where the prime  means differentiation with respect to $q$.
\item 
Introducing the variable $Y=-\ln X \in [0,\infty[$ one can see that
the moments are the Laplace transform of the function
$e^{-Y}{\cal P}(X(Y);\lambda)$,
\begin{equation}
	\BRA X^q\KET_{\lambda} = \int\limits_0^{\infty} dY\, \LBK e^{-Y} {\cal
	P}(X(Y);\lambda)\RBK e^{-Yq} \, .\label{6.48}
\end{equation}
Thus, one may reconstruct the distribution from the moments by
inverse Laplace transformation (if the latter can be carried out).
\item
The case of strictly linear ${\tilde{\tau}}(q)=\gamma q $ ($\gamma=0$ is not
interesting)
allows for  a {\em  global} rescaling of the distribution function
by its first moment.
\begin{equation}
	{\cal P} (X;\lambda) = \frac{1}{\lambda^\gamma}
	\tilde{\cal P} \LK \frac{X}{\lambda^\gamma}\RK \, ,\;\;
	\BRA X\KET_\lambda = c_1\lambda^\gamma \, ,\label{6.49}
\end{equation}
and the coefficients $c_q$ are determined by the scale independent
distribution $\tilde{\cal P}(Z)$
\begin{equation}
	c_q=\int\limits_0^1 dZ\, \tilde{\cal P} (Z) Z^q \, .\label{6.50}
\end{equation}
A linear behavior of ${\tilde{\tau}}$ seems to be  natural for power law
scaling; all moments scale as powers of the first and only one
critical exponent describes the scaling of all of them, including that
of the distribution function. The rescaled distribution function is
now scale independent and can take any form, e.g. it can be a broad
distribution in the sense defined above. 
Examples are the Rayleigh and Porter-Thomas distributions found for the
box-probability in the ideal delocalized phase
(see~Eqs.~(\ref{5.25},\ref{5.26})).  
   We refer to the
situation of linear $\tilde{\tau}(q)$ as the {\em  gap scaling} situation. 
\item
 The linear case is not the most general one. 
For finite curvature in ${\tilde{\tau}}(q)$, ${\tilde{\tau}}''(q) <
 0$,
the
 coefficients
$c_q$ can be treated as constant since the essential $q$ dependence
 that dominates for small $\lambda$ comes from $\tilde{\tau}(q)$. 
We will see in the following that  
the distribution function ${\cal P}(X,\lambda)$ can be essentially
reconstructed from ${\tilde{\tau}}(q)$ by a Legendre transformation and it is
 an almost log-normal distribution with special features.
One can write 
\begin{equation}
	{\cal P}(X;\lambda) dX = {\cal N}  \lambda^{-{\tilde{f}}
	(a)} da  \, ,\;\; a:= \frac{\ln X}{\ln \lambda} \, 
	\label{6.51} 
\end{equation}
where the normalization ${\cal N}$ has only a weak $\lambda$
dependence (e.g.~$\sim \ln \lambda$).  The essential
information is now contained  in the function $\tilde{f}(a)$ which
is the Legendre transform of $\tilde{\tau}(q)$,
\begin{equation}
	{\tilde{f}}\LK a(q)\RK = a(q) q - {\tilde{\tau}}(q) \, ,\;\; 
	a(q):= \frac{d {\tilde{\tau}(q)}}{dq} > 0 \, .\label{6.52}
\end{equation}
Before we outline a derivation of the statement we make further comments.
\item
In contrast to the gap scaling situation with arbitrary form of the scale
independent distribution we have now the situation that the form
of the distribution is fixed by a spectrum of scaling exponents.
 This
situation has been phrased (in a bit more restricted context)
{\em  multifractal scaling} situation (for a review see \cite{Pal87}). 
The spectrum has several analytical properties following from the
Legendre transformation. It has a unique maximum at $(a(0),0)$ and 
a slope of $1$ at $a(1)$. The  derivatives fulfill
\begin{equation}
	\frac{d{\tilde{f}}}{da}=q(a) \, ,\;\;
	\frac{d^2{\tilde{f}}}{da^2}= \frac{1}{
	{\tilde{\tau}}''(q(a))} < 0\, . \label{6.53}
\end{equation}
Close to the typical value
\begin{equation}
	X_{\rm t}(\lambda)= \exp \LK \BRA \ln X \KET_\lambda \RK
	\propto \lambda^{a_0} \, ,\;\; a_0=a(0) \, ,\label{6.54}
\end{equation}
the function $\tilde{f}(a)$ is well approximated by a parabola and 
${\cal P}(X;\lambda) $ is approximated by a log-normal distribution
\begin{equation}
	{\cal P}(X;\lambda) d X \approx {\cal N} \exp \LBK -\frac{
	\ln^2  \LK X/X_{\rm t}\RK 
	}{2\sigma^2 (-\ln \lambda)}\RBK d\ln X
	\label{6.55}
\end{equation}
where $\sigma^2 = -1/\tilde{\tau}''(0)$ is a positive constant.
\end{enumerate}
The derivation of Eq.(\ref{6.51}) relies on the fact that a finite
curvature
of $\tilde{\tau}(q)$ allows for a stationary point evaluation of the 
integral  in Eq.~(\ref{6.45}) for each moment $\BRA X^q\KET_\lambda$
in the limit $\lambda \to 0$. Writing $X=\lambda^a$ and transforming 
to the variable $a$, with the  distribution  $\tilde{\cal P} (a,\lambda)d a
= {\cal P}(X,\lambda) dX $, yields the integral relation
\begin{equation}
	 c_q \exp\LBK {\LK\ln \lambda \RK{\tilde{\tau}}(q)} \RBK =
	\int\limits_0^\infty da\, \exp\LBK -\ln \lambda \LK 
	F(a,\lambda) - a q  \RK\RBK \, \label{6.56a}
\end{equation}
with 
\begin{equation} 
	F(a,\lambda):=-\frac{\ln \LK \tilde{\cal P}(a,\lambda)\RK}{\ln
	\lambda}
	\, .\label{6.56b}
\end{equation}
The function $F(a,\lambda) -a q$ can have a maximum with respect to
$a$ (for $\lambda\ll 1$) that is only weakly dependent on $\lambda$.
It yields a sharp maximum in the
integrand (due to the large pre-factor -$\ln \lambda$). Then, the integrand
is determined up to ${\cal O}(1/(-\ln \lambda))$ by the value of the
integrand
at this maximum and one finds that ${\tilde{\tau}}(q)$ is the Legendre
transform of $F(a,\lambda)$. Consequently, $F(a,\lambda)$ must have
a very weak $\lambda$ dependence (e.g. $\sim \ln |\ln \lambda|$).
In case that $F(a,\lambda) -a q$ has no   maximum with respect to
$a$ that is only weakly dependent on $\lambda$
 the corresponding ${\tilde{\tau}}(q)$ will not have
finite
curvature, but the gap scaling situation applies. 

The phrase {\em  multifractality} has its origin in the interpretation of
 $X(\lambda)$ as a local
 measure defined for
a geometrical object embedded in a  $d$-dimensional cube of linear size $L$.
The scale $\lambda$ is given by the ratio
  $(L_b/L)^d$
where $L_b$ is the linear  size of a box on which the measure 
is calculated. Due to normalization $X(1)=1$ the average of $X(\lambda)$ 
is the inverse of the number of non-empty boxes and thus
yields, for $\lambda\ll 1$, the fractal dimension $D_F$ of
the geometrical object
\begin{equation}\label{6.57}
	\BRA X\KET_{\lambda} \propto \LK \frac{L_b}{L}\RK^{D_F}=
	\lambda^{D_F/d} \, . 
\end{equation}
The fact that higher moments $\BRA X^q\KET_\lambda$ may scale not
simply as  $\propto \lambda^{(D_F/d)q}$  but as 
$\propto \lambda^{(1/d) (D_F  + \tau(q))}$ where $\tau(q)$ defines a
set of $q$-dependent {\em  generalized dimension} $D(q)$ by
 $\tau(q)=: (q-1)D(q)$,  has 
 led  to the notion of multifractality. Especially, one has
$\tau(0)=-D_F$, $\tau(1)=0$. 
The Legendre transform of $\tau(q)$ then defines the so-called
multifractal $f(\a)$-spectrum. It  can be  interpreted as 
the fractal dimension of that subset of the  object where the
measure scales locally as $\LK L_b/L\RK^\a$ (see~\cite{Hal86}).
The general analytic properties of multifractal exponents
are visualized in Fig.~22

We see that, for measures  $X(\lambda)$,  we have
the following identification
\begin{equation}
	\tilde{\tau}(q)= \frac{1}{d}\LK D_F + \tau(q)\RK \, , \;\; 
	\tilde{f}(a) = \frac{1}{d}\LK -D_F + f(\a) \RK \, , \;\; a =
	\frac{\a}{d} \, .\label{6.58}
\end{equation}
Under quite general conditions, $\tilde{f}(a)$ (or $f(\a)$)
is  also bounded from below by 
$-D_F/d$ (or $\tau(1)=0$) and terminates with infinite slope at the endpoints
of the finite interval $[a(\infty), a(-\infty)]$ (or 
$[\a(\infty),\a(-\infty)]$).
In many cases the parabolic approximation of Eq.~(\ref{6.55})
extends to values $\alpha \geq \alpha(1)$ and the constraint on the
slope of $f(\a)$ at $\alpha(1)$ allows to write a parabolic
approximation (PA) for $f(\a)$ that depends only on  one-parameter (besides
$D_F$) $\alpha_0$
\begin{equation}
	f(\alpha)= D_F - \frac{\LK \alpha -\alpha_0\RK^2}{4\LK\alpha_0
	-D_F\RK}\, .\label{parabol}
\end{equation}
For example, the 
 box-probability $P(E;L_b)$ considered in the LD-transition problem
(Eq.~(\ref{5.13b})) is of this  type of variable, since the squared
 amplitude
of the 
 wave function defines a local measure.  The
fractal
dimension $D_F=d$  is always trivial.
 In the idealized delocalized phase we found the gap scaling class
 and $\tau(q)=
(q-1)d$. The interesting information is contained in the shape of the
corresponding  scale
independent distribution. 

For the LD transition, however, it is known from
a calculation by Wegner\cite{Weg80,Cas86}  for $2+\epsilon$ dimensions that    
the multifractal scaling class applies and $D(q)\not\equiv d$.
The PA he calculated yields 
\begin{equation}
	\alpha_0-d = \epsilon= 1/g^* \, .\label{Wegneralpha0}
\end{equation}

\medskip

In summary, 
we found  that the restriction to power-law scaling
leads to two {\em  distinct} classes of distribution functions.
The gap scaling class  is characterized by a single scaling
exponent
and  a related scale independent distribution of arbitrary shape.
The multifractal scaling class  is characterized by a spectrum
of scaling exponents and a shape of the distribution close to
log-normal. The distribution of local box-probability in the ideal
delocalized phase falls into the gap scaling class.
 The possibility of the multifractal scaling class is
interesting for the LD transition point, because it opens a
possibility
to find an appropriate order parameter in terms of the LDOS.
If, as it can happen in the multifractal scaling class,
 average and typical DOS scale differently
with respect to system size, the inability of the average  DOS
to work as an order parameter must not be shared by the typical DOS.
In Sect.~\ref{sublocden} we will see that, indeed, 
this is the case.
\section{Statistics in Realistic Models}\label{starea}
In this section we will review known facts about the statistics of
conductance, energy levels and local quantities for  the delocalized
phase with large typical conductance $g\gg 1$ and for the localized
phase with finite typical localization length $\xi \gg l_e$. 
The onset of broad distributions with log-normal tails 
in the delocalized phase is most
interesting
here. In 2D it leads to multifractality of the LDOS.
In order to interpret this phenomenon
we consider the conductance distribution for quasi-1D systems for
which
the number of channels is large, but the width is still of the order
of
a microscopic scale, e.g. the mean free path. Under these conditions
a Fokker-Planck approach works that reproduces the UCF phenomenon, but
the
long tails  are absent. We  come to the conclusion that the log-normal
tails are 
 a precursor of the LD transition
and do only occur if the relevant scaling variable is not infinite.
To study more realistic quasi-1D systems the method of Finite Size
Scaling (FSS) will be outlined. In this method a scaling variable is
defined
which is close to the definition of typical conductance.

\subsection{Onset Of Broad Distributions}\label{subons}
In the delocalized phase of a finite mesoscopic conductor 
where the average  conductance is large $\BRA g\KET \gg 1$ the 
B\"uttiker formula tells  that many (${\cal O}(g)$) open modes are occupied.
Thus, the conductance is a sum of many random numbers 
${\cal T}_i\ \approx 1 $.
Naive application of the central limit theorem suggests that
the conductance distribution will be a Gaussian centered around $\BRA
g\KET$ with a relative deviation $\sqrt{\BRA (\delta g)^2\KET}/\BRA g\KET \sim
1/\sqrt{\BRA g\KET}$ where $\delta g = g- \BRA g\KET$. 
 However, this reasoning ignores the strong
correlations between the transmission eigenvalues ${\cal T}_i$. This
correlation
leads to the occurrence of universal conductance fluctuations (UCF)
(cf.  Sec.~\ref{subuni}), 
i.e.
the variance is $\BRA (\delta g)^2 \KET \sim {\cal O}(1)$ and does
 neither depend on the
actual value of $\BRA g\KET$ nor on the system size $L$.
Taking UCF into account the conductance
distribution seems to follow one-parameter scaling: it
is Gaussian with scale dependent average value (which is also 
typical here) and a universal variance.
\begin{equation}
	{\cal P}(g;L) \propto \exp \LBK - \frac{
	\LK g-\BRA g\KET_L \RK^2}{2\BRA (\delta g)^2 \KET}\RBK \, .\label{7.1}
\end{equation}
 This was the common belief until the
work of Al'tshuler {\em et al.} \cite{Alt86-89}.
 In this work it was shown by a
combination of diagrammatic perturbation theory and
renormalization
group techniques that the conductance distribution deviates from a
Gaussian
for $|\delta g|\gg \sqrt{\BRA g\KET}$. For these tails the
distribution crosses over to a log-normal form (see~Fig.~23).
 This was concluded
from the calculation of high moments  which showed the characteristic
dependence 
\begin{equation}
	 \BRA (\delta g)^n\KET \propto  e^{un^2}\, .\label{7.1b}
\end{equation}
Here $u\sim {\cal O} (1/\BRA g\KET)$ indicating that a finite
value of conductance is responsible for the effect.

The calculations were carried out for 2D (with localization lengths
much larger than system size) and $d=2+\epsilon$ (in the delocalized  phase). 
In the work by Al'tshuler {\em et al.} it was conjectured that their finding
will
penetrate to the LD transition point, i.e. that the conductance
distribution
will become very broad.

The observations of these log-normal tails
led to questioning the whole idea of one-parameter scaling. Later,
Shapiro and coworkers \cite{Sha88-92} showed that these tails are not in
conflict with one-parameter scaling theory as defined in Secs.~\ref{subscadis}.
On the basis of the $2+\epsilon$ results they concluded that a
critical
conductance distribution will appear with long power-law tails.
Such tails lead to divergence of low moments and exclude any
one-parameter
scaling theory in terms of moments of the conductance.
 
Later works \cite{Myz95,Fal95,Mir95} 
confirmed the calculation of Al'tshuler {\em et al.}
 and  showed its  principle validity
for 3D as well. The 1D and quasi-1D cases need special attention since
 strong localization dominates.

In the localized phase the B\"uttiker formula tells that the
conductance
is dictated by the smallest value  $\nu_1$ corresponding to
eigenvalues of
$M^{\dagger}M$
(cf.~Eq.~(\ref{3.51})). This value is known to be Gaussian
distributed
and thus it is no surprise that the conductance distribution,
for $\xi\ll L$, $g\sim \exp\LK - \xi/L\RK \ll 1$,  is of the log-normal form.
This reflects   the Gaussian distribution of inverse
localization lengths. 
The typical localization length ($\BRA \nu_1 \KET^{-1}2L$) defines 
the typical conductance as the geometric mean $g_{\rm t}=\exp\LK\BRA \ln
g\KET\RK$.
Here, in principle a second parameter besides $g_{\rm t}$ might occur
in the distribution, namely the variance of $\nu_1$.
However, as happened already in the 1D case (cf.~Sect.~\ref{subscadis}), all
known results indicate that this variance is simply related to the
average value (cf.~e.g.~\cite{Mar93})
 and thus only one-parameter  seems to describe 
the distribution accurately,
\begin{equation}
	{\cal P}(g;L) d g  \propto \exp
	\LBK -\frac{\LK \ln g  - \BRA \ln g\KET_L  \RK^2}{2
	\sigma_L^2} 
	\RBK d\ln g
	\, ,\label{7.2}
\end{equation}
with
\begin{equation}
	\sigma_L^2  \propto \BRA \ln g \KET_L \, .
	 \label{7.3}
\end{equation}
The behavior of the conductance distribution close to the LD
transition
will be discussed in Sect.~\ref{subcond}.

Concerning the level statistics we restrict our discussion to the
two-level
correlation function $R(s)$. 
The energy separation $s\Delta$ can be
associated with an inverse time scale. The behavior of $R(s)$ is related
to the question how a wave packet evolves in time $\h/(s\Delta)$ 
\cite{Arg93,Cha96}.
For times larger than the diffusion time $t_D$ the wave packet explores
the whole system several times and it is not surprising that here
the standard random matrix theory results are approximately valid.
In the standard random matrix theory for the idealized delocalized phase
$R(s)$  only depends on the level distance $s$ measured in average
level spacing units. A finite, but  large, conductance is related to
the diffusion time  via the
 Thouless energy $E_{\rm Th}=h/t_D \gg \Delta$,
  the relevant transport energy scale. Consequently, the
dimensionless
conductance $g$ (here $g$ can be taken as   the average value)
 sets a scale on the $s$-line. Indeed, one finds
different level correlation functions for $s\ll g$ and for $s\gg g$.
In the first case, as expected, the function $R(s)$ is approximately
given by
the standard random matrix theory result and only small  corrections 
(proportional to 
powers of $1/g$) occur.
For example, the result  for systems with broken time reversal symmetry
reads 	 \cite{KraMir94}
\begin{equation}
	R(s) = - \LK \frac{\sin \pi s}{\pi s}\RK^2 + 
	\frac{A_d}{\pi^2 g^2}\sin^2\LK \pi s\RK \, \label{7.4}
\end{equation}
($A_d$ a constant depending only on   dimension and symmetry).

The  case $s\gg g$ has been investigated by Al'tshuler and Shklovskii
\cite{AltShk86}
and the finding is (for $s$ not exceeding the inverse of the microscopic
mean free time $\tau$ and possible
oscillations on the scale $s\sim 1$  are ignored)
\begin{equation}
	R(s)=\frac{C_d}{g^{d/2}|s|^{2-d/2}}\label{7.5}
\end{equation}
 where $C_d$ is a numerical coefficient that depends only on d and on the
symmetry class. The result can be understood by supposing 
diffusive motion of wave packets in the system.

In 2D there exists a broad cross over regime between the two
asymptotic regimes 
of Eqs.~(\ref{7.4},\ref{7.5}) \cite{KraLer95}.
 This can be understood as follows.
The 2D metal resembles a critical system in as much as the
corresponding
$\b$-function is close to $0$. The strong localization that will
finally dominate comes along with very large localization lengths.
It has to be expected that the level statistics will change
 on approaching the LD transition\footnote{We will discuss some of the 
consequences in
Sect.~\ref{subcond}.}.
 A similar behavior can be expected to occur in  the 2D
case.

The level-correlation function for the localized phase with 
 vanishing localization length is given by the Poisson
statistics, Eq.~(\ref{5.10}), i.e. no-correlation for $s\not=0$.
To take a finite localization length into account which is, however,
much smaller than the system size $L$, $\xi\ll L$,
it is
helpful to think of dividing the system into subsystems
of volume $\xi^d$. The level-correlation function
R(s;L) will then obey a scaling form \cite{Aro95,Alt95}
\begin{equation}
	R(s;L)= \LK \frac{\xi}{L}\RK^d
	F (s \LK \xi/{L}\RK^d) \, .\label{7.6} 
\end{equation}
In the thermodynamic limit the level correlations vanish,  
$\lim\limits_{L\to \infty} R(s,L) =0$, in accordance with the Poisson
statistics  being the limiting case.

The scaling form is in agreement  with the one-parameter scaling
theory since the variable $\xi/L$ is related to the typical
conductance in the localized phase, $g_t=e^{-2L/\xi}$.

The form of the scaling function $F(x)$ can be obtained for
 $x$ being
large or small compared to $1$. 
For $x\gg 1$,  the corresponding wave-packet dynamics probes only the
region
within one localization volume and is, thus, diffusive. Consequently
one recovers the Al'tshuler-Shklovskii behavior \cite{Alt95}
\begin{equation}
	F(x)\propto x^{-(2-d/2)} \, .\label{7.7}
\end{equation}
For $x\ll 1$  but $|\ln x|< L/\xi$  
 regions much larger than one localization volume are
probed
and the only correlation between levels comes from weak couplings
between such volumes \cite{Alt95} resulting in a logarithmic dependence
 on $x$
\begin{equation}
	F(x) \propto \LK \ln x\RK^d \, .\label{7.8}
\end{equation}
 
\medskip

We  now turn  to the statistics of local quantities.
Again, for a good metal with $g\gg 1$ the distribution of local
quantities is essentially given by the standard random matrix
theory (cf.~Eqs.~(\ref{5.25},\ref{5.26})). 
However, the long tails are different and the finite value of $g$
becomes important.
We concentrate on the box-probability $P(E;L_b)$ in the following.
This generalizes to the reduced LDOS $\tilde{\rho}(E;L_b)$.
We introduce the  normalized quantity
\begin{equation}
	t:=\frac{P(E;l)}{\BRA P(E) \KET_{l/L}} \, ,\;\; \BRA P(E) \KET_{l/L}=
	(l/L)^d  \label{7.9}
\end{equation}
and consider the case $g\gg 1$. The scale $l$ stands for microscopic box size.
The threshold for the standard random matrix result turns out to be
$t\sim 1/\sqrt{g}$.
For $t\ll 1/\sqrt{g}$ the Rayleigh and Porter-Thomas distributions are
valid
up to small corrections of ${\cal O} (1/g)$ (cf.~\cite{MirR95}).

For $t\gg 1/\sqrt{g}$   a number of authors \cite{Myz95,Fal95,Mir95,Smol96}
have found the onset of broad distributions
which in 2D and 3D develop  long tails
of the form
\begin{equation}
	{\cal P} (t) \sim \exp \LK -A_d \ln^{B_d} (t/t_d) \RK \, .\label{7.10}
\end{equation}
Here $A_d$ depends on the dimensionality $d$ and on the  disorder strength 
and $B_2=2$. For $B_3$ values of $2$ and $3$ have been
predicted. The value $t_d$ is a typical value serving as reference point.
 The precise expressions 
are  currently under debate (because the  authors of
\cite{Myz95,Fal95,Mir95,Smol96}  
 use  different
approximative calculation schemes). However, what seems to be clear is
the occurrence of long-tails of the general form Eq.~(\ref{7.10}).
For the 2D case there is also agreement that $A_2 \propto g/(\ln
L/l)$ and also $t_d\propto g^{-1} \ln (L/l)$.  
Due to this scaling behavior the
 result can (for $l\approx l_e$) 
be  identified  with a multifractal distribution
within the parabolic approximation (PA, cf.~Eq.~(\ref{parabol}))  
\begin{equation}
	{\cal P} (t) \sim \exp \LK -\frac{\LK \ln t -
	(\a_0-2)\ln(l/L)\RK^2}{-\ln(L/l)4(\alpha_0-2)}\RK
\end{equation}
where
the essential parameter $\alpha_0-d$ is here (d=2) proportional to the
inverse of the conductance
\begin{equation}
	\alpha_0-d \propto g^{-1}\, .\label{7.11}
\end{equation}
Thus,  
 states in a good 2D metal  (with the localization length being much
larger than the  system size)  are
multifractally
distributed. Since the metallic  situation in 2D resembles
 a critical regime of
 a LD transition
the  finding is, after Wegner's  pioneering 
work \cite{Weg80} and the work by Al'tshuler {\em et al.} \cite{Alt86-89}
 a further
analytical hint at the multifractality of critical eigenstates.

The case of 1D and quasi-1D conductors need special attention. 
In 1D the localization length $\xi$ is of the order of the mean free
path
$\xi \approx l_e$  and thus there is no room for diffusive behavior.
For the quasi-1D case one has to be a bit more precise in fixing the
sizes. The length $L$ is treated variable and the cross-section
$L_t^{d-1}$
 is kept fixed. However, the dimensionality is  in general essential
for the determination of physical quantities such as the localization
length.
A strong simplification arises for the case when $L_t\approx l_e$
but a huge number $N_c\gg 1$ of channels is occupied. Under these
conditions the dimensionality becomes inessential and the only kinematic
quantity besides $L$ is the channel number $N_c$.
 We refer to this case as the {\em  idealized} quasi-1D
case.
For the idealized quasi-1D system one can show that 
\begin{equation}
	\xi(N_c)=A N_c l_e \label{7.12}
\end{equation}
with $A$ a numerical constant of order $1$. We have anticipated this
result already in Sect.~\ref{subloc} and will sketch a more reliable 
derivation in the
following subsection. The fact that one can reach a metallic
regime
by 
\begin{equation}
	l_e \ll L \ll \xi (N_c) \label{7.13}
\end{equation}
makes the idealized quasi-1D system more interesting than strictly 1D.
 On the other hand,
the idealized quasi-1D system is still tractable by analytical methods
(cf.~e.g.~\cite{Sto91,Fyo91-95}).

In  metallic idealized quasi-1D systems one has an average
conductance
 (which
is also typical) 
\begin{equation}
	g  = A' \xi/L \, ,\;\; A' \sim {\cal O} (1)\, . \label{7.14}
\end{equation}
The distribution of $t$ is, for $t\ll 1/\sqrt{g}$, again given by the
standard random matrix theory with small corrections of ${\cal O}
(1/g)$ (cf.~\cite{MirR95}).
However, for $t\gg 1/\sqrt{g}$ the situation is different as compared
to the 2D and 3D case.
One finds no log-normal tails but stretched exponential tails
(at least when the box is shrinking to a point)
\begin{equation}
	{\cal P} (t) \sim \exp \LK -B \sqrt{C t g}\RK  \, ,\;\; 
	B,C      \sim {\cal O} (1)  \, .\label{7.15}
\end{equation}
Note, that $g= A \xi/L$ is the quasi-1D conductance. 

In the localized regime of idealized quasi-1D systems, $
	 \xi \ll L \, ,
$
 where
the typical conductance is $g \propto e^{-2L/\xi}$, 
one finds a similar behavior for $t \gg (\xi/L)^{-1}$
\begin{equation}
	{\cal P} (t) \sim \exp \LK -B \sqrt{C t \LK \xi/L\RK }\RK  \, .
	\label{7.17}
\end{equation}
Note, that $\xi/L$ is related to the typical conductance here.
Both equations, (\ref{7.15},\ref{7.17}), have $\xi/L$ as the reference
point. In order to have $t \gg (\xi/L)^{-1}$ one has to have $\xi\gg
l_e$. Thus, the stretched exponential tail does occur when the
box-size
is much less the localization length.
 In the  case where the box-size
becomes much larger than 
the localization length the previously obtained results
of the idealized localized phase apply (cf.~Sect.~\ref{sublocpha}). 

The tails of the distribution ${\cal P}(t)$ that have been observed
in idealized quasi-1D (stretched exponential), in 2D (log-normal) and
in 3D (log-cube) have all been interpreted in terms of {\em  anomalously
localized states} (or {\em  pre-localized states})
\cite{Alt86-89,Myz95,MirR95}. The notion refers to anomalously rare states with
sharp peak in the amplitude on top of an extended backround (in 3D
delocalized phases) or  localized states with anomalously short
localization radius (in $d\leq 2$). 
 The question arises what the tails might say in view of
the LD transition. The 2D case seems to be clear. Here the log-normal
tails correspond to multifractal scaling to be expected 
once the correlation length is much larger than system size.
This interpretation is in accordance with the finding of log-normal
tails in $d=2+\epsilon$. The distributions become multifractal 
when the conductance is close to its critical value.
In 2D the correlation length is the localization length which is much
larger than the system size in the $g\gg 1$ regime, and thus the
states are {\em  critical} in the sense that the system size is less
 the
correlation
length.
In 3D with $g\gg 1$ the correlation length is (see Eq.~(\ref{6.43}))
microscopic and there is no reason to expect  multifractal scaling.
In quasi-1D with $g\gg 1$ the quasi-1D localization length is much
larger than the system length $L$, however due to the strong difference
between
$L_t\approx l_e$ and $L$, this length cannot be considered as the 
correlation length. In strict 1D the correlation length is the 1D
localization
length which is again microscopic and no {\em  critical} behavior of
eigenstates
can be expected on scales larger than $l_e$. 

In this context it is worthwhile to discuss the findings about the
LDOS for {\em  open} systems. These are defined by surrounding the system 
with an ideally conducting medium. 
In that case we cannot talk about eigenstates
of the system. Still the LDOS as given by the imaginary part of the Green's
function is a meaningful object. The system itself does not conserve
total probability which formally breaks hermiticity of the effective
Hamiltonian for the system. For the 2D and 3D case, however, the
results for the distribution of LDOS are qualitatively the same as for
the distribution of the restricted LDOS in the closed systems. 
A qualitative difference occurs for the quasi-1D case.
In the open quasi-1D conductor the LDOS shows log-normal tails, 
\begin{equation}
	{\cal P} (\rho) \sim \exp \LK - A g \ln^2 (\rho/\nu)\RK \, ,
	\label{7.18}
\end{equation}
where $\nu$ is the average DOS, $A$ is of order $1$, and $\rho/\nu \gg
1$. This equation is also valid in strict 1D.  The log-normal tails
are  {\em not} due to the
statistics of the eigenstates of the corresponding closed system, but
can rather be interpreted as reflecting the distribution of
current-relaxation times 
for  the open system. The latter have a similar log-normal
distribution.
The broadness of this distribution is due to the presence
of anomalously long-living states \cite{MirR95}.

In order to better understand the idealized quasi-1D system  we will
focus in the following subsection on the description of transport
in idealized quasi-1D systems.
One particular question we like to address\cite{ChaJanPol95}:
 is it possible to obtain
similar log-normal tails for the conductance distribution in the
metallic regime as those 
found in the
$d=2+\epsilon$ case by Al'tshuler {\em et al.} \cite{Alt86-89}? 
\subsection{Statistics in Quasi-One-Dimension}\label{substaqua}
The transport properties of a quasi-1D system can be obtained within a
transfer matrix modeling as outlined in Sect.~\ref{subqua}. The
characteristic parameters are the length $L$, the number of channels
$N_c$ related to the cross-section $L_t^{d-1}$ 
and the mean free path $l_e$. 

In close analogy to the procedure that led to a FPE for the intrinsic
resistance in 1D (Eq.~(\ref{6.33}) one can derive a FPE for
the whole transfer matrix $M$. The essential assumptions for this to be
possible are:
\begin{enumerate}
\item
 The locally weak scattering condition
\begin{equation}
	\frac{1}{l_e} = \lim\limits_{\delta L\to 0}
	 \frac{N_c^{-2} \sum_{\a\b}\BRA
 	R_{\a\b}(\delta L)\KET}{\delta L} \, .\label{7.19}
\end{equation}
\item
The statistical independence of strip transfer matrices defined for adjacent
strips.
\end{enumerate} 
Fixing the statistical properties of a single strip transfer matrix 
yields a FPE for ${\cal P}(M;L)$.
One particular model turned out to be analytically tractable
to a large extent, the so-called {\em  isotropic} quasi-1D model.
In this model \cite{DMPK} it is assumed that forward and backward
scattering
within a strip $S$-matrix is equal for all channels.
This crucial simplification means that there is no 
diffusion in transversal direction. 
The modeling thus restricts the width $L_t$ to be less than the mean
free path $l_e$ \cite{MelAkk88}.
Therefore,  an idealized
quasi-1d
model in the sense of  the foregoing section results.
Furthermore, the most explicit  results can be obtained in the 
limit of large channel numbers, or more precisely when
 $N_c \to \infty$  but $N_c l_e/L$ remaining fixed
and finite. 

In terms of the polar parameterization Eq.(\ref{3.52}) 
the isotropy assumption is equivalent to the assumption
that the unitary matrices $u^{i}$ are distributed isotropically within
the unitary group. Similar to the standard random matrix theory for
Hamiltonians the statistics of eigenvectors is treated as being trivial and
decoupled from the statistics of the radial parameters $\lambda_i$.

A consequence of the isotropy assumption is the fact that a closed FPE
can be found for the distribution of the radial parameters $\lambda_i$
alone.
These radial parameters determine the conductance (Eq.~(\ref{3.51}))
\begin{equation}
	g=\sum_i {\cal T}_i = \sum_i \frac{1}{1+\lambda_i}
	\, .\label{7.20g}
\end{equation}

The derivation follows the same principle steps as outlined in
Sect.~\ref{subscadis} and the final result is the well-known
DMPK-equation \cite{DMPK}
\begin{eqnarray}
	l_e\frac{\partial {\cal P}\LK\LB \lambda_i\RB\RK}{\partial
	L}&=&\frac{2}{\gamma}\sum_i 
	\frac{\partial}{\partial \lambda_i} \lambda_i \left(
	1+\lambda_i\right) J \frac{\partial}{\partial \lambda_i}J^{-1}
	 {\cal P}\LK\LB\lambda_i\RB\RK\, ,\nonumber\\
J&=&\prod_{n<m}|\lambda_m -\lambda_n|^{\beta}\; ,\;\; \gamma=\beta N_c
+2 -\beta\, .\label{7.20}
\end{eqnarray}
Here  $\beta=1,2,4$ characterizes different symmetry classes  (systems
with time reversal symmetry  $\beta=1$, systems with broken time
reversal symmetry
$\beta=2$, systems with time reversal symmetry and spin-orbit
interaction
 $\beta=4$) and $J$ is the Jacobian for the transformation from the 
 Cartesian parameterization $M_{kl}$ to the
polar parameterization of the transfer matrix.

The DMPK equation has been solved  exactly by Beenakker and Rejaei
\cite{BeeRej93} for $\b =2$ and for $\b=1,4$ formal solutions
exist \cite{Cas95}.
Common to the solutions is that they represent a Gibbs-ensemble of the
form
already encountered in the standard random matrix theory for energy
levels, Eqs.~(\ref{5.19a},\ref{5.19b}),
\begin{eqnarray}
	{\cal P} \LK\LB\lambda_i\RB\RK&=&\exp\LK 
{-\beta {\cal H}\LK\LB\lambda_i\RB\RK}\RK\, ,
	\nonumber\\
	{\cal H} \LK\LB\lambda_i\RB\RK&=&  \frac{1}{2}\sum_{m\not=n}
	 U(\lambda_n,\lambda_m) +\sum_i
	V(\lambda_i) \, .\label{7.21a}
\end{eqnarray}
For $\b=2$ the explicit form of the two- and one-body potentials read
\begin{eqnarray}
	U(x,y) &=& -\frac{1}{2}\ln(|x -y|) 
	-\frac{1}{2}\ln(|{\rm arsinh}^2\sqrt{y}
	-{\rm arsinh}^2\sqrt{x}|)\, ,\nonumber\\
	V(x) &=&\frac{N_cl_e}{2L}{\rm arsinh}^2\sqrt{x} + {\cal O}(1/N_c)\,
	.\label{7.22a} 
\end{eqnarray}

It is worth mentioning that, for $\lambda_i \ll 1$ (open channels
${\cal T}_i \approx 1$),
 the two-body potential reduces to
\begin{equation}
	U(x,y) = - \ln(|x - y|)\label{7.23a}
\end{equation}
which has been used to model quasi-1d wires, before the exact solution
was found (see e.g.~\cite{Sto91}).
A general feature of the Gibbs-ensembles is that physical parameters,
such as $N_c,l_e$ and $L$ do only occur in the one-body potential
$V(x)$ in the combination $N_cl_e/2L$.
In contrast, the two-body potential  is a {\em  universal
function} reflecting universal level repulsion.

Without solving the  DMPK equation explicitly one can study the
 resulting
equations for the moments $\BRA  g^n \KET$ by integrating
 on both sides of the DMPK equation over $\LK\sum_i (1+\lambda_i)^{-1}\RK^n$.
A perturbative analysis  as a series expansion 
in  powers of the single parameter
 $N_cl_e/L$  is possible in the limit $N_cl_e \gg L  \gg l_e$,
 i.e. in the diffusive limit (Mello {\em et al.} in Ref.~\cite{DMPK}).
This yields the result for the average conductance
\begin{equation}
	\BRA g\KET = N_cl_e/L + C_\beta + {\cal O}\LK (N_cl_e/L)^{-1}\RK\, .
	\label{7.23b}
\end{equation}
The constant $C_\b$ contains  corrections to the
classical result  (for $\b=1$ these  are negative weak localization
corrections). Thus, the leading term is the one we have already
anticipated in Eq.~(\ref{3.59}).
The leading term for the variance $\BRA (\delta g)^2\KET$ turns out to
be universal
\begin{equation}
	\BRA (\delta g)^2\KET =\frac{2}{15 \b}\label{7.24b}
\end{equation}
and proofs the UCF effect for isotropic quasi-1D.
The effort to derive results for higher moments by the perturbative
method increases drastically.

In the limit $L\gg N_cl_e$ it is advantageous to transform
to the parameters $\nu_i$ (Eq.~(\ref{3.51})) since they become the
Lyapunov exponents in the thermodynamic limit. Their interaction can then
be neglected  and one ends up with an effective joint
probability
density of the form \cite{Mac92}
\begin{equation}
	{\cal P} \LK \LB \nu_i\RB\RK = C_{\b,N_c} \prod_{i=1}^N \exp
	\LBK -\frac{(\b N_c + 2-\b) l_e}{2 L} \LK \nu_i - \frac{L
	(\b N_c +1 -\b i)
	}{l(\b N_c +2 -\b}\RK^2 \RBK\, .
	\label{7.25b}
\end{equation}
demonstrating again the self-averaging feature of Lyapunov exponents.
In that limit the conductance is $g=2 e^{-\nu_1}$ since all higher
$\nu_i$ contribute much less. The conductance distribution is then
found to be of the log-normal form.  The typical value and the
log-variance are related by ($N_c \gg 1$)
\footnote{Notice, that  the so-called global transfer matrix
approach \protect\cite{Pich90,Sto91} yields a factor 
\protect$1$ between  variance and average
value
instead of a factor \protect$2$ occuring in Eq.~(\protect\ref{7.25c}). 
}
\begin{equation}
	 \BRA (\delta \ln g)^2 \KET = - 2 \ln g_{\rm t} =
	-2 \BRA \ln g \KET=
	\frac{4L}{N_cl_e\b}\, .\label{7.25c}
\end{equation}
The localization length is then found  to be
\begin{equation}
	\xi= \b N_c l_e \, .\label{7.25h}
\end{equation}

Before we look at the conductance distribution in more detail 
we report
on the findings concerning the distribution of transmittances
$
	T_{\a\b}= |t_{\a\b}|^2
$ and 
$
	T_\a=\sum_\b T_{\a\b}
$.

They are related to the eigenvalues ${\cal T}_i=(1+\lambda_i)^{-1}$ of
$t^{\dagger}t$ by unitary matrices $u$ and $v$
\begin{equation}
	T_{\a\b} = \sum_{kl} u_{\a k} u^{*}_{\a l} 
	\sqrt{{\cal T}_k {\cal T}_l} v_{k \b} v^{*}_{l \b} \, ,\;\; 
	T_\a =\sum_{l} |u_{l \a}|^2 {\cal T}_l \, .\label{7.24}
\end{equation}
These equations point out that  transmittances are analogs 
 of wave function amplitudes $|\psi_\a (k)|^2 = |U_{k\a}|^2$.
However, the transmittances do  depend also explicitly on the
 eigenvalues ${\cal T}_l$. The statistics of the unitary matrices
was assumed to be isotropic in the DMPK approach and the eigenvalues
 follow the DMPK equation.
 
It turned out that the distribution of transmittances can be
calculated exactly \cite{VanLan96}. Actually, only
${\cal P}(T_n)$ needs  to be calculated in the  large $N_c$ limit.
In a work by Kogan and Kaveh \cite{Kog95} it was shown that,
for isotropic scattering in the large $N_c$ limit, 
the distributions of $T_\a$ and $T_{\a \b}$ are connected by
\begin{equation}
	{\cal P} (T_{\a\b}) = \int\limits_0^\infty d T_\a \, 
	T_\a^{-1} \exp \LK -T_{\a\b}/T_\a\RK {\cal P}(T_\a)\, . \label{7.25}
\end{equation}
 In the metallic regime $N_cl_e \gg L$, $\BRA g\KET \gg 1$, the
 distribution
of $T_{\a\b}$ is of the Rayleigh type for the bulk of the distribution
 ($\b=2$)
and develops stretched exponential tails (first obtained by
 Nieuwenhuizen and van Rossum \cite{Nie95})
\begin{equation}
	{\cal P} (T_{\a\b}) \sim \exp \LK -2 \sqrt{T_{\a\b}} \RK \, ,
	\;\;  T_{\a\b} \gg \LK N_cl_e/L\RK^2 \, .\label{7.26}
\end{equation}
It is therefore completely analogous to the distribution of wave
function
amplitudes in {\em  closed} quasi-1D systems (see~Eq.~(\ref{7.15})).
The corresponding distribution of $T_\a$ is Gaussian in the bulk and
develops
exponential tails
\begin{equation}
		{\cal P} (T_{\a}) \sim \exp \LK - {T_{\a}} \RK \, ,
	\;\;  T_{\a} \gg  N_cl_e/L  \, .\label{7.27}
\end{equation}
These results depend heavily on the isotropy assumption and they
indicate that it is very unlikely that the distribution of conductance,
$g=\sum_\a T_\a$, in the metallic regime 
can develop log-normal tails in isotropic quasi-1D.
 However, so far 
the extension to conductance is not easy since the correlation between
eigenvalues ${\cal T}_i$ have to be taken into account. 

In the localized regime $L\gg N_cl_e =\xi/\b$,
the statistics is dominated again by the smallest value $\nu_1$ of $\nu_i$,
i.e. by the largest value ${\cal T}_1$ of ${\cal T}_i$. 
This value is log-normally distributed (see~Eq.~(\ref{7.25c}))
and accordingly the transmittances are log-normally distributed, too.

Since the conductance is a linear statistics of the transfer matrix
radial coordinates $\lambda_i$
\begin{equation}
	g=\sum_i f(\lambda_i) \, ,\;\; f(x) = (1+x)^{-1} 
\label{7.28}
\end{equation}
the problem of a complete statistical description
 of $g$ reduces to the problem of a
linear statistics $X=\sum_i f(x_i)$ in a Gibbs-ensemble of a classical
gas determined by a 
universal two-body potential (which can be chosen to be symmetric)
and a parameter-dependent one-body
(confining)
potential 
\begin{eqnarray}
	{\cal P}(X)&=& \int d^{N_c}x \,  {\cal P}\LK\LB x_i\RB\RK \delta
	\LK X -\sum_i f(x_i)\RK \, , \;\; 
	{\cal P}\LK\LB x_i\RB\RK = Z^{-1} 
	\exp{-\beta {\cal H}\LK\LB x_i\RB\RK}
	\nonumber\\
	{\cal H}\LK\LB x_i\RB\RK&=& \frac{1}{2}\sum_{m,n} U(x_n,x_m) +\sum_i
	V(x_i)\, .\label{7.29}
\end{eqnarray}   

Beenakker pointed out \cite{Bee93} that the one-body potential
can be viewed as a source term in the {\em  partition sum}
\begin{equation}
	Z\LBK V\RBK = \int d^{N_c} x \exp\LBK -\beta {\cal H} \LK\LB
	x_i\RB\RK\RBK \, .\label{7.30}
\end{equation}
All cumulants \footnote{Cumulants $\BRA\BRA X^n\KET\KET$ are linear
combinations of moments of order $k\leq n$. While moments can be
generated from a partition sum $Z$, the corresponding cumulants are
generated by $\ln Z$.  The Gaussian distribution is characterized by
vanishing cumulants for $n\geq 3$ } of the level-density, $
{\rho}(x):=\sum_i\delta(x-x_i)\, ,\label{7.31} $ can be obtained by
functional derivatives
\begin{equation}\label{7.31b}
	\BRA\BRA {\rho}(x_1)\ldots 							{\rho}(x_k)\KET\KET = \frac{\delta^k
	\ln Z}{\delta\LK -\beta V(x_1)\RK\ldots\delta\LK-\beta V(x_k)\RK} \, .
\end{equation}
Cumulants of a linear statistics $X=\sum_i f(x_i)$ 
are given by integration 
\begin{equation}\label{7.32}
	\BRA\BRA X^k\KET\KET = \int dx_1\ldots dx_k \BRA\BRA
	{\rho}(x_1)\ldots {\rho}(x_k)\KET\KET f(x_1)\ldots f(x_k)\, .
\end{equation}
The whole distribution ${\cal P}(X)$ can be obtained from a modified 
partition sum  $Z(\kappa)$,
\begin{equation}\label{7.33}
	{\cal P}(X) =\int d\kappa \, e^{i\kappa X} Z(\kappa)\, ,
\end{equation}
where  $Z(\kappa)$ follows from  $Z$ through a simple shift in the
one-body potential
\begin{equation}\label{7.34}
	V_\kappa(x):=V(x)+i\frac{\kappa}{\beta} f(x)\, .
\end{equation}

In order to work with functional derivatives one has to know 
$Z$ (or $Z(\kappa)$) as an explicit  functional of the
one-body-potential. Alternatively, 
the knowledge of the average level density
$\nu (x):= \BRA {\rho} (x)\KET$ as a functional of the one-body-potential
would be enough. 

Such representation of the average level-density $\nu(x)$ was
developed by Dyson \cite{Dys72}.
Here we outline a derivation \cite{ChaJanPol95}
which allows for a systematic check of
the range of validity of the results.

We start from the partition sum $Z$, or $Z(\kappa)$  and proceed by
the following steps:
1. The Hamiltonian is expressed by the level-density $\rho(x)$, replacing sums
by integration. 
Here the absence of self-interaction is ignored, i.e 
$\sum\limits_{n\not=m}U(x_n,x_m)$ is replaced by $\int dx\, dy\, \rho(x)
\rho(y) U(x,y) $.
 2. With the help of a  $\delta$-functional we
introduce a field $\phi(x)$ that takes the role of the level-density. 
3. We represent the $\delta$-functional by its Fourier representation
which introduces a conjugate field $\psi(x)$.
4. Now the original integration over the set $\LB x_i\RB$ can be carried out
leaving a field theoretical partition function in terms of two field
degrees of freedom, $\phi(x)$ and $\psi(x)$. 5. Due to the two-body
character of the original ${\cal H}$ the field $\phi(x)$ can be
integrated out by a Gaussian integration.

The final result is a path integral representation of the partition
sum $Z$ or of the complete distribution function ${\cal P}(X)$, where
the integration runs over field configurations of $\psi(x)$.
We concentrate on ${\cal P}(X)$ which reads
\begin{eqnarray}
	{\cal P}(X) &=&\int D\LBK\psi\RBK \exp -S\LBK \psi; X \RBK\nonumber\\
	S\LBK\psi;X \RBK &=& \frac{-1}{2\beta}\LK \psi\right| K\left|
	\psi\RK + \frac{1}{2} \LK f \right| K \left| f \RK Q^2(\psi;X) 
	+ F_{N_c} \LBK \psi +\b V\RBK \,  .\label{7.35}
\end{eqnarray}
Here we use a scalar product notation,
\begin{equation}
 	(f|A|g):=\int dx \,dy \, f(x)A(x,y)g(y)\, ,\;\;  (f|g):=\int dx\,
	f(x)g(x)\, ,\label{7.35b}
\end{equation}
 $K$ denotes the inverse operator of $U$ ($U(x,y)=\LK x \right|
 U\left| y\RK$), the functional $Q(\psi;X)$ is defined as
\begin{equation}\label{7.36}
	Q(\psi;X):= \frac{X - \b^{-1} \LK f \right| K\left| \psi \RK}{\b^{-1}
	\LK f \right| K\left| f\RK}\, ,
\end{equation}
and the functional $F_{N_c}$ is a free energy of $N_c$ independent
particles with one-body-potential $\psi +\b V$, 
\begin{equation}
	F_{N_c} \LBK \psi +\b V\RBK := -N_c \ln \LBK \int dx\, \exp\LBK
	-\LK \psi(x) + \b V(x)\RK\RBK \RBK \, .\label{7.37}
\end{equation}

The omission of the self-interaction can be cured  on an effective
potential level. By shifting the one-body-potential $V(x)$ to
$\tilde{V}(x)= V(x) -\frac{1}{2} U(x,x+\Delta (x))$ where $\Delta (x)$
is the average level spacing at $x$, the theory is able to account for
those quantities that are smooth on the scale of $\Delta (x)$.
For example, with the logarithmic interaction, $U(x,y)=-\ln \mid
x-y\mid $ the effective potential reads $\tilde{V}(x)=V(x)
-\frac{1}{2}
\ln \nu(x) $. In the following, we will denote $\tilde{V}$ by $V$.

The path integral in Eq.~(\ref{7.35}) cannot be calculated exactly.
However, for large $N_c$ one can use the method of stationary point.
To be more explicit we will now use the conductance $g$ as linear statistics
variable and recall that $V(\lambda)$ contains the pre-factor
$g_0= N_cl_e/L$	which we assume to be large: $ g_0\gg 1$ since we are
looking for the conductance distribution in the metallic regime.

Introducing the {\em  mean-field} level density $\nu^{0}_g$
as
\begin{equation}
	\nu^0_g:= \frac{N}{Z^0_g}\exp \LBK -\LK \psi^0_g +\b V\RK\RBK 
	\, ,\;\; Z^0_g:= \int d\lambda \, \exp\LBK -\LK \psi^0_g
	(\lambda) + V(\lambda)\RK\RBK \label{7.38}
\end{equation}
corresponding to the stationary point,
\begin{equation}
	\frac{\delta S}{\delta \psi}{\Bigg|}_{\psi^0_g} =0 \, , \label{7.39}
\end{equation}
the mean-field equation reads\footnote{This mean-field equation is in
agreement with the one obtained by Dyson \cite{Dys72} and Beenakker
\cite{Bee93} for the special two-body-potentials used in that works.
}
\begin{equation}
 	\left.\mid \nu^0_g\RK = -\tilde{K} \left| V+\b^{-1} \ln
 	\nu^0_g \RK +\b^{-1} Q(g) \tilde{K} \left| f \RK +
 	\frac{N_c K \left| 1\RK}{\LK 1 \right| K \left| 1\RK} \, .\label{7.40}
\end{equation}
Here the kernel $\tilde{K}$ is defined as
\begin{equation}
	\tilde{K} = K -\frac{K\left| 1\RK \LK 1 \right| K}
	{\LK 1\right| K \left| 1\RK} \, ,\;\; \tilde{K} \left|
	1\RK = \LK 1\right| \tilde{K}  =0 \, ,\label{7.41}
\end{equation}
and $Q(g)$ as
\begin{equation}
	Q(g):= \frac{g-\overline{g}_g}{\b^{-1} \LK f \right| \tilde{K} 
	\left| f \RK}\label{7.42}
\end{equation}
with 
\begin{equation}
	\overline{g}_g:= -\LK f\right| \tilde{K} \left| V+ \b^{-1} \ln
	\nu^0_g\RK +\frac{N_c \LK f\right| K \left| 1\RK}{\LK 1\right| K \left|
	1\RK} \, .\label{7.43}
\end{equation}
In deriving these equations it has been used that the mean-field level
density is normalized $\LK 1\mid \nu^0_g\RK =N_c$ and yields the
current $g$ as an expectation value $\LK f \mid \nu^0_g\RK =g$.

Now one can draw the following conclusions:
\begin{enumerate}
\item
For $|V(\lambda)| \gg \b^{-1}\ln \nu^0_g$ 
the expression $\overline{g}_g$ equals the average value of $g$,
independently of current $g$. Since $V$ contains the large factor
$g_0$
the inequality is satisfied as long as 
\begin{equation}
	|\delta g| \ll \BRA g\KET \, ,\;\; \BRA g\KET \gg 1 \, .
	\label{7.44}
\end{equation}
\item
The stationary point of $S$ is then given by
\begin{equation}
	S\LBK \psi^0_g;g\RBK =\frac{1}{2} \frac{\LK g -\BRA g\KET
	\RK^2}{\b^{-1} \LK f\right| \tilde{K}\left| f\RK} + S_{\BRA g\KET}
	\label{7.45}
\end{equation}
where $S_{\BRA g\KET}$ is independent of current $g$.
\item
One can also analyze fluctuations around the stationary point and show
that they give sub-leading contributions to the path integral.
Finally, one arrives at the conclusion, that the distribution of
conductance is, for $|\delta g|\ll \BRA g\KET$,
\begin{equation}
	{\cal P}(g;L) = {\rm const.} \exp \LBK - \frac{\LK g -\BRA
	g\KET_L \RK^2}{2\b^{-1} \LK f \right| \tilde{K}\left| f\RK}\RBK
	\label{7.46}
\end{equation}
up to ${\cal O} (\ln g)$ corrections in the exponent. 
\end{enumerate}
Therefore, our conclusion is that the conductance distribution
in isotropic quasi-1D does not show long-tails in the regime 
$|\delta g|\ll \BRA g\KET$, but is Gaussian with universal variance.
The universality is due to the universal two-body potential $U(x,y)$
which reflects essentially the universal level repulsion expressed by
the Jacobian in Eq.~(\ref{7.20}).
Recall, that the occurrence of log-normal
tails in $d=2+\epsilon$
happened  already for $|\delta g|\gg  \sqrt{\BRA g\KET}$.

The absence of log-normal tails in the conductance distribution for
isotropic
quasi-1D can be interpreted as being caused by the isotropy
assumption.
Relaxing from the isotropy assumption the resulting FPE for
the radial parameters $\lambda_i$ is not closed. Rather the diffusion
function depends on the statistics of eigenvectors of $M^{\dagger}M$.
The latter is presently unknown, but one can conclude in general
\cite{Chapriv95}
that long tails in the distribution of the eigenvectors will lead to
long-tails in the distribution of $g$. In addition, one can conclude
that in such case the probability density of the radial parameters 
is no longer given by a Gibbs-ensemble with only one-body and two-body
potentials.
Also  the results for
transmittances  indicate  that the isotropy assumption
expels  any log-normal tails for the conductance distribution.

On the other hand, the isotropy assumption can only  be justified for
transversal width $L_t$ being less the mean free path $l_e$.
It is instructive to look at the correlation length $\xi_c$ relevant
to such idealized quasi-1D systems.
Recall, that the correlation length, in the spirit of one-parameter
scaling theory, is defined as that fictitious system size $\xi_c$ for
which the conductance of a $d$-dimensional cubic system is of order
$1$.
To attach a correlation length to the quasi-1D system requires 
a meaningful definition of a conductance in a cubic system.
The easiest way is to use a parallel composition law and
define
\begin{equation}
	g_{\rm cube}(L)= g_{\rm qua}(L,L_t) \LK L/L_t\RK^{d-1}\label{7.21}
\end{equation}
where $g_{\rm qua}(L)$ is the conductance of the quasi-1D system with
length $L$ and cross-section $L_t^{d-1}$.
For the isotropic quasi-1D system with $ g_{\rm qua}(L) \approx N_c l_e/L$
and $L_t= Al_e$, $A\leq 1$, we thus have
\begin{equation}
	g_{\rm cube} (L) \approx {N_c}\LK \frac{L}{l_e}\RK^{d-2}\label{7.22}
\end{equation}
which is extremely large. Consequently, the correlation length
is
\begin{equation}
	\xi_c \approx l_e N_c^{-1/(d-2)} \, .\label{7.23}
\end{equation}
Thus, even for $d=2$ the isotropic quasi-1D model corresponds  to
microscopic correlation lengths. One should note that it is {\em  not}
the quasi-1D localization length $\xi=\b N_c l_e$ which plays the role
of the correlation length, but $\xi_c$. Both coincide only for strict
1D where $N_c=1$.

We interpret  the occurrence of log-normal tails in the
conductance distribution as precursors of the LD transition.
For $d=2+\epsilon$ they are  controlled by 
the one-parameter scaling variable $g_t$. With
this interpretation 
 the absence of such tails in isotropic quasi-1D is no surprise
since the appropriate 
scaling variable $\propto N_c (L/l_e)^{d-2}$  is sent to infinity.

\subsection{Finite Size Scaling}\label{subfin}
Giving up the isotropy assumption allows to study more realistic
systems by a  quasi-1D set-up. True dimensionality effects should
enter the physical properties.
An analytical treatment becomes much more difficult, but 
a numerical treatment turns out to be  very efficient. 
The Green's function can be
studied by recursive methods \cite{Mac81}, putting strips of
cross-section $L_t^{d-1}$ together. Alternatively, strip  transfer
matrices can be multiplied \cite{Pich81}. That quantity which
 can be obtained with high precision (due to its self-averaging property)
is the quasi-1D localization length 
$\xi(L_t)$. 
By studying $\xi(L_t)$ as a function of $L_t$ one should be able 
to study the localization behavior in  a  large (cubic)
$d$-dimensional system.

Our first task is to construct  a suitable scaling variable
which can be
derived from  $\xi(L_t)$. When we increase $L_t$
and  the system flows towards the insulating (localized)
phase, $\xi(L_t)$ will stop growing as soon as $L_t$ is
comparable with the (finite!) localization length of the infinite
$d$-dimensional 
system: 
\begin{equation}\label{locinfty}
	\xi(\infty)=\lim\limits_{L_t \to \infty}\xi(L_t)\, .
\end{equation}
 On the other hand, $\xi(L_t)$
will grow forever, if the system  floats to   the metallic
(delocalized)
phase.

At the critical point, however, we expect the system to be scale
invariant (diverging	correlation length), i.e. $\xi(L_t)$ should
scale exactly as $L_t$ as soon as finite size effects have died out.
So one  tries 
\begin{equation}\label{renormloc}
	\Lambda (L_t):=\xi({L_t})/{L_t}
\end{equation}
 as a scaling variable.

A further reason for this choice comes from the interpretation
that,  in the metallic regime,
an appropriate conductance for the corresponding cubic system 
is given by
Eq.~(\ref{7.21}) where $g_{\rm qua}(L;L_t)\approx \xi(L_t)/L$
yielding
\begin{equation}\label{ohm1}
	g_{\rm cube}(L=L_t) \approx \Lambda(L_t)
\end{equation} 	
In the localized regime, where $\xi(L_t)$ approaches $\xi(\infty)$,
$\Lambda(L_t)$ represents the inverse logarithm of the corresponding
conductance 
\begin{equation}\label{loc2}
	-\ln g_{\rm cube} (L=L_t) \approx  \Lambda^{-1} (L_t)
\end{equation}
Now, one-parameter
scaling means that away from the critical point (where $\Lambda({L_t})\equiv
\Lambda^*$ is a constant) $\Lambda({L_t})$ does not depend on
the correlation length $\xi_c$ (which can be identified with 
$\xi(\infty)$ in the localized phase)  and ${L_t}$ 
separately, but only on their dimensionless ratio
\begin{equation}
	\Lambda({L_t}) = 
f\left(\frac{\xi_c}{{L_t}}\right).\label{brelf}
\end{equation}
Here  $f$ is a called the  scaling function.
In the presence of LD transitions, $f$ is a two-valued function.
Close to a critical point (where linearizing of the corresponding
 $\beta$-function is
appropriate) 
 the  scaling function is of the  form (cf.~Eq.~(\ref{6.19b}))
\begin{equation}\label{fcrit}
	f(x) = \Lambda^\ast \pm A_{(\pm)}  x^{-\frac{1}{\nu}}
\end{equation}
where the positive (negative) sign marks the delocalized (localized)
branch of $f$: $\beta(\ln \Lambda)>0$ ($\beta(\ln \Lambda) <0$).

The general
procedure  to extract the scaling function from a large number
 of  calculated localization lengths
$\xi(L_t)$ is to   find a quantity $\xi_c$
independent of ${L_t}$ such that the entire data set
$\xi({L_t}; \tau)$
(where $\tau$ is a system parameter related to e.g. the disorder or Fermi
energy with critical value $\tau^\ast =0$) collapses onto a single
curve.
This curve  then defines  $f$.
 In practice, this can be done by plotting the
$\ln \Lambda (L_t;\tau)$ 
curves on transparencies and moving them around by hand
until they fall on top of each
other or by a least squares fitting procedure.
A typical example for 
 a scaling function is shown in 
Fig.~24. After determining $\Lambda^\ast$ one can determine the
critical
exponent $\nu$ from a fit to Eq.~(\ref{fcrit}).

It may, however, turn out to be difficult to find the critical value
$\Lambda^\ast$. For example, it can happen that the curves
$\Lambda(L_t;\tau)$, for
different  
choices of $L_t$, 
do not intersect at one point $\tau=0$.
This can be due to the fact that the system size is not large enough
to reach the asymptotic scaling regime, i.e. it can be a finite size effect.
Assume that, at the {\em  true} critical point  $\tau=\tau^\ast=0$,
the scaling variable still shows a dependence on $L_t$ of the form
\begin{equation}
	\Lambda (L_t;\tau=0) - \Lambda^\ast =  B  L_t^{-y_{\rm
	irr}}\label{corrsc}
\end{equation} 
Here $y_{\rm irr}>0$ is a  critical exponent.
More convenient is  $ B  L_t^{-y_{\rm irr}}={\tilde{B}} \LK
L_t/\xi_{\rm irr}\RK^{-y_{\rm irr}}$ with $\tilde{B}$ of order $1$ and
an explicit scale $\xi_{\rm irr}$ called {\em  irrelevant} length.
Since $\xi_{\rm irr}$ stays finite the correction to $\Lambda^\ast$
decreases when $L_t$ becomes larger than $\xi_{\rm irr}$.
\begin{equation}
	\lim\limits_{L_t\to \infty} \Lambda (L_t;\tau=0) = 
	\Lambda^\ast \, .\label{8.10a}
\end{equation}
 In this sense, the length $\xi_{\rm irr}$ becomes irrelevant in the large 
$L_t$
limit. However,
in a finite size calculation one can never be sure that microscopic scales
are already much smaller than $L_t$
 and one has to face the fact that corrections
to the ideal scaling, Eq.~(\ref{fcrit}), can occur. They can then be analyzed
by means of irrelevant length scales, Eq.~(\ref{8.10a}).

The best studied system is the quantum Hall system
in 2D.
 Although no metallic phase exists, the position of the critical
point
on the energy scale is precisely known for certain models due to
spectral symmetries. This fact and the
two-dimensionality enables to get results for critical quantities with
high
precision (see e.g.~\cite{HucR,JanB}).
 Also a 3D analog of the quantum Hall system has been
studied
recently \cite{ChaDoh95}.
 Further results are known for 2D systems with spin-orbit
scattering (see e.g.~\cite{Eva87,Fas91,Schw95}) 
and for  a number of 3D systems (see e.g.~\cite{Hof93,Hen94,Kaw96}).
\section{Statistics At The Transition}\label{statra}
The finite size scaling method made it possible to analyze a
self-averaging
quantity serving as scaling variable in the sense of one-parameter
scaling theory. It demonstrates not only that one-parameter scaling
does work, but also allows for explicit calculations of the critical
exponent
of the correlation length.
 However, not very much information comes out of this for the
statistical properties of non-self-averaging quantities like
conductance
and LDOS. At least, by Eqs.~(\ref{ohm1},\ref{loc2}) one can
expect that the scaling variable $\Lambda$ of the finite size scaling
method
is related to the typical conductance $g_t$. The latter
can be identified as the mean value in the delocalized phase and
as the geometric mean in the localized phase. Since, in the
delocalized
phase the geometric mean is very close to the mean (the distribution
is Gaussian with $\sqrt{\BRA (\delta g)^2\KET}/\BRA g\KET \approx
1/\BRA g\KET \ll 1$) it is reasonable to expect that the geometric
mean
\begin{equation}
	g_t:=\exp \LK \BRA \ln g\KET \RK \label{8.0}
\end{equation}
can serve as a typical conductance which in turn is a  
scaling variable.

In this section we will firstly consider the statistics of critical
eigenstates. After the findings of the previous section
 we already expect them to  show the multifractal scaling property.
We will discuss the implications of this for the LDOS. Several arguments
will point out that the typical conductance will determine
the distributions of local quantities. After a brief discussion of
critical energy level statistics we collect results for the
critical conductance distribution.

\subsection{Multifractality Of Critical Eigenstates}\label{submul}

In finite size systems the correlation length $\xi_c$ 
of the electronic states
is larger than the system size $L$ for a certain parameter  range, 
$\Delta \tau $,
 around the
critical parameter value $\tau=0$.
 These states are called {\em  critical states}.  In the
thermodynamic limit 
$\Delta \tau \propto L^{-1/\nu}$ where $\nu$ is the critical exponent
of $\xi_c$.

After  the pioneering works by Wegner \cite{Weg80}
and  Aoki \cite{Aok83Aok86} it became clear
that the critical wave functions have a  multifractal structure
(for a review see \cite{JanR} and references therein). 
The entire distribution of local amplitudes and its scaling behavior
is encoded in the multifractal $f(\alpha)$ spectrum, as outlined in
Sect.~\ref{subscadis}.   The distribution ${\cal P}(P;L_b/L)$
is broad on all length scales and close to a log-normal distribution.
The most important 
quantity is the maximum position, $\alpha_0$, of $f(\alpha)$.
It
 describes the scaling behavior of the geometric mean 
of what serves as a {\em  typical} amplitude of a critical wave function.

 In early works Aoki \cite{Aok83Aok86}
 gave a nice argument for the multifractal behavior of
critical wave functions (although at that time the phrase
{\em  multifractality}
was not yet common). His argument goes as follows.
Consider  the inverse participation
number defined by 
\begin{equation}
	{\cal{P}} = \int\limits_{\Omega} {d}^d r\,
	|\psi ({\bf r})|^4  \label{8.1}
\end{equation}
where $\Omega$ denotes a $d$-dimensional
region with linear size $L$.
If the wave function $\psi ({\bf r})$ is uniformly
distributed -- like in the delocalized  phase -- then ${\cal{P}}\propto
L^{-d}$ and
the participation ratio $p=({\cal{P}}L^d)^{-1}$ is constant.
In the localized regime ${\cal{P}}\approx \xi^{-d}$ and $p$ vanishes
in the thermodynamic limit.
At the transition point where the wave function is extended
the participation ratio still has to
vanish in the thermodynamic limit if the LD phenomenon is similar
to a second order phase transition for which
the participation ratio acts as an order
parameter.
Consequently, ${\cal{P}}$ scales with a power $d^{*}<d$.
Wegner had already  calculated the whole spectrum of exponents
for generalized inverse participation numbers within the non-linear
sigma-model \cite{Weg80}. This spectrum  was interpreted as a multifractal
spectrum by Castellani and Peliti  \cite{Cas86}.
After extensive numerical work (e.g.~\cite{Poo91,Schr91,Huc92,Schw95})
the following description of the
statistics
of critical wave functions $\psi(\bf r)$ is now established:

At the LD transition the  distribution function
${\cal P}(P;L_b/L)$  gives rise to the power law scaling for the
moments,  
\begin{equation}
	\left\langle \lbrack P(L_b)\rbrack^q \right\rangle_L 
\propto (L_b/L)^{d + \tau(q)} \, ,\label{2}
\end{equation}
where $d+\tau(q)$ is a non-linear function of $q$.
This non-linearity is a direct
consequence
of Aoki's observation that $d+\tau(2)= d+d^{*}\not=d+d$.
In practice it turns out that, to a good accuracy,
the disorder average can often  be
substituted by the spatial average over one wave function for a given
configuration. Within  numerical
 accuracy the resulting spectra  are identical.
  
 The corresponding (universal) distribution function 
can be described in terms of the
single-humped, positive  $f(\alpha)$ spectrum,
\begin{equation}
	{\cal P}(P;L_b/L)\, dP \propto (L_b/L)^{d-f(\alpha)} \, d\alpha \, ,
 \label{8.3}
\end{equation}
where $\alpha:=\ln P/\ln (L_b/L)$;
As shown in Sect.~\ref{subscadis}
  $f(\alpha)$ is   related to   $\tau(q)$ by a Legendre transformation
\begin{equation}
	f(\alpha(q))=\alpha(q) q -\tau(q) \; ,\;\;
	\alpha(q)=d\tau(q)/dq\, .\label{8.4}
\end{equation}
The parabolic approximation (PA), Eq.~(\ref{parabol}), 
 contains  $\alpha_0$ as the only
parameter besides $d$.  This is due to the assumed validity of
the PA  at least up to $|q|\leq 1$.
Equation (\ref{parabol}) 
   corresponds to a log-normal
distribution centered around the typical value $P_{\rm t} =\exp <\ln P>
 \propto
(L_b/L)^{\alpha_0}$ with log-variance proportional to $\alpha_0-d$.
A simple one-parameter approximation for
$f(\alpha)$ 
 which takes into account that the support
$[\alpha(\infty),\alpha(-\infty)]$
of $f(\alpha)$ is finite, is the semi-elliptic
approximation 
(SEA)
\begin{equation}
	f(\alpha)\approx d\sqrt{1-\frac{(\alpha -\alpha_0)^2}{\alpha_0^2-d^2}}
\, .
\label{8.6}
\end{equation}

To demonstrate that the  distribution of local amplitudes of
critical eigenstates is encoded in $f(\alpha)$ we discuss
 numerical results
for a quantum Hall system (QHS) \cite{Prac96}. The quantum
Hall system is the most convenient to study critical eigenstates
since one knows the critical point exactly and it is only
two-dimensional which keeps the numerical effort low.

The 
wave functions are calculated for the 
 model of independent (spin-less) electrons
subject to strong magnetic field and  disorder. The disorder was
implemented by a set of $\delta$-impurities with random positions and
random strengths   symmetric around zero.
The microscopic length scale of the problem is the magnetic length $l_B$
 defined by  the size of an area
penetrated by a single flux-quantum, $2\pi l_B^2$.
The representing Hamiltonian matrix was worked out
 in the Landau representation diagonalizing the clean system (no disorder).
The clean case has the famous Landau level spectrum of highly
degenerate and equally spaced levels at discrete energy values. The
spacing is given by the cyclotron frequency $\h \omega_c = \h eB/m$.
Disorder (symmetric around zero energy)
broadens the Landau levels to Landau bands of non-degenerate
states.
 For strong
magnetic fields the broadened Landau bands  are still
separeted, the band-width $\Gamma$ being smaller than $\h\omega_c$ (see
Fig.~25).

Restricting to one Landau band  (it is most convenient to take the
lowest) the LD transition takes place
precisely
at the center of the symmetric Landau band (cf.~\cite{JanB},\cite{HucR}). 
A system of area $L^2$ is represented by a finite matrix of dimension
$N=L^2/(2\pi l_B^2)$.  The numerical
diagonalization yields the eigenvalues and eigenstates for any
desired energy window within the lowest Landau band.
 
In Fig.~26
 the squared amplitudes of a wave function from the center
of the Landau band are shown together with
the $f(\alpha)$ spectrum calculated from these amplitudes.
 The corresponding histogram of
the logarithm of amplitudes (measured on a box of size $4l_B^2$)
is displayed in Fig.~27 together
with the distribution function calculated from the $f(\alpha)$ spectrum
using Eq.~(\ref{8.3}).  These figures demonstrate that the distribution
of amplitudes is (i) encoded in the $f(\alpha)$ spectrum and (ii)
is close to a log-normal distribution characterized by one critical
exponent
$\alpha_0=2.28\pm 0.03$ (the average over 130 critical states).

 Figures~26, 27 correspond to a model of
$\delta$-impurities  resulting in a short range potential correlation.
Similar calculations for a long range potential correlation (the potential
correlation
length is much larger than the magnetic length) have been performed
within the Chalker-Coddington network model \cite{Kles95}.
A corresponding critical wave function and its histogram are shown in
Figs.~28, 29. They demonstrate the
universality
of the multifractal properties with respect to the potential
correlation length.

To  study the
spatial correlations of amplitudes for a fixed energy 
 consider the $q-$dependent correlations
\begin{equation}
	M^{[q]}(r,L_b,L):=\left\langle \lbrack P_i(L_b)\rbrack^q
\lbrack P_{i+s}(L_b)\rbrack^q\right\rangle_L\label{8.7}
\end{equation}
where the average is to be taken over all pairs of boxes with fixed distance
$r=sL_b$.

For critical states where the microscopic scale ($l_B$ in our case)
 and the macroscopic scale (the localization
length $\xi$ in our case)  are separated, one can expect power law
behavior of $M^{[q]}$ 
in the regime 
\begin{equation}
	l_B\ll L_b,r,L \ll \xi \, .\label{8.8}
\end{equation} 
Usually, in critical phenomena one studies correlations for infinite
system size (and $L_b$ being microscopic)
as a function of $r$ alone. This is justified if, for large enough system 
sizes $L$, the correlation function is independent of $L$ (for simplicity
 we neglect 
 any trivial $L$-dependence due to pre-factors in the
definition of the observable $P$).
 However,
this is not true in the multifractal case.
 Multifractality reflects broadness of the
distribution function ${\cal P}(P,L_b/L)$ on all length scales. 
 The local box observable $P_i(L_b)$
depends on a large number of conditions for the entire system of
linear size $L$, simultaneously.
In the context of the LD transition coherence at zero temperature
is due to  quantum mechanical phase
coherence of the electron's wave function, and disorder introduces a
huge number of parameters, e.g. the position of point-scatterers.
In the multifractal scaling case one has to face the fact that 
$M^{[q]}$  depends non-trivially on $L$, even for $L\to\infty$. 
We have incorporated such behavior already in our general discussion
on correlation functions in Sect.~\ref{subrelbet}. 

Therefore, we consider the regime $l_B\ll L_b < r < L \ll \xi_c$ 
and make the ansatz 
\begin{equation}
	M^{[q]}(r,L_b,L)\propto L_b^{x_2(q)}L^{-y_2(q)}r^{-z(q)}\,
	.\label{8.10} 
\end{equation}

The task is now to find the scaling relations between the
 set of exponents $x_2(q),y_2(q),z(q)$ and  the $\tau(q)$
function.
These scaling relations can be derived by requiring consistency with 
 the limiting situations (i) where $r$ is of the
order of
 $L_b$ and (ii) where $r$ is of the order of $L$. We find
 \cite{Cat,JanR} 
\begin{eqnarray}	
	y_2(q) &=& d+\tau(2q) \label{8.11a}\\
	x_2(q) &=& 2d + 2\tau(q) \label{8.11b}\\
	z(q)&=& d +2\tau(q) - \tau(2q)\, .\label{8.11c}
\end{eqnarray} 
It is worth mentioning that the sum $x_2(q)-y_2(q)-z(q)$
vanishes due to the normalization of the wave function.

The analytic behavior of   $z(q)$
according to Eq.~(\ref{8.11c}) is shown in  Fig.~30. In
general, it is non-negative  and 
asymptotically bounded by the dimension $d$ .
To check on the validity of Eq.~(\ref{8.11c})
Pracz {\em et al.} took  $100$ critical states of a system with $L=200 l_B$
and calculated $M^{[q]}(L_b,r,L)$ with fixed values $L_b=l_B, 4 l_B$;
$L=200l_B$. 
Within the errors  the validity of the
 scaling relations in Eqs.~(\ref{8.11c}, \ref{8.11b})  could  be confirmed. 

To include aspects of local energy statistics as well, let us consider
 the $q$-dependent correlation of box probabilities corresponding to 
two different eigenstates with energies $E$ and $E+\omega$
\begin{equation}
	M_{\omega}^{[q]}(r,L_b,L):=\left\langle \lbrack P_i(E;L_b)\rbrack^q
	\lbrack P_{i+s}(E+\omega;L_b)\rbrack^q\right\rangle_L \, .\label{8.13} 
\end{equation}
To understand the correlation behavior of non-localized
states with respect to the
energy separation one has to
compare the relevant energy scales of the problem.
These are the average level spacing $\Delta$ and the (frequency
dependent)
Thouless energy
energy $E_{\rm Th}(\omega)$  corresponding to the time a wave packet
(formed from states within an energy window of width $\omega$) 
needs to diffuse
through the system, $L^2=(\hbar/E_{\rm Th}(\omega))D(\omega)$.
Here $D(\omega)$ is the corresponding diffusion constant.
 According to Chalker \cite{ChaPA},
 these scales give rise to the definition of two
length scales depending on  the energy separation $\omega$:
\begin{eqnarray}
	{\tilde{L}}_\omega & := & (\omega/E_{\rm Th}(\omega))^{-1/2}L \label{8.14}\\
	 L_\omega & := &
(\omega/\Delta)^{-1/d}L \, .\label{8.15}
\end{eqnarray}
The first length scale, $\tilde{L}_\omega$,
is the typical distance a wave packet will travel diffusively in a time
$\hbar/\omega$. From this it is natural to assume
that correlations between $P_{i}(E;L_b)$ and $P_{i+s}(E+\omega;L_b)$
will be present at least for distances $r \ll {\tilde{L}}_\omega$ whereas for
 larger distances the amplitudes are  uncorrelated.
Such uncorrelated behavior  is typical for the standard
 random matrix theory approach to extended states in random systems.
Recall, that in this theory  it is assumed  that
the unitary matrices that diagonalize the Hamiltonian are distributed
uniformly in the unitary group and no correlations (apart from the unitarity
property) between different
matrix elements occur. Thus, the
presence of correlations here  is an explicit
  breakdown of  the {\em  no-preferential
basis} 
assumption.
In electron systems with spatial
disorder  a preference
to some basis is always given. This preference
 cannot be seen  in $M^{[q]}_\omega$ for distances $r\gg \tilde{L}_\omega$.

The second length scale,  $L_\omega$, is the linear  size of a system
with
level spacing $\omega$. Two
  wave functions with energetic separation smaller
than
 the level spacing show a spatial correlation behavior
of its amplitudes similar to
that corresponding to  one of those wave functions, i.e.
they are statistically
indistinguishable. 

 At the critical
point of the LD transition the typical conductance becomes a size independent
quantity, $g^\ast$,   and with the help of the Einstein relation
between conductivity and diffusion one finds 
${\tilde L}_\omega= (g^{\ast})^{1/d}L_\omega$ \cite{ChaPA}.
Since $g^{\ast}$ is of ${\cal O}(1)$ the two length scales coincide at the LD
transition. Therefore, one can   focus 
on the role of $L_\omega$.

Asking for the scaling properties of $M^{[q]}_\omega$
in the regime $L_b < r < L_\omega \leq L$
we make the  ansatz (cf.~Eq.~(\ref{8.10}))
\begin{equation}
	M^{[q]}_\omega\propto
L_b^{X_2(q)}r^{-z(q)}L_\omega^{Z(q)}L^{-Y_2(q)}\, .\label{8.16} 
\end{equation}
Here we have already anticipated that the exponent with respect to $r$
is  $z(q)$, as given before.
As in the case of $M^{[q]}$ for a fixed energy by considering limiting
situations one finds
\begin{eqnarray}
       X_2(q)=2d+2\tau(q)=Y_2(q) \label{8.21}\\
       z(q)=Z(q)=d+2\tau(q)-\tau(2q)\, . \label{8.22}
\end{eqnarray}
 Now the following conclusions can
be drawn:
1.  The result
for $z(q)$ is the same as in the case of  zero energy separation.
2. The energy separation $\omega$ is not an independent scaling parameter
 but appears only in the combination $L_\omega/r$. 
3.  The exponent
corresponding to the box size, $x_2(q)=X_2(q)$, is not affected by a finite
energy separation but the exponent
corresponding to the system size $L$ (which is $y_2(q)$ for zero energy
separation) 
 splits up into the exponents $Z(q)$ (corresponding to $L_\omega$)
and   $Y_2(q)$ (corresponding to $L$ for finite energy separation).

In the work by Pracz {\em et al.}  it was   
 demonstrated that  the
scaling relations are consistent with numerical results.
Similar findings have  been obtained by Metzler \cite{MetDoc}
within the scattering matrix network model (Chalker Coddington model).

\subsection{Local Density Of States As Order Parameter}\label{sublocden}
Having established the role of $r/L_\omega$ as the relevant scaling parameter
for correlations of eigenstate amplitudes 
with universal exponent $z(q)$ (related  to $\tau(q)$ by a scaling relation)
let us now discuss the consequences of this  for the interpretation
of the local density of states being an order parameter of the LD transition.
 With the smearing-out
of the $\delta$-functions  the 
LDOS  is given by (see Eq.~(\ref{3.17}))
\begin{equation}
   \rho(E,{\bf r}) = \Gamma(E)^{-1}|\psi(E,{\bf r})|^2 \, \label{8.23}
\end{equation}
where $|\psi(E,{\bf r})|^2$ stands for the microcanonical average of
squared amplitudes at a given energy $E$.
Since $\Gamma(E)$ 
 behaves as $L^{-d}$,  the  scaling behavior of the LDOS
is determined by that of the wave function.
Consequently, we have
\begin{equation}
    \left\langle \lbrack \rho(E,{\bf r})\rbrack^q
	\right\rangle_L \propto L^{(q-1)d -\tau(q)}\, \label{8.24}
\end{equation}
and for  the typical value
\begin{equation}
	\rho_{\rm t} = \exp [\left\langle  \ln (\rho({\bf r}))\right\rangle_L] 		\propto L^{d-\alpha_0}
\label{8.25}
\end{equation}
which {\em  does} reflect the LD transition.
 Scaling $L$ with the correlation  length
$\xi_c\propto \tau^{-\nu}$  the typical LDOS
vanishes on approaching the critical point with exponent 
$\beta_{\rm t}=\nu(\alpha_0-d)$ ($\approx 0.7$ in quantum Hall systems).

The behavior of the typical LDOS is, according to the argument
presented in the beginning of this section, similar to that of the
participation
ratio introduced by Wegner \cite{Weg80}.
 The participation ratio was based on the
second moment of local probability. Higher moments will work as well
and our notion of typical LDOS is just a bit more convenient
since it focuses on a typical value rather than on some specific moments.

Our  interpretation is also consistent with the findings for the LDOS
in the localized and delocalized phase as discussed in Sect.~\ref{staide}.
Therefore, we summarize the findings:
\begin{enumerate}
\item
In the localized phase the typical local density of states  vanishes.
\item
In the delocalized
 phase the typical local density of states is finite and positive.
\item
 Approaching the LD transition from the
 delocalized phase the typical LDOS vanishes with a positive critical
 exponent,
\begin{equation}
	\b_{\rm t}=\nu(\a_0-d)\, . 
\label{betatyp}
\end{equation}
\end{enumerate}
A schematic picture for the typical LDOS that stresses its role as
 order parameter 
is shown in Fig.~31.

The unconventional feature as compared to ordinary critical phenomena
lies in the facts that 
\begin{enumerate}
\item
 The order parameter field has a broad distribution
resulting in a non-linear dependence of exponents on the degree of
moments considered (multifractality).
\item
The average value
shows a  vanishing scaling exponent  while the typical value  gives rise to
a positive scaling exponent
\end{enumerate}

The scaling relations that we derived for the wave functions
amplitudes transform to scaling relations of the LDOS since
each box amplitude has to be multiplied by a constant factor of $L^{d}$,
\begin{equation}
	\left\langle (\rho(E,{\bf r}_1))^q 
	(\rho(E+\omega,{\bf r}_2))^q\right\rangle_L \propto
	(r/L_{\omega})^{-z(q)} L^{-\tilde{z}(q)} 
	\label{8.26}
\end{equation}
with 
\begin{equation}
	z(q)=d+2\tau(q)-\tau(2q)\; , \;\; \tilde{z}(q)=2(1-q)d + 2\tau(q)\, 
.\label{8.27}
\end{equation}
These  scaling relations form, in the sense of Sect.~\ref{subrelbet},
 the appropriate
scaling relations connecting the spatial correlations
of the local order parameter field to its scaling dimensions
(cf. Eqs.~(\ref{8.24}),(\ref{8.26}),(\ref{8.27})). 

We  mention    that
 $\eta:=z(1)=d-D(2)\approx 0.4\not= \tilde{\eta}:=\tilde{z}(1)= 0$ 
(for the correlator of the density
of states in quantum Hall systems)
 with $L_\omega/r$ forming the scaling parameter is equivalent
(cf.~\cite{ChaPA,HucSCH94})
to the phenomenon of  {\em  anomalous diffusion}  found by
Chalker and Daniell \cite{ChalDan88}.
 As pointed out in \cite{ChalDan88},
the anomalous character of diffusion lies
in the non-Gaussian dispersion of a wave packet in time $t$ despite the fact
that the average diameter grows like $\sqrt{t}$. This non-Gaussian
time dispersion is caused by the multifractal character of eigenstates.

A further support for the order parameter 
 interpretation comes from conformal mapping
  ideas for 2D systems.
In ordinary critical phenomena theory,
the following assumption for critical correlation functions 
$\chi(r)\propto r^{-\tilde{\eta}}$
 seems
plausible: Scale invariance, reflected by power laws, should hold
also for local scale transformations which preserve angles but may change
scales locally (called conformal mappings) \cite{Car84}.

Thus, the main idea behind conformal mapping arguments in scaling
 theory
 is an 
extension of a  homogeneity
law for correlation functions with respect to rescaling. Such a 
law exists also in  the multifractal  case, since any rescaling of {\em  all}
length scales in the LDOS correlator (for a fixed energy)
by the same scaling factor $s$ leads to
\begin{equation}
 	\label{9.6}
	\langle \rho^q(s{\bf r}) \rho^q(s {\bf r'}) \rangle_{sL} = 
	s^{-\tilde{z}(q)}
	\langle \rho^q({\bf r}) \rho^q({\bf r'}) \rangle_{L} \, . 
\end{equation}
where $\tilde{z}(q)$ is given by Eq.~(\ref{8.27}).

Extending  this law
 to conformal mappings
of a geometry $\Omega$ 
to  geometry $\tilde{\Omega}$ leads,
for large but finite 2D systems, to
\begin{equation}
	\frac{\langle \rho^q(w(z_1)) \rho^q(w(z_2)) \rangle_{\tilde{\Omega}}}
	{\langle \rho^q(z_1) \rho^q(z_2) \rangle_{\Omega}} =
	|w'(z_1)|^{-\frac{\tilde{z}(q)}{2}}|w'(z_2)|^{-\frac{\tilde{z}(q)}{2}}
	\, , 
	\label{9.7}
\end{equation}
where $w(z)$ is any holomorphic function of complex coordinate $z$
and $w'(z)$ denotes the derivative.

By choosing $w(z)= (L_T/2\pi)\ln z$ which maps the plane onto
a strip one can show  (cf.~\cite{Car84})
that the correlator in the strip is  characterized by a $q$-dependent
quasi-1D
localization length $\xi(q;L_t)$ which is related to $\tilde{z}(q)$
by 
\begin{equation}
\label{conf1}
	\xi (q;L_t)/L_t  =2/(\pi \tilde{z}(q))\, .
\end{equation}

In FSS calculations it is rather the average of the logarithm of the
correlation function which is calculated by knowing that this quantity
defines a typical localization length.
 Therefore, the  result of Eq.~(\ref{conf1})  leads
in particular (consider $q$ close to $0$) to the relation 
\begin{equation}\label{conf2}
	\Lambda^{\ast} = \frac{2}{\pi (\alpha_0 -d )}\, ,\;\; d=2 \, .
\end{equation}
Here $\Lambda^\ast$ appears since it is just the typical localization
length at criticality divided by the width $L_t$.
Equation~(\ref{conf1}) has been confirmed in numerical calculations by
Dohmen {\em et al.} \cite{Doh96}, and Eq.~(\ref{conf2}) is, so far,
 in accordance
with all known numerical results.
\subsection{Conductance As Scaling Variable}\label{subcond}
The fact  that the critical value $\alpha_0-d$ is related to the
critical value of the scaling variable (at least in 2D) is an
important
hint for the typical conductance $g_{\rm t}$ being a one-parameter 
scaling variable.
The reasoning behind this statement goes as follows:
as pointed out in Sect.~\ref{subfin} the scaling variable $\Lambda (L_t)$
of finite size scaling in quasi-1D can be related to the typical
conductance of a cube $g_{\rm t}(L=L_t)$. 
From the $2+\epsilon$ field theoretic calculations at the LD
transition \cite{Weg80} as well as from the multifractal LDOS found in
2D \cite{Fal95} we can conclude:
if the correlation length is larger than system size 
and the typical conductance is
still large, the 
eigenstates
behave critical and are multifractal. The multifractal exponent
$\alpha_0-d$
is determined by $g_{\rm t}$ and 
\begin{equation}\label{alph-g}
	\a_0-d \propto 1/g_t\, . 
\end{equation}
At a generic critical point with $\Lambda^\ast \sim g_t\sim {\cal O}(1)$
the conformal mapping relation tells, that also here the 
value of $\alpha_0-d$ is determined by $g_t$.
This means that the bulk of the LDOS distribution function is, in the
critical
regime, already
determined by the value of typical conductance.

Before we collect results for the conductance 
distribution  at criticality we discuss a similar object related to the level statistics. Not only the LDOS and conductance distributions
 should be universal
functions  at criticality (see Eq.~(\ref{limitdis})), but also e.g. the
level spacing distribution $P(s;L)$ and the two-level correlation function
$R(s;L)$.
To study these quantities in view of the one-parameter scaling picture
was put forward in a work by Shklovskii {\em et
al.} \cite{Shk93}.
These authors pointed out that the study of the level spacing
distribution
$P(s;L)$ is, starting from a Hamiltonian modeling,
numerically much easier to perform than calculating conductances.
Only the spectrum has to be calculated. According to the one-parameter
scaling picture $P(s;L)$ should obey the following scaling behavior
\begin{equation}
	\lim\limits_{L\to\infty} P(s;L)=
	\lim\limits_{L\to\infty} F (s;\alpha_L )=  P_{\rm L,D,C}(s)\, .
	\label{10.1}
\end{equation}
Here $F(s;\alpha_L)$ is a function that does no longer  depend on  
microscopic details,  but only on one scale dependent parameter
$\alpha_L$.
Depending on the initial value of $\alpha_L$ the system flows to either
the localized fixed point (L), the delocalized fixed point (D) or it
stays at the critical fixed point (C). In terms of the typical
conductance, $\alpha_L = g_{\rm t}(L)$, the fixed points are
characterized
by 
\begin{equation}
	\lim\limits_{L\to\infty} g_{\rm t}(L) = \LB \begin{array}{cc}
	0 & ({\rm L})\\
	g_{\rm t}^\ast & ({\rm C}) \\
	\infty & ({\rm D}) \end{array} \RB\, .
	\label{10.2}
\end{equation}
The corresponding asymptotic 
level-spacing distributions $P_{\rm L,D,C}(s)$
 are given by the Wigner surmise, Eq.~(\ref{5.22}), 
\begin{equation}
	P_{\rm D}(s)= A_\b\, s^\b \exp \LK - B_\b\, s^2\RK \, ,\label{10.3}
\end{equation}
the Poisson distribution, Eq.~(\ref{5.11}), 
\begin{equation}
	P_{\rm L}(s)= \exp \LK -s\RK \, ,\label{10.3b}
\end{equation}
and by a, yet unknown, critical distribution $P_{\rm C}(s)$.

Numerical studies by several authors (e.g.~\cite{Shk93}, \cite{Eva94},
\cite{Hof94},
\cite{Zha95}, \cite{Fei95}) have demonstrated that a
size independent 
critical  distribution $P_{\rm C}(s)$ distribution exists, once the
parameter $\tau$ triggering the LD transition is put to its critical
value $\tau=0$.
It is also known that, qualitatively,
 the distribution resembles the Wigner surmise
for
values $s\leq s_0\approx 2$ and the Poisson law for $s\geq s_0$. At
$s_0$
the distributions intersect. 
Furthermore, the way the critical distribution is approached
allows for a calculation of the critical exponent of the correlation
length $\nu$.
This can be done in several ways. The authors of \cite{Shk93}
took as a scaling variable the integral over $P(s;L)$ up to $s_0$
(one could use also some other point)
\begin{equation}
	\alpha(L;\tau) := \int\limits_0^{s_0} ds\, P(s;L;\tau) \, .\label{10.4}
\end{equation}
and analyzed it according to the general ansatz, (Eq.~(\ref{6.19b}), 
\begin{equation}
	\alpha(L;\tau)= \alpha^\ast + A \tau L^{1/\nu} + {\cal
	O}\LK \tau^2\RK  \, . \label{10.5}
\end{equation}
The values for $\nu$ found by this method do agree with those found
in FSS
analysis within the errors. 

As to the precise form of $P_{\rm C}(s)$ there exists no conclusive
theory, except that the very small $s$ behavior is again dictated by
the
level repulsion $\propto s^\b$. For example, the tail of $P_{\rm
C}(s)$
is not exactly known. It seems now
to be established that there is a leading exponential tail with
sub-leading
stretched exponential tail contributions. The question about the
form of the 
tail is related \cite{Alt88} to the question about the 
relation between the number variance $\Sigma^2$ and the average level
number $\BRA n\KET$, (see Sect.~\ref{staide}).
Based on one-parameter scaling arguments Aronov {\em et al.} \cite{Aro95}
had predicted (for the LD transition point without magnetic fields)
the following scaling law, for large $\BRA n\KET$, 
\begin{equation}
	\Sigma^2 \propto \LK \BRA n \KET\RK^{1-\frac{1}{d\nu}} \,
	\, .\label{10.6} 
\end{equation}
However, in this work the anomalous diffusion at criticality (or in other
words
the multifractality) was not taken into account. Correcting this
omission leads to an additional linear term, characterized by
a finite level compressibility
\cite{Cha96eta}
\begin{equation}
	\frac{d \Sigma^2}{d  \BRA n \KET}= \frac{d-D(2)}{2d}\, .
	\label{10.7}
\end{equation}
In view of the finite level compressibility Eq.~(\ref{10.6}) describes
 sub-leading contributions to $\Sigma^2$.

 It is very interesting
that 
the analysis of $\Sigma^2$ as a function of $\BRA n\KET$ seems to
allow for
 computing
the correlation length exponent $\nu$ and the fractal dimension $D(2)$
related to the order parameter exponent $\beta_{\rm t}
 =(\alpha_0-d)\nu$
(within the parabolic approximation to $f(\alpha)$ $D(2)\approx
 d - 2(\alpha_0-d)$). 

 As to the  critical  conductance distribution
the general considerations formulated in Eqs.~(\ref{10.1},\ref{10.2}) 
for the level spacing distribution do apply as well.
The precise form of the critical conductance
distribution ${\cal P}^\ast (g)$  is not known.
Based on the $2+\epsilon$ calculations by Al'tshuler et
al. \cite{Alt86-89}, 
Shapiro and coworkers \cite{Sha88-92} 
predicted that power law tails should occur.
So far, this could not neither be proved nor disproved by numerical
means.
However, in a work by Markos and Kramer \cite{Mar93} it was shown that
the critical conductance distribution is universal for a 3D system
without magnetic field, i.e. it does not depend on system size $L$
and does not depend on which parameter $\tau$ is taken to its critical
point
value. The distribution showed fluctuations $\sqrt{\BRA (\delta
g)^2\KET}$
that are of the same order (${\cal O}(1)$) as  the mean  value of the
distribution. The precise form of the tails could not be determined.

The  scaling of the conductance
distribution close to criticality  
follows  from expanding the scaling variable around criticality which
brings the exponent $\nu$ into play,
\begin{equation}
	{\cal P} (g;L;\tau) = F(g;\alpha_L(\tau)) = {\cal
	P}^\ast (g) + F'(g;\alpha^\ast) A \tau L^{1/\nu} + \ldots \, ,
	 \label{10.10}
\end{equation}
where $F'$ denotes the derivative of the function $F$ with respect to
the scaling variable $\alpha_L$.
To analyze the behavior of ${\cal P}(g;L;\tau)$ at one particular
point is numerically not advantageous. Therefore some integrated
quantities work better. However, as explained earlier, 
the chosen quantity should not be sensible to the
tails
of the distribution. Therefore, one should refrain from taking moments. 
Unfortunately, not much has been done in this direction.

Fastenrath {\em et al.} \cite{Fas92} have investigated a 
quantity which is similar to the dissipative conductance, the so-called
Thouless number, for quantum Hall systems and studied
the distribution function.
Also in this case it turned out that, at the critical point, the distribution
function was scale independent with fluctuations of the same order
as the typical value. They explicitly took the typical Thouless number
$g_{\rm t}=\exp \BRA \ln  g \KET$ and analyzed it according to 
\begin{equation}
	g_{\rm t}(L;\tau)  = g_{\rm t}^\ast + A \tau L^{1/\nu} +\ldots
	\, .\label{10.8}
\end{equation}
They found a value for $\nu$ ($\approx 2.3$)
which is in good agreement with the value
obtained by FSS. Thus, their observation gives further support that
the typical conductance is an appropriate scaling variable.

Cho and Fisher \cite{Cho96} have performed a calculation of a true
two-probe conductance within  the Chalker-Coddington network model 
for the quantum Hall
effect.
There findings are consistent with  the two-probe experiment by Cobden
and Kogan  \cite{Cob96} mentioned in the introduction: the
distribution is almost uniform between $0$ and $1$. This means in
particular
that the fluctuations are of the same order as the average value.
For this uniform distribution on a finite interval the average value
is already a convenient choice for the scaling variable and
 Cho and Fisher were able to show that the average value
 shows scaling of
the form written in Eq.~(\ref{10.8}) with the correct critical exponent
$\nu\approx 2.3$. Similar results were also obtained by Wang {\em et al.}
\cite{Wan96}. 

Thus, although the precise form of critical conductance distributions
is not known, there are a number of results which confirm that it is
universal
and that the notion of typical conductance allows for studying
the critical exponent of the correlation length.
\section{Summary}\label{sum}
In this review we have discussed a few fundamental  experiments that show
the importance of quantum mechanical interference effects for
transport measurements in mesoscopic systems. Static disorder leads to
complicated interference patterns  of electronic wave functions which
can cause large fluctuations in physical quantities. In addition, they 
 can lead to
localization of states to a finite volume within the mesoscopic
system. Mesoscopic systems can be experimentally realized at very low
temperatures ($\leq 1$K) and with small devices ($\leq 1\mu$m)
such that the phase coherence length $L_\phi$ is larger than system
sizes $L$.

We introduced the basic physical quantities which are the local and
global
density of
states (LDOS and DOS), the global conductance which is (in atomic units
$e^2/h$) given by a dimensionless quantity $g$, and the conductivity
tensor
$\sigma_{\mu\nu} (\r,\r';\omega)$ (or equivalently the diffusion
function  $D_{\mu\nu}(\r,\rp;\omega)$), related to local charge transport. 
The relevant physical energy scales are the quantum kinematic scales of
level spacing $\Delta$ and
 Fermi energy $\EF$, and the transport energy scale set by the
Thouless energy $E_{\rm Th}$.
The conductance is given by the Thouless formula, $g=E_{\rm
Th}/\Delta$.
 A quantum kinematic length scales is 
the Fermi wavelength $\lambda_F$, and   transport length scales
are the microscopic mean free path $l_e$ and 
 the localization length $\xi$ for localized states.

We discussed   models for disordered mesoscopic electron systems in
terms of Hamiltonians and scattering $S$-matrices. 
From the Hamiltonian models energy eigenvalues $\varepsilon_\a$ 
and eigenvectors $\psi_\a$
can be calculated. Their statistical properties determine the
statistics
of the most important quantity, the Green's function which determines
all of the relevant quantities and scales.
Within the $S$-matrix modeling the transmission strengths $T_{\a\b}$
are the most important quantities. They directly determine the
conductance via the B\"uttiker formula, $g=\sum_{\a\b} T_{\a\b}$.
In addition, certain network $S$-matrix models also allow for a
determination of energy eigenvalues and eigenvectors.
 
For the statistical and scaling properties of mesoscopic electron
systems
we collected the following insights.
\begin{enumerate}
\item 
Ideal localized systems (with vanishing localization
 length $\xi$, i.e. vanishing conductance $g=0$)
 are characterized by a
compressible spectrum of uncorrelated energies and a LDOS that
typically vanishes, although the average DOS is finite.  
\item
Ideal delocalized systems (with infinite Thouless energy $E_{\rm Th}$,
i.e. infinite conductance $g=\infty$)
are characterized by an incompressible spectrum of correlated energies
and Gaussian distributed wavefunctions which leads to
 an exponential  tail  for the LDOS probability distribution.

\item 
Changing control parameters, $\tau$, 
of the mesoscopic system like particle density,
disorder strength, pressure or applied fields can drive the system
through a disorder induced localization-delocalization (LD) transition
(at
$\tau=0$)
which resembles a critical phenomenon. 
Due to the randomness of disorder the transition has to be described
in terms of distribution functions the flow of which with respect to
system size (scaling) forms the object of interest in a critical
phenomenon
description.

\item
A one-parameter scaling theory for distribution
functions of global quantities like 
conductance and  energy-level spacing seems to work well.
It states that the distributions are essentially determined by
one scale dependent parameter $\alpha_L$, called scaling variable, that
 determines the flow,
${\cal P}(X;L) \approx F(X; \alpha_L)$.

\item
A possible  scaling variable for the LD transition is given by 
the geometric mean of the conductance distribution  which we called typical
conductance
$g_{\rm t}$. With increasing system size $g_{\rm t}$ can approach the
idealized
localized phase ($g_{\rm t}=0$), or it can approach the idealized
delocalized
phase ($g_{\rm t}=\infty$). Corrections to distribution
functions as compared to the ideal situations are controlled by finite
values of $g_{\rm t}$.
Right at the LD transition $g_{\rm t}$
will become a size independent universal quantity $g_{\rm t}^\ast$
that is  of order $1$ in generic LD transitions.
The corresponding distribution functions of global quantities like
conductance
and level-spacing are universal scale-independent functions.

\item
The fictitious system size $\xi_c$ for which $g_{\rm t}(\xi_c) \approx
g_{\rm t}^\ast$ defines the correlation length of the LD transition.
In the localized phase it can be identified with the localization
length $\xi$. Close to the transition point, where it diverges,
it scales with a critical exponent, $\xi_c \propto \tau^{-\nu}$.

\item
 While in the localized phase the typical (not the average)
 LDOS vanishes, the LDOS shows  power law scaling in non-localized
phases. The average value is always scale-independent.
In the ideal delocalized phase one exponent, simply related
to dimensionality $d$, controls the scaling of the whole distribution
function. At the LD transition a spectrum of
scaling exponents is necessary to describe the scaling of the
distribution which  is close to a log-normal
distribution. The spectrum has an interpretation in terms of a
spectrum of fractal dimensions (multifractal spectrum).
Still, there is one particular exponent, related to the
typical value of the LDOS, which determines the log-normal
approximation for the distribution.
It is called $\alpha_0-d >0$ and is determined by the value of the
scaling variable at criticality.
Thus, multifractality is not in conflict with one-parameter scaling
theory.

\item
On the contrary, the multifractal scaling of the LDOS opens the
possibility
to consider the typical value of the LDOS as an appropriate order
parameter
for the LD transition. The corresponding critical exponent
is given by $\b_{\rm t}=\nu(\a_0-d)$. Further support to this idea
comes from the observation that a number of scaling relations
for the LDOS can be formulated that 
are in close analogy to similar relations in
ordinary critical phenomena. 
\end{enumerate}

In short,  the occurrence of  broad
distributions and multifractality in disordered mesoscopic electron
systems
is by no means in contradiction to the one-parameter scaling
theory, but points out  that only typical rather than average
quantities can  serve as  order parameter and scaling
variable
in this theory.

\bigskip
\centerline{{\bf Acknowledgments}}

I gratefully acknowledge helpful discussions with A. Altland, Y.
Avron, M. Backhaus, J. Chalker, A. Dohmen, P. Freche,  
J. Hajdu, B. Huckestein, R. Klesse, I. Lerner, M. Metzler, D.  Polyakov,
K. Pracz, B. Shapiro
  and M.
Zirnbauer.

This work was supported by SFB 341 of the
Deutsche Forschungs\-gemein\-schaft and the MINERVA foundation.  I thank
the Department of Theoretical Physics in Oxford, U.K., and the
Institute of Theoretical Physics at the Technion, Haifa, for the kind
hospitality during visits in which part of the work was carried out.

\begin{appendix}
\section{Brief Account Of Linear Response Theory}\label{bri}
We start  with a   general outline of how to calculate linear response
quantities when the system is specified by a Hamiltonian. We then
specialize to additive systems where the resolvent $G(z):=(z-H)^{-1}$
of the corresponding one-particle Hamiltonian (taken at complex
energies $z$) yields all the basic physical quantities. Studying
transport in multi-probe geometries will lead to the concept of the
scattering
matrix.

Consider a many-particle system characterized by 
 the Hamilton operator $H$ coupled
to a linear increment of a  force field, $F$,
\begin{equation}
	{\cal H}_F=H-X F \label{4.1}
\end{equation}
where $X$ is a hermitean operator of the system. 
The system is assumed to be in an equilibrium 
state at temperature $T$ described by the (grand) canonical density
 operator
$\rho_F$.
The   thermodynamic susceptibility
$\chi_{_{YX}}$ is defined as
\begin{equation}
	\chi_{_{YX}}:=\frac{\partial{\BRA Y\KET}}{\partial F}\, .\label{l4.2}
\end{equation}
Linearizing $\rho_F$ with respect to the  increment $ F$
 the thermodynamic  susceptibility reads 
\begin{equation}
	\chi_{_{YX}}=\b\kubosp{\Delta X}{\Delta Y}\, .\label{4.3}
\end{equation}
where the bilinear expression (Kubo scalar product)
\begin{equation}
	\kubosp{X}{Y} := \frac{1}{\b} \Int_0^\b d\lambda\,
	\BRA XY(i\hbar\lambda )\KET = \kubosp{Y}{X}^{*}\label{4.4}
\end{equation}
 is     called
the  canonical correlator of $X$ and $Y$ with $\b=(k_BT)^{-1}$,
$\Delta X := X- \BRA X\KET$, and
$X(t)=e^{(i/\hbar)Ht}Xe^{-(i/\hbar)Ht}$ is the  time
 evolution of $X$ in the Heisenberg picture. 
The isothermal susceptibility describes an idealized linear response
 situation where the 
force field 
does not disturb  equilibrium, but
is just a parameter field of the system.

We  turn now  to the dynamic linear response problem 
\cite{KuboII,Fick90} 
where the following
(Kubo) process is considered: the increment in an applied force field 
$ F(t)$  depends on time
$t$ and is switched on adiabatically at $t=-\infty$.
The total Hamilton operator is
\begin{equation}
	{\cal H}_F(t)=H-XF(t)e^{\eta t}\, ,\label{4.5}
\end{equation}
where the  slowness parameter
$\eta$ has to be much smaller than the inverse of a typical positive time
scale $t_m$ at which the response to the field is measured
The system is assumed to be in equilibrium for $t=-\infty$ described
by the density operator $\rho(F=0)$ and isolated for
$t>-\infty$. 
Restricting our consideration to one Fourier component of the
force field,
\begin{equation}
	F(t)=Fe^{-i\w t}\, , \label{4.6}
\end{equation}
the resulting  linear increment in the Fourier component of
the mean value of an observable $Y$ is  
\begin{equation}
	\BRA Y\KET_F = \chi_{_{YX}}(z) F \label{4.7}
\end{equation}
where the  dynamic susceptibility
$\chi_{_{YX}}(z)$ 
is given by the  Kubo formula
\begin{equation}
	\chi_{_{YX}}(z)  =
	  \b\dcorrz{\dot{X}}{Y}=\b\dcorrz{\Delta \dot{ X}}{\Delta Y}\, .
	  \label{4.7b}
\end{equation}
Here $\dot{X}$ denotes the time derivative of $X$ and
 the dynamic correlator
\begin{equation}
	\dcorrz{X}{Y}:=\Int_0^{\infty}d t\, e^{izt}\kubosp{X}{Y(t)}\label{4.8}
\end{equation}
appears as  the Laplace transform ($z=\w +i\eta$) 
of the time dependent canonical
correlator. 

Integrating by parts in Eq.~(\ref{4.7b}) yields the fundamental 
equation of motion
for dynamic correlators
\begin{equation}
	\dcorrz{\dot{X}}{Y} = \kubosp{X}{Y} +iz\dcorrz{X}{Y} =
	-\dcorrz{X}{\dot{Y}}\, .\label{4.9}
\end{equation}
Another important equation of motion, referred to as Kubo relation, holds for 
the canonical correlator
\begin{equation}
	\kubosp{\dot{X}}{Y}=\frac{-i}{\b\h}\BRA\LBK X, Y\RBK\KET\label{4.10}
\end{equation}
where  $\LBK X,
Y\RBK $ denotes 
the commutator of $X$ and $Y$.
It is worth mentioning that the second term in Eq.~(\ref{4.9})
can be finite in the case of zero frequencies,
$\w =0$, since it defines the  long time average of
$\kubosp{X}{Y(t)}$.
\begin{equation}
	\overline{\kubosp{ X}{Y}} := \lim_{\eta \to +0} \eta
	\dcorr{X}{Y}\LBK i\eta \RBK \, .\label{4.11}
\end{equation}

Equations~(\ref{4.3},\ref{4.7b},\ref{4.9},\ref{4.10}) 
form the backbone
of any linear response theory  (cf.~\cite{KuboII,Fick90}). 
With these we can now write down the expressions for the 
 physical quantities considered
in Sect.~\ref{bas}.

The local charge density, $q(\r)$, can be related  to local
electric   
potentials, $U(\r)$, by the charge response function,
$\Pi(\r,\rp;\w)$\footnote{For simplicity
we drop the dependence on the time dependent vector potential.
},    
\begin{equation}
	q(\r)=	\int d^dr'\, 	\Pi(\r,\rp;\w) U(\rp) \, .\label{2.8}
\end{equation}
Charge response and conductivity (Eq.~(\ref{2.7})) 
are related due to the continuity
equation
\begin{eqnarray}
	{\rm div}\, \j(\r)-i\w q(\r)&=&0		\, ,	\label{2.9a}\\
	\sum_{\mu\nu}
	\frac{\partial}{\partial r_\mu}\frac{\partial}{\partial r'_\nu} 
	\sigma_{\mu \nu}(\r ,\rp;\w)& =& i\w \Pi(\r,\rp;\w) \, .\label{2.9b}
\end{eqnarray}

The particle density $n(\r)$ couples to the local chemical potential
$\mu (\r)$
in the same way as the local charge $q(\r)=-en(\r)$ couples to the
local electric potential $U(\r)$
and thus the thermodynamic charge response reads
\begin{equation}
		\Pi(\r,\rp)=-\b\kubosp{\Delta q(\rp)}{\Delta q(\r)}
		\, ,\;\; \int d^dr'\,
		\Pi(\r,\rp)=e^2\rho(\r;\mu)	\label{4.12}
\end{equation}
while the dynamic charge response reads
\begin{equation}
		\Pi(\r,\rp;\w)=-\b\dcorrz{\Delta \dot{q}(\rp)}{\Delta q(\r)}	
			      =	\Pi(\r,\rp) -i\b z
			      \dcorrz{\Delta q(\rp)}{\Delta q(\r)}  \, .
				\label{4.13}
\end{equation}

Note, that  the equilibrium charge response $\Pi(\r,\rp)$
is not equal to the $\w\to 0$ limit of the dynamical charge response
$\Pi(\r,\rp;\w)$.
This can be also seen when considering the response to homogeneous potentials.
While a  homogeneous electric potential
does not create a charge response, and hence $\int d^dr'\, 
\Pi(\r,\rp;\w)=0$,  shifting the overall  chemical potential by a
constant usually leads to a change in the particle density, i.e.~$\int d^dr'\,
\Pi(\r,\rp)=e^2\rho(\r;\mu)$.

According to the general scheme described above,
The  conductivity tensor appears as the dynamic current correlator
\begin{equation}
		\sigma_{\mu \nu}(\r ,\rp ; \w) = 
		\b \dcorrz{j_\nu (\rp)}{j_\mu (\r)}		\label{4.14}
\end{equation}
and the relation in Eq.~(\ref{2.9a}) can be viewed as a special case of
Eq.~(\ref{4.9}). 
The corresponding global conductivity tensor reads in the
thermodynamic limit $L\to \infty$
\begin{equation}
		\sigma_{\mu \nu}(\w) = \frac{e^2}{L^d}
		\b \dcorrz{v_\nu}{v_\mu}		\label{4.14b}
\end{equation}
where ${\bf v}$ denotes the velocity operator of electrons.

More delicate is the representation of the global conductance
coefficients $G_{km}(\w)$, since one has to know the local potential
distribution within the conductor as a functional of applied
electro-chemical potentials $U_m$ and we do not go into the detail of
this interesting problem. However, due to the continuity equation the
situation simplifies for the d.c. situation: The current in any lead
is
independent of the actual potential distribution $U(\r)$ within the
conductor
(cf.~\cite{Jan91}) and $G_{km}$ is determined by the dynamical
 correlator of asymptotic current operators  which are
the time derivatives of the total charge operator, $I_k=\dot{Q}_k$,
\begin{equation}
		G_{km} =  -\beta 
	\dcorr{\dot{Q}_m}{\dot{Q}_k}
		\label{4.15}
\end{equation}
where omission of the argument $[z]$  refers to  the d.c. situation.

Still, the calculation of the correlators requires the knowledge
of the many particle density operator $\rho_{F=0}$ and the machinery
of many-body physics is necessary to get quantitative results.

Assuming additive electron Hamilton operators like the ones introduced
in Sect.~\ref{mod} the correlators can be written in closed form
with the help of the one-particle resolvent (Green's function 
in real space representation)
\begin{equation}
	G^{\pm}(E) := (E\pm i\epsilon -H)^{-1}\, ,\;\; \epsilon \to +0
			\label{4.16}
\end{equation} 
of the corresponding one-particle Hamiltonian $H$.
It  defines  a spectral resolution via 
\begin{equation}
	\delta \LK E-H \RK = (2\pi i)^{-1} \LK \GM (E) - \GP (E)\RK\, .
		 \label{4.17}
\end{equation} 
With the help
of
the Fermi distribution function
\begin{equation}
	f(E)=\LBK 1 + e^{\b\LK E-\mu \RK} \RBK^{-1}\, , \label{4.18}
\end{equation}
(where $\mu$ denotes the chemical potential)
 average values can be calculated by 
\begin{equation}
	\BRA X \KET  = \frac{1}{2\pi i}\INTF 	
	\Tr\LB
	\GM (E) X -\GP(E) X\RB \, .\label{4.19}
\end{equation}
The canonical correlator reads
\begin{equation}
\kubosp{\Delta X}{\Delta Y}  =  
	\frac{i}{2\pi\b}
	\INTF   \Tr\LB
	 X\GM Y \GM - X\GP Y \GP\RB (E) \, ,	 
	\label{4.20}
\end{equation}
and identifying $\eta$ with $2\epsilon\h^{-1}$ yields
\begin{equation}
	\dcorrz{X}{Y}  =  
    	 \frac{\hbar}{2\pi \b}\INTF  {\rm Tr}\LB \LK \GM -\GP\RK_E\LK
    	X \GM_{E-\hbar \w}\GM_E Y 
	-  Y \GP_{E+\hbar \w}\GP_E X  \RK  \RB 
	\, .\label{4.21}
\end{equation}
for the dynamical correlator,
provided the equilibrium mean values of $X$ or $Y$ vanish.

Since the Fermi function is a step function at $T=0$ the general
structure
$\int dE \, f(E) A(E)$ occurring in Eqs.~(\ref{4.19}--\ref{4.21}) 
allows to conclude  e.g. for the canonical correlator
\begin{equation}
	\b \kubosp{X}{Y}\Bigg|_{T,\mu} = - \int d E\, 
	\frac{\partial f}{\partial E}
	\, \b \kubosp{X}{Y} \Bigg|_{T=0,\mu =\EF =E}\, .\label{4.22}
\end{equation}
A similar equation holds for the dynamic correlator.
Therefore, for additive (non-interacting) Fermion 
systems
we can always restrict the  discussion to the zero temperature limit.

In the zero  frequency limit  further simplifications 
are possible. For example, any auto-correlator is
essentially determined by states at the Fermi level $\EF$,
\begin{equation}
	\b \dcorr{X}{X} (T=0) = \pi\h\; \Tr\LB X\delta (\EF -H)X\delta 
	(\EF -H)\RB  \label{4.23}
\end{equation}
which shows that the longitudinal
conductivity  
 is a Fermi level quantity.

To simplify other correlators 
   it is 
convenient to use the equation of motion on the level of the one-particle
resolvent operator 
\begin{equation}
	G(z){\dot X} G(z') = \frac{i}{\h} \LB G(z)X - XG(z')
	+\LK z-z'\RK G(z)XG(z')\RB \, .\label{4.24}
\end{equation}
An important example for an (essentially)  Fermi
level quantity is the zero frequency dynamic correlator of two
operators one of which is a time derivative
\begin{equation}
	\b \dcorr{\dot X}{Y} (T=0) = \b \kubosp{\Delta X}{\Delta Y} (T=0)
	+ \lim_{\epsilon \to +0}
	\frac{\epsilon}{\pi}\; \Tr \LB X\GM(\EF)Y \GP(\EF)\RB\,
	.\label{4.25}  
\end{equation}

The d.c. conductance coefficients 
at $T=0$ read for $n\not= m$
\begin{equation}
	G_{km}(\w=0,T=0) = \frac{-\h}{2\pi}\; \Tr \LB 
	G^{-}(\EF)\dot{Q}_mG^{+}(\EF)\dot{Q}_n\RB\,
   \label{4.26} 
\end{equation}
and it is obvious from Eq.~(\ref{4.25}) that the conductance coefficients are
Fermi level quantities, too.

By applying Eq.~(\ref{4.25}) to the longitudinal conductivity 
one finds
\begin{equation}
	\sigma_{xx}(\w=0, T=0)= \frac{4e^2}{hL^d} {\epsilon}^2
	\; \Tr \LB G^{-}G^{+}x^2 -G^{-}x G^{+}x\RB(\EF )\, .\label{4.27}
\end{equation}

From Eq.~(\ref{4.26}) one can conclude that the conductance
coefficients
are determined by   asymptotic Green's functions, 
and it is tempting to formulate  a scattering
theoretical  description. This is indeed possible by
decomposing the total Hamiltonian into two parts, one of which
describes the leads, $H^{0}_{\rm leads}$, and a second part which
describes the conductor plus its coupling to the leads, $H_{\rm
scatter}$,
\begin{equation}
	H=H^{0}_{\rm leads} + H_{\rm scatter}\, .\label{4.28}
\end{equation}
The total resolvent $G(E)$ can then be decomposed according to Dyson's
formula on introducing the $T$-matrix of scattering theory
\begin{equation}
	G(E)=G^{0}(E)+G^{0}(E)T(E) G^{0}(E)\, ,\;\; T(E)=H_{\rm
	scatter}G(E)(G^{0}(E))^{-1} \, .\label{4.29}
\end{equation}
The eigenstates of the leads are scattering states, denoted as
$\left| k,\a\KET$,
where $k$ labels the lead and $\a$ the quantum number (channel) of a scattering
state at energy $E$. The matrix elements
\begin{equation}
	t_{km}^{\a\b}(E):=\BRA m,\b\mid T(E)\mid k,\a\KET \label{4.30}
\end{equation}
define transmission amplitudes to scatter from channel $\a$ in lead
$k$ to channel $\b$ in lead $m$ and form the {\em  matrix of transmission
amplitudes} 
$t_{kn}$. Applying Eq.~(\ref{4.25}) and  Eq.~(\ref{4.26})
and using the fact that the asymptotic charge operators are nothing
but
$-e$ times the projector onto the Hilbert space of the leads, the
important B\"uttiker  formula ($m\not=k$)\cite{Buet86}
\begin{equation}
	G_{km}(\w =0,T=0)=
	 \frac{e^2}{h}\; \Tr \LB {t}_{km} t^{\dagger}_{km}
\RB (\EF)\, \label{4.27b}
\end{equation}
can be obtained. 
\end{appendix}

\newpage

\centerline{\bf\large Figure Captions}

\bigskip
\bigskip

\noindent
Figure 1:
Schematic view of a mesoscopic conductor. Electrons experience
elastic scattering events on their way from the current source to the
current sink.

\bigskip
\noindent
Figure 2: A mesoscopic ring-shaped conductor. A flux $\phi$ is put
through
the ring.

\bigskip
\noindent
Figure 3:
Resistance oscillations for a ring-shaped conductor with flux
$\phi$. The period is $\phi_0=h/e$ as shown in the inset where the
Fourier spectrum is displayed. (after \protect\cite{Tim88}).

\bigskip
\noindent
Figure 4: Universal conductance fluctuations as a function of applied magnetic field. Different curves correspond to gold-nanowires
of different length. The average value of conductance was shifted for
convinience (after \protect\cite{Heg96}).

\bigskip
\noindent
Figure 5:
 Hopping transport in amorphous Silicon. The logarithm of the
conductivity  is plotted versus \protect$T^{-1/4}$ showing Mott's
law. The different curves are for different deposition temperatures
(after \protect\cite{Bey74}).

\bigskip
\noindent
Figure 6:
The resistance of thin disordered film of coupled fine
Cu-particles   as function of
logarithm of the temperature (after \protect\cite{Koba80}).

\bigskip
\noindent
Figure 7:
 {Diffusion path in a disordered system. 
The electron propagates in both directions (time reversal symmetry).
Due to constructive interference the quantum return probability is
twice 
as great as the classical return probability.}

\bigskip
\noindent
Figure 8:
 { The resistance curves of a thin Mg-film (upper set of curves) as
function of magnetic field (magneto-resistance). The weak localization
occurs for zero field.
 After a superposition with \protect$1/100$
atomic layer of Au the magneto-resistance changes drastically. The Au
introduces a rather pronounced amount of spin-orbit scattering which
rotates the spin in the complementary scatterd waves. This changes the
interference from constructive to destructive (after \protect\cite{Ber84}).}

\bigskip
\noindent
Figure 9: {Conductivity as a function of electron concentration
indicating strong localization (after \protect\cite{Stu94}). }

\bigskip
\noindent
Figure 10:
 {The quantum Hall effect: The Hall resistance
\protect$\rho_{xy}$ and the magnetoresistance \protect$\rho_{xx}$ as
function of the magnetic field measured in medium mobility GaAs
heterostructures at \protect$60$mK (after \protect\cite{Tsui89}).} 

\bigskip
\noindent
Figure 11:
 {Schematic view of a multi-probe conductor. Currents $I_k$
entering
the system from  probe $k$  are counted positive.}

\bigskip
\noindent
Figure 12:
 {Schematic view of a box shaped conductor in d
dimensions characterized by length $L$ and cross section $L_t^{d-1}$.}

\bigskip
\noindent
Figure 13:
 {Schematic view of energy branches for semi-infinite leads.
Due to transversal quantization a discrete number of modes exist
(characterized by a wave number $k$) for a given value of the Fermi
energy
\protect$E_F$.}

\bigskip
\noindent
Figure 14:
{Scattering in a quasi-1D system.}

\bigskip
\noindent
Figure 15:
 {Adding two 1D conductors in series.}

\bigskip
\noindent
Figure 16:
 {The  scattering element of a 2D network. For unit amplitude on
the
ingoing link from the left and all other ingoing links being
empty the amplitudes on the outgoing links are
\protect$r_L$, \protect$r_R$ and \protect$r$, respectively.}

\bigskip
\noindent
Figure 17:
{The network resulting from the elements of
Fig.~16.}

\bigskip
\noindent
Figure 18:
 {The basic scattering elements of the Chalker-Coddington
model.}

\bigskip
\noindent
Figure 19:
 {The Chalker-Coddington network model for 2D disordered electrons
in strong magnetic fields. For strong reflection
 to the left (right) the electrons perform
chiral loops.}

\bigskip
\noindent
Figure 20: {Qualitative behavior of the \protect$\b$-function
for disordered electrons. For \protect$d=2$
the full line corresponds to
absence of magnetic fields and spin-orbit scattering while the dotted line
corresponds to the presence of spin-orbit scattering.
In \protect$d=3$ a LD transition occurs quite generally at some critical value
\protect$g^\ast$ which forms an unstable fixed point of the flow
determined by the \protect$\b$-function.}

\bigskip
\noindent
Figure 21:
 {The \protect$\b$-function as following from Eq.~(\protect\ref{6.28}).
The curves correspond to (from below) \protect$d=1,2,2.2,3$.}

\bigskip
\noindent
Figure 22: {Qualitative behavior of multifractal exponents. 
The top figure shows the exponents \protect$\tau(q)$
corresponding to moments of a multifractal 
distribution, the middle shows the resulting 
generalized dimensions \protect$D_q=\tau(q)/(q-1)$ and the bottom shows the
\protect$f(\a)$-spectrum being the Legendre transform of \protect$\tau(q)$.
\protect$f(\a)$ describes the scaling of the whole distribution.}

\bigskip
\noindent
Figure 23:
 {Qualitative behavior of the conductance distribution
in a metal. Within a range of width  \protect$d\approx \sqrt{\BRA g\KET}$
around the average value \protect$\BRA g\KET$  the
distribution is Gaussian with universal variance of order $1$.
Outside this region the distribution develops log-normal tails.}

\bigskip
\noindent
Figure 24:
 {The scaling function for a 3D quantum Hall
 system showing a LD transition. \protect$\Lambda(E,M)$
denotes the scaling variable as a function of energy \protect$E$ and system
 size \protect$M$. \protect$\xi(E)$ 
denotes the correlation length obtained from
a fitting procedure. Different data symbols correspond to different
energies and system sizes (after \protect\cite{DohDip}).}

\bigskip
\noindent
Figure 25:
 {Qualitative picture visualizing the
 explanation of the quantum Hall effect being due to a
sequence of localization-delocalization
 transitions occurring at the Landau energies.
\protect$\rho,\sigma_{xx},\sigma_{yx}$ denote the density of states,
 the dissipative conductivity and the Hall conductivity, respectively.
 In finite
systems the range of extended states (grey) on the energy scale has finite
width.
This width is believed to shrink to zero in the thermodynamic limit.}

\bigskip
\noindent
Figure 26:
 {Squared amplitudes of a critical wave function for a system of linear
size \protect$200$ magnetic lengths. The corresponding 
\protect$f(\alpha)$ spectrum ($\protect\bullet$)
 together with the parabolic
approximation ($\protect\cdots$)
 and a semi-elliptic approximation  
(--)  are also shown (after \protect\cite{Prac96}).}

\bigskip
\noindent
Figure 27:
 {Histogram of the logarithm of squared amplitudes shown in
Fig.~26. The continuous curve is the distribution function
following from the corresponding \protect$f(\alpha)$ spectrum with
\protect$\alpha_0=2.28\pm0.02$ 
(after \protect\cite{Prac96}).} 

\bigskip
\noindent
Figure 28: {Squared amplitudes of a critical wave function 
in the Chalker-Coddington network model.(after \protect\cite{KleDoc}).}

\bigskip
\noindent
Figure 29:
 {Histogram of the logarithm of squared amplitudes shown in
Fig.~28. The continuous curve is the distribution function
following from the parabolic approximation to the \protect$f(\alpha)$
spectrum
with \protect$\alpha_0=2.27\pm0.02$.
The inset shows the histogram for the squared  amplitudes itself (after
 \protect\cite{MetDoc}).}

\bigskip
\noindent
Figure 30:
 {The function of critical exponents $z(q)$ following from the
the scaling relation
Eq.~(\protect\ref{8.11c}) (after \protect\cite{Prac96}).}

\bigskip
\noindent
Figure 31:
    {Qualtative behavior of the typical local density of states
reflecting
a localization-delocalization transition at critical Fermi energy
\protect$E^\ast$.}

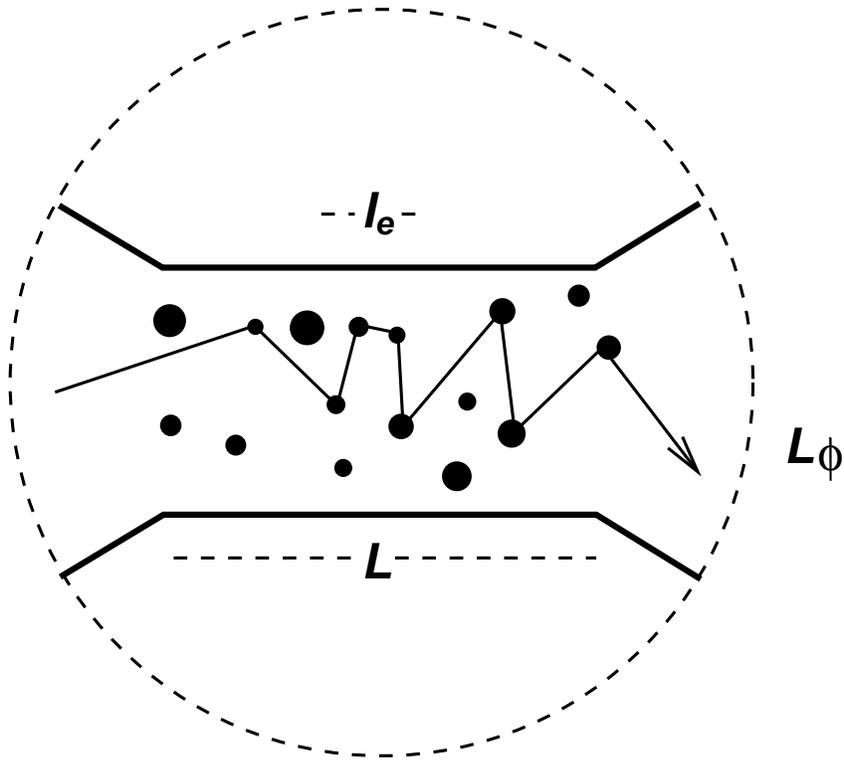
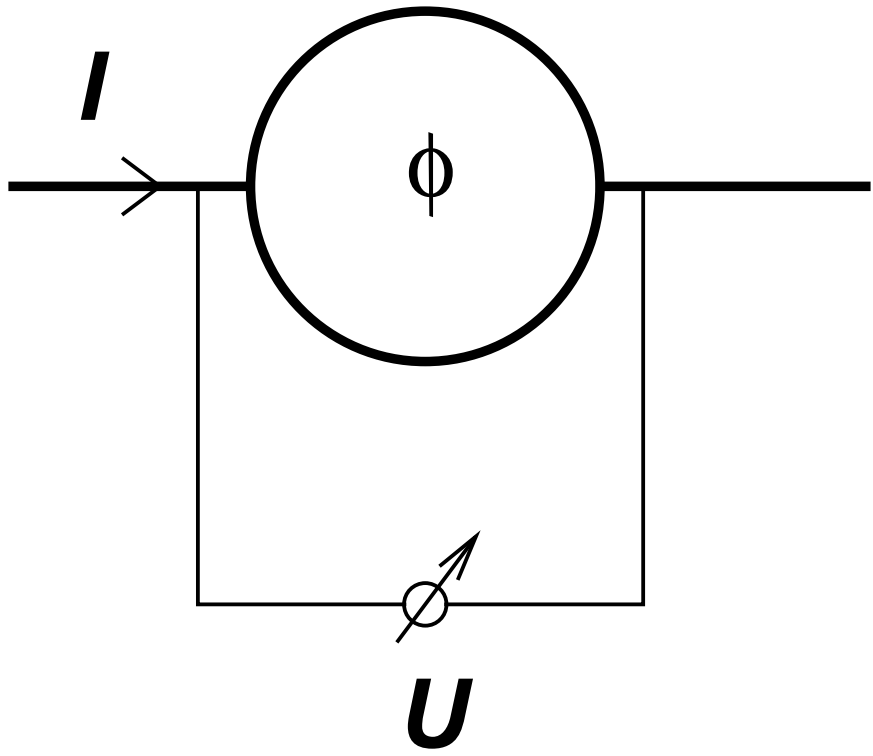
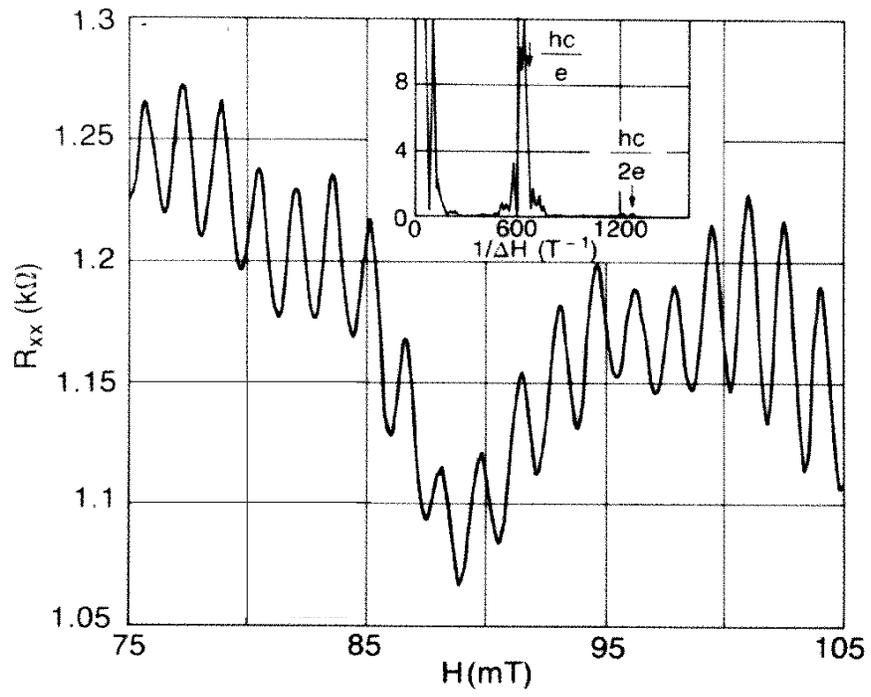
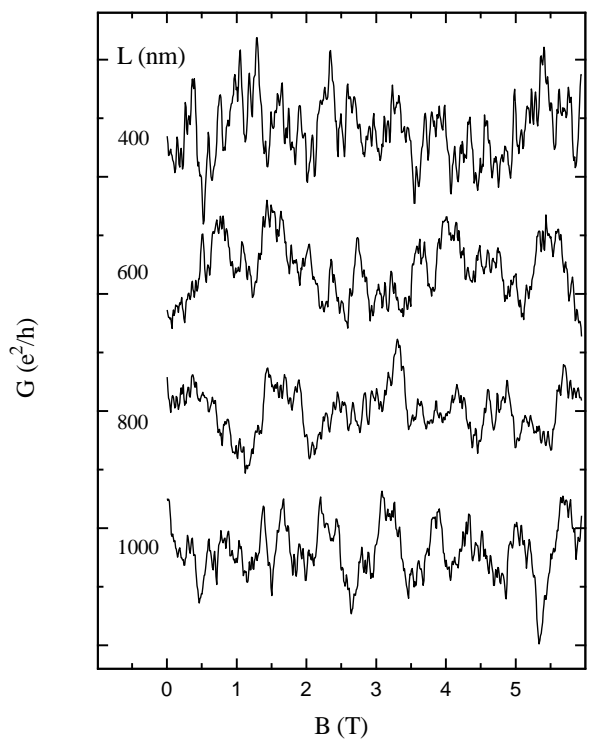
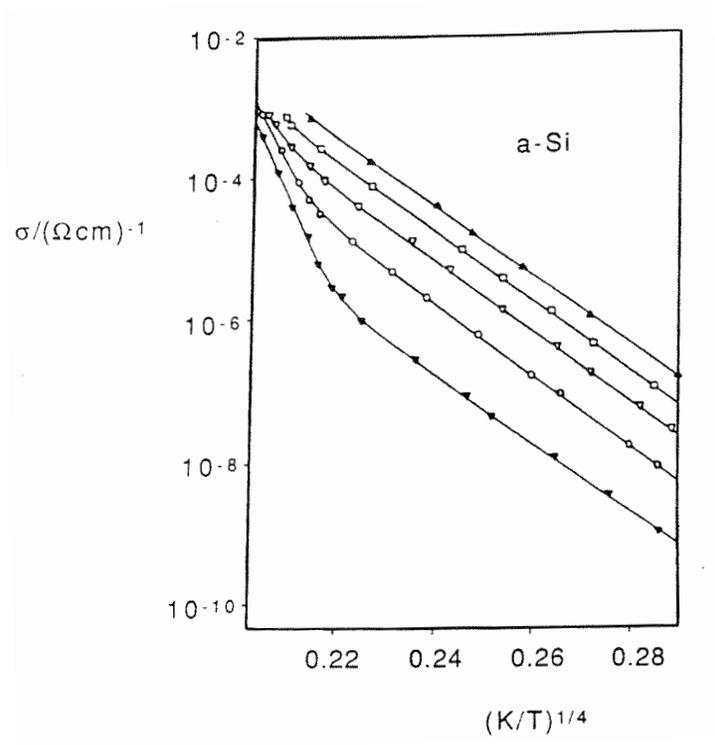
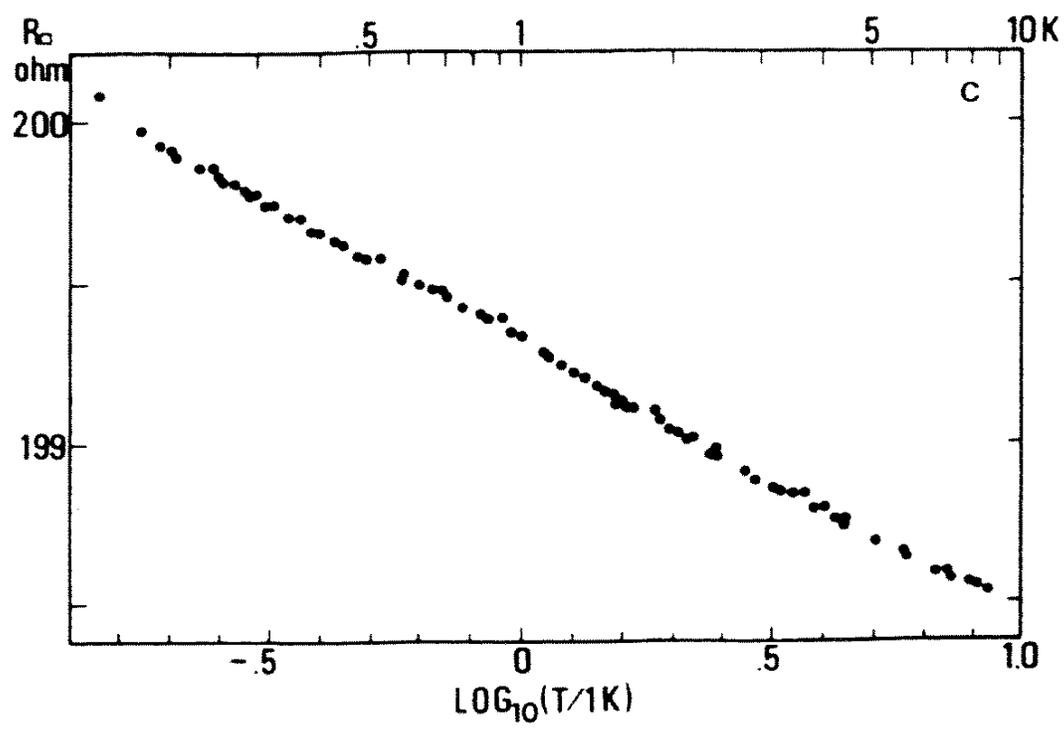
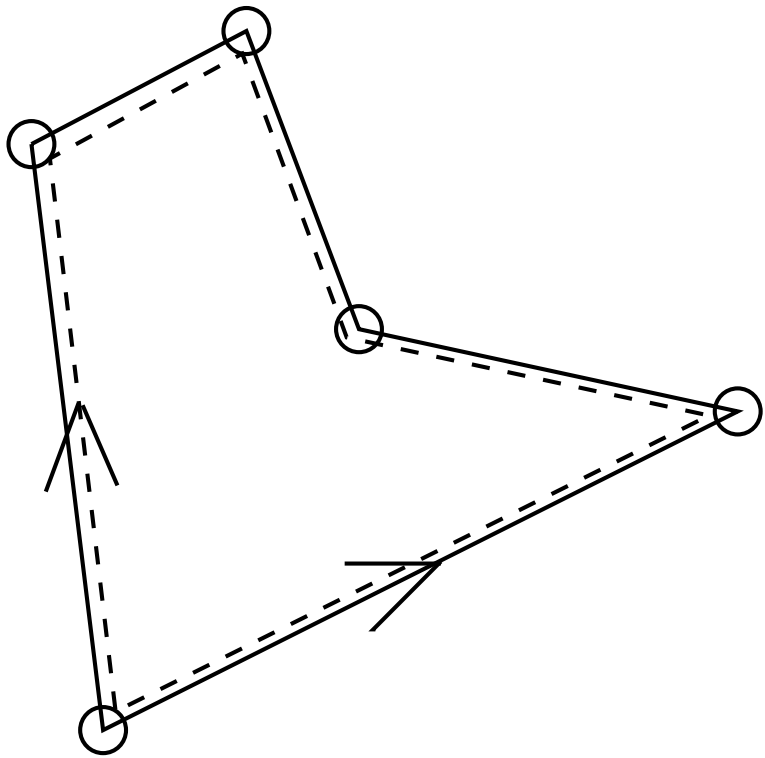
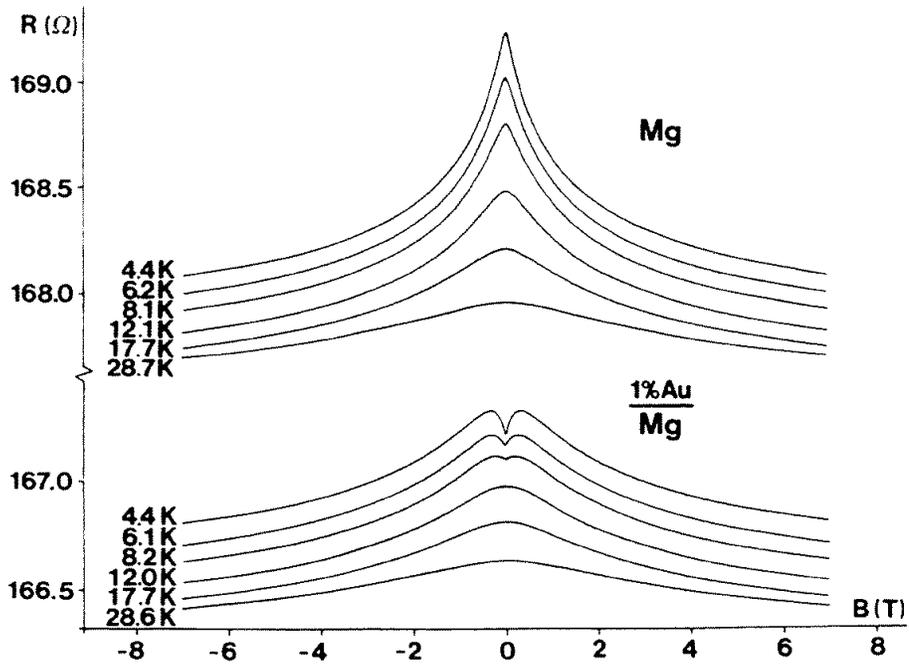
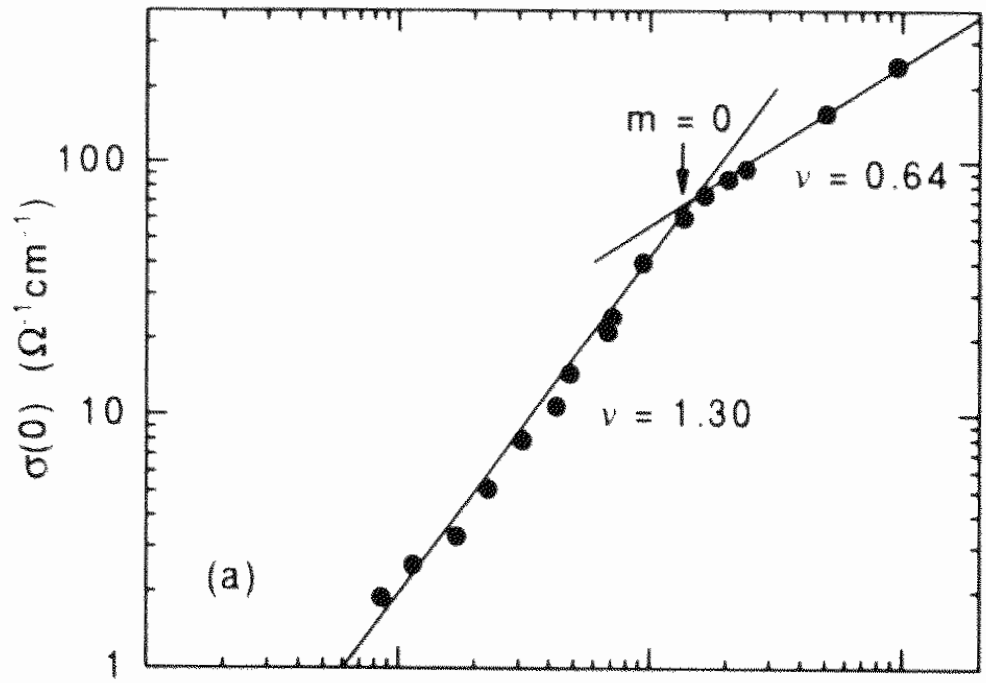
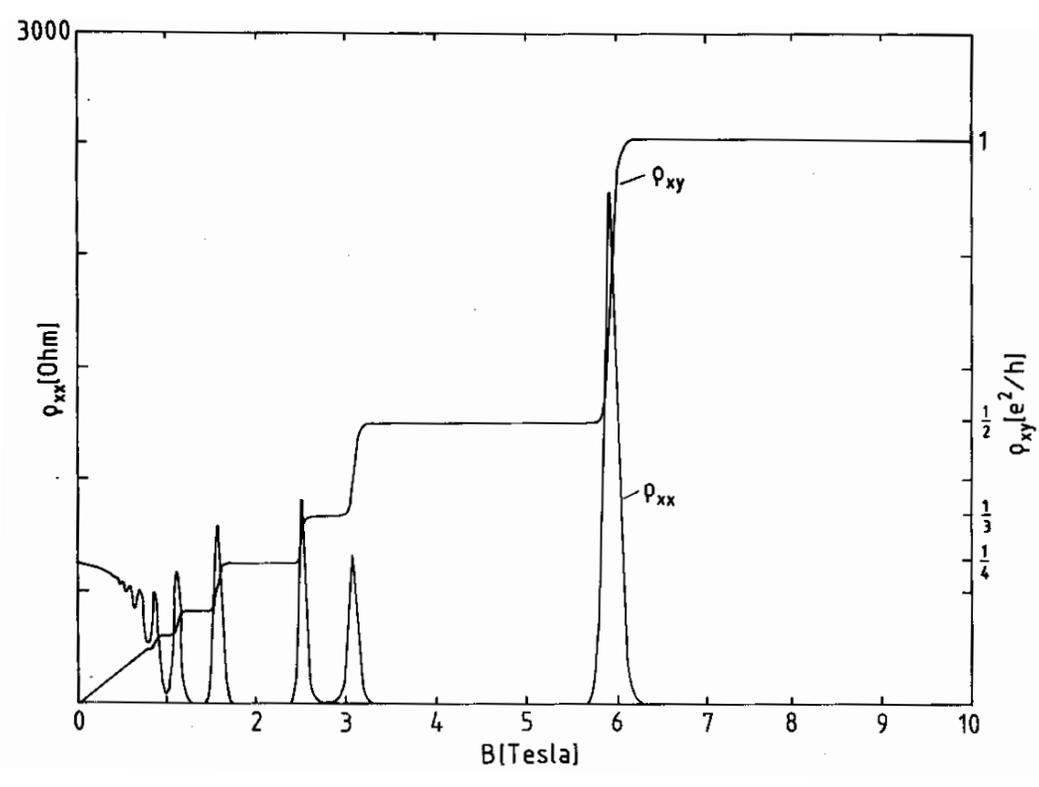
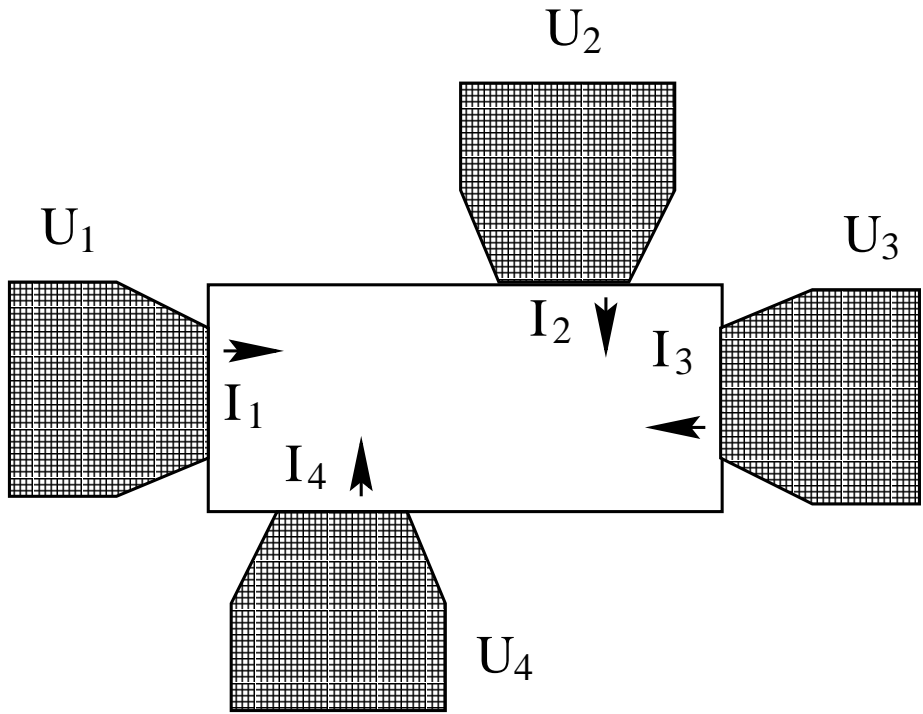
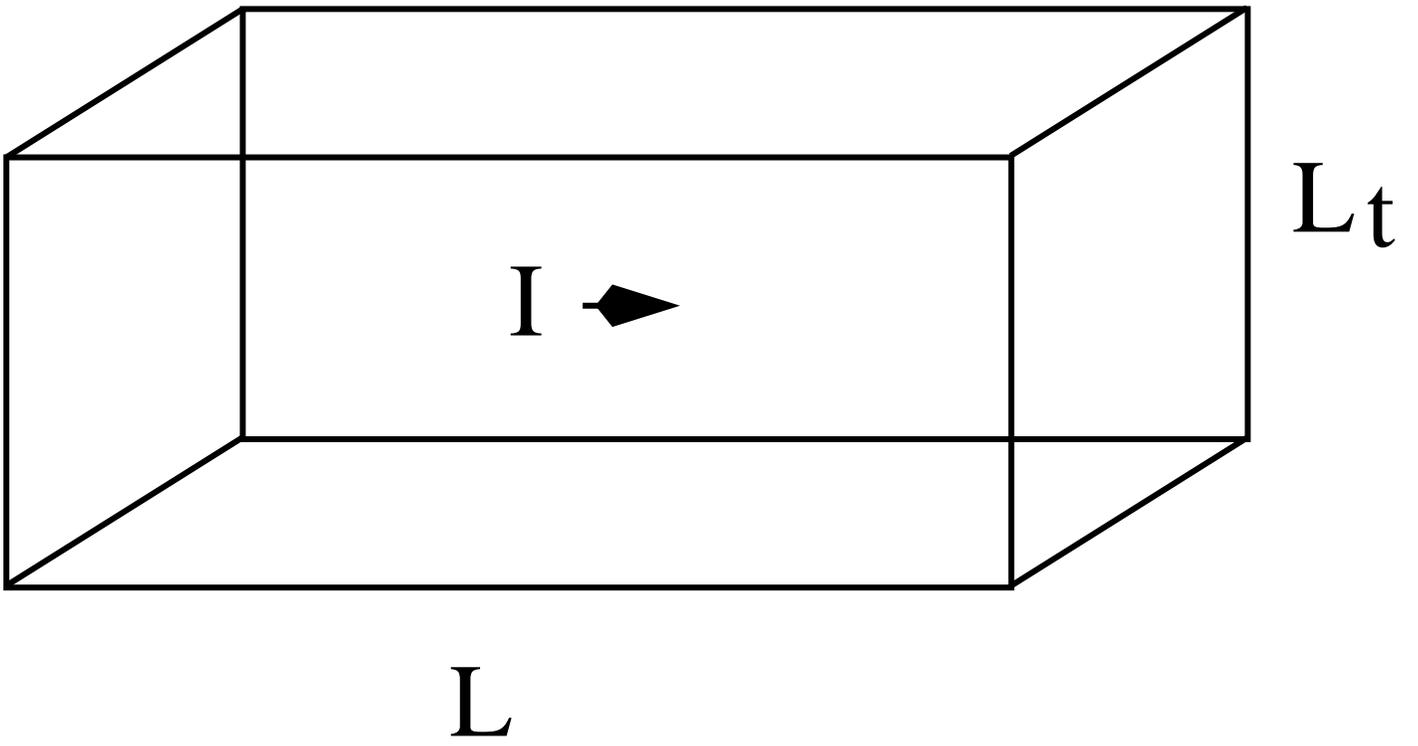
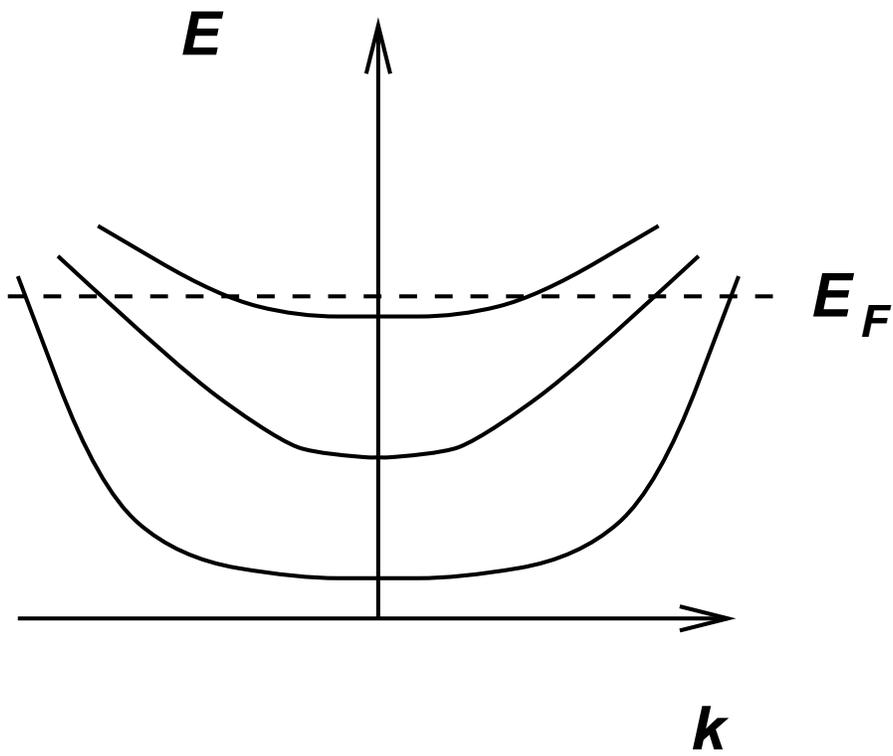
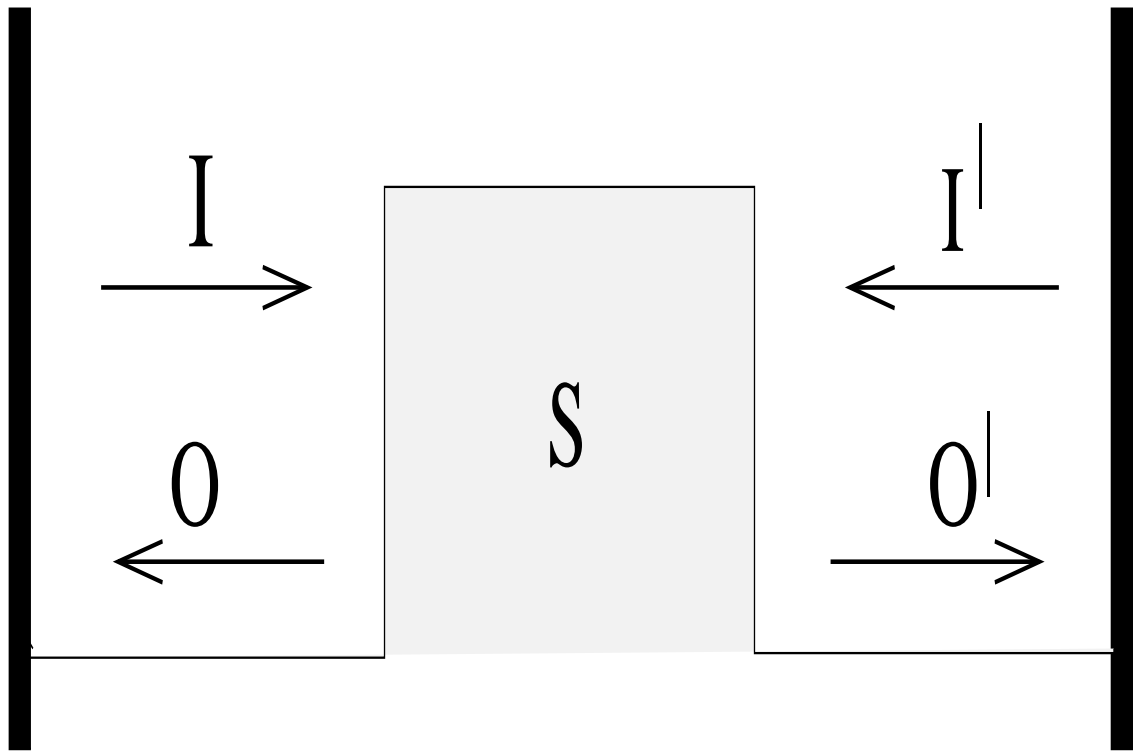
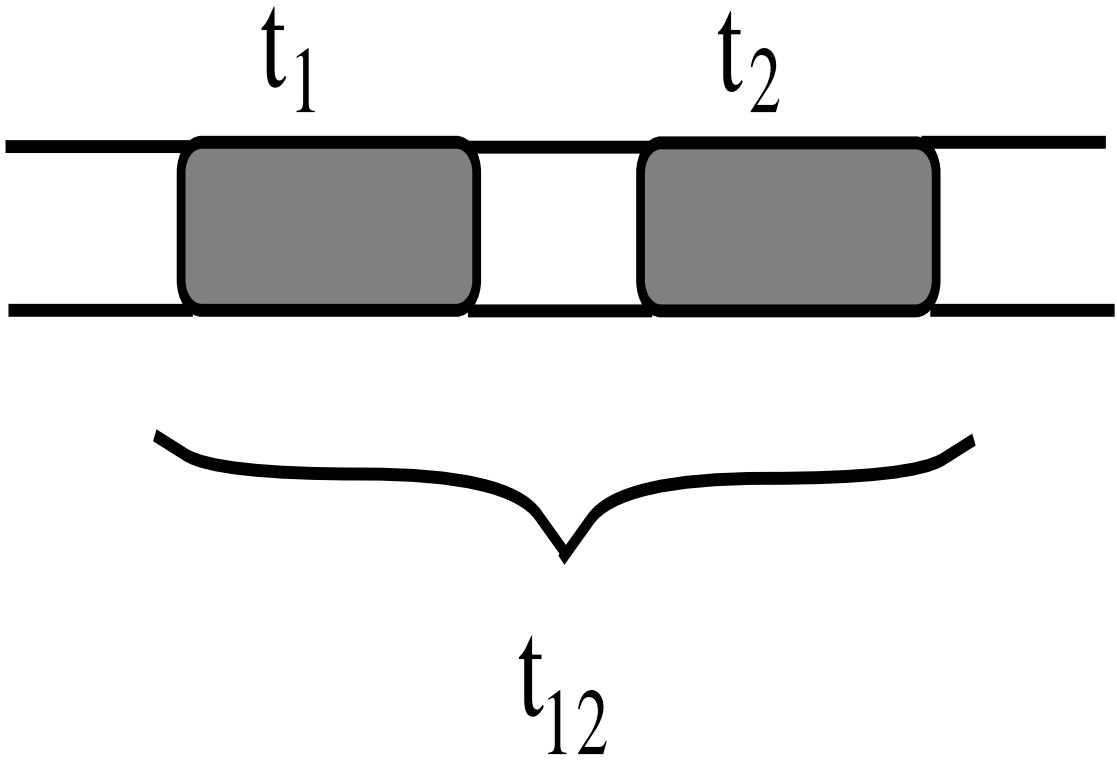
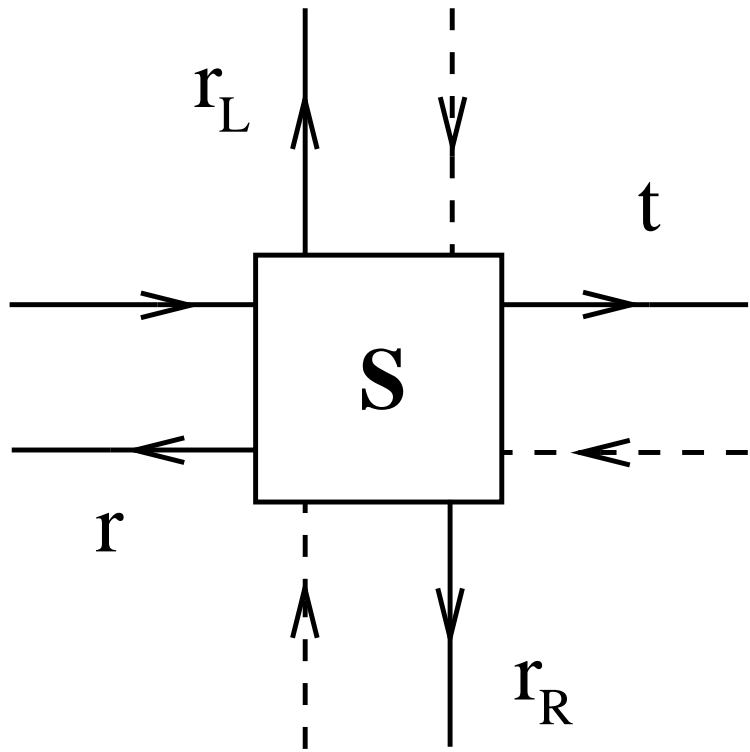
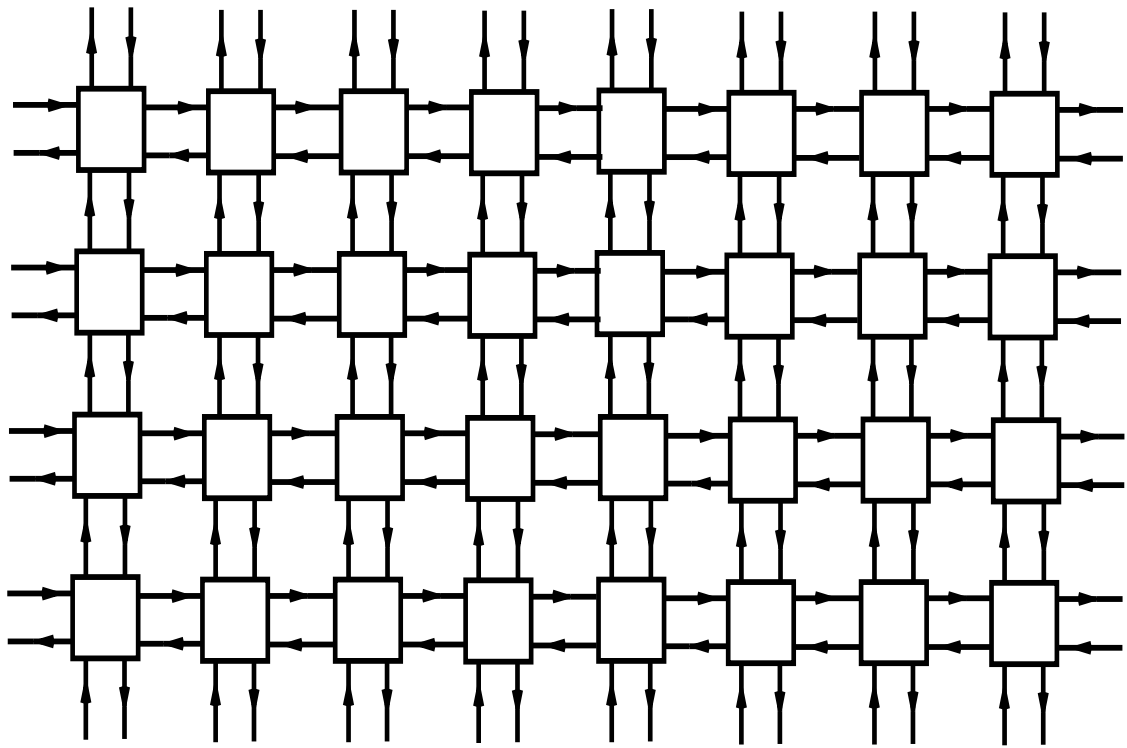
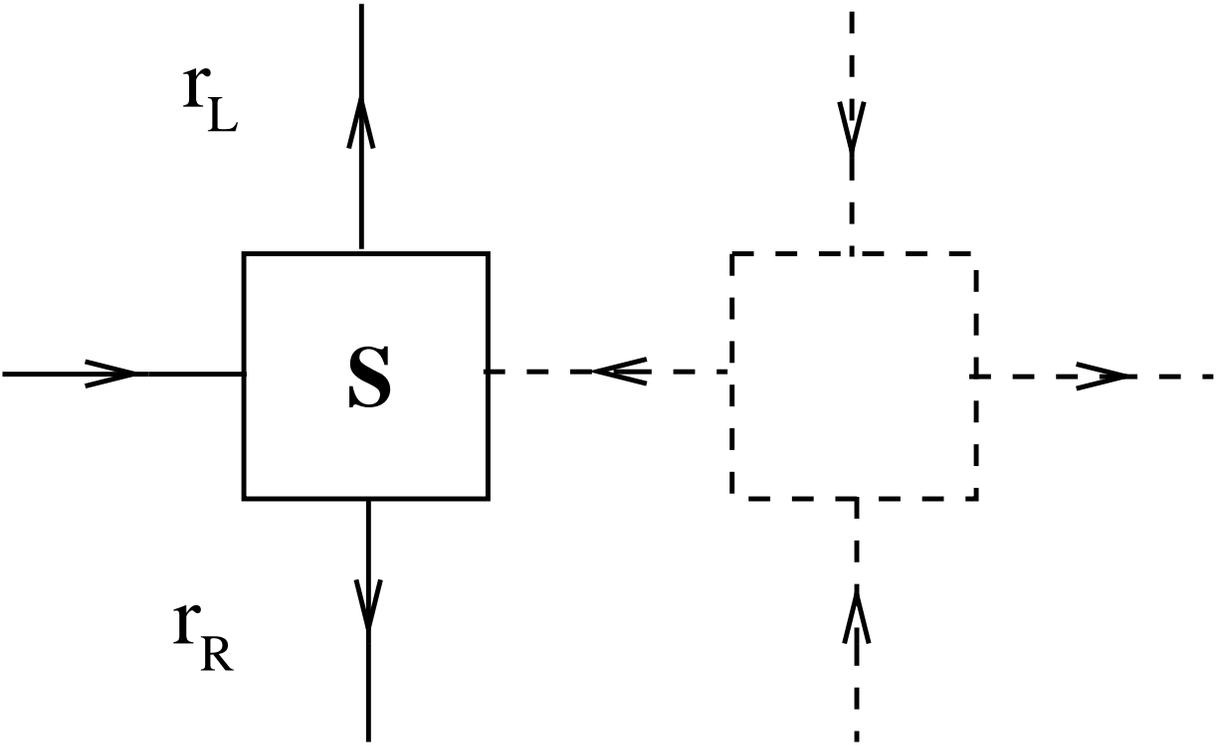
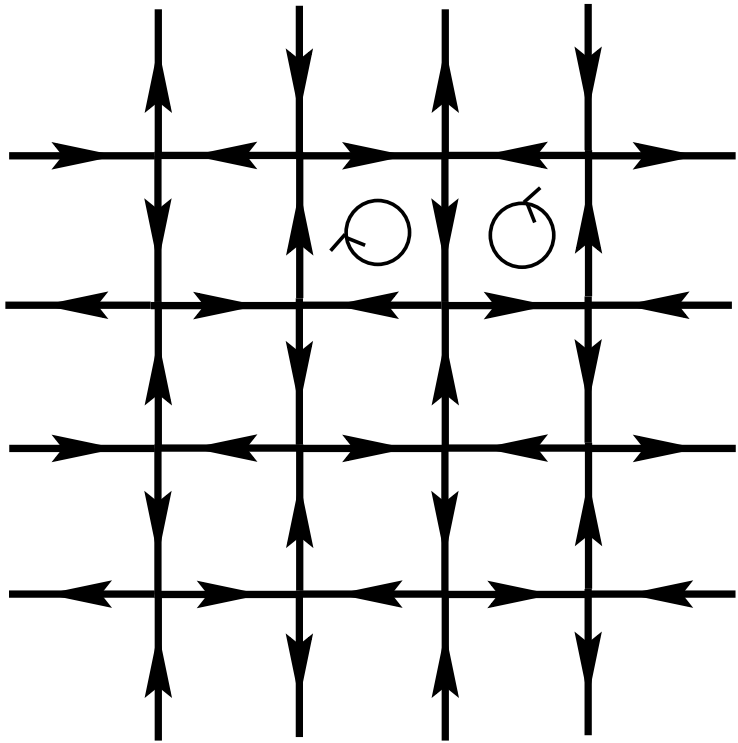
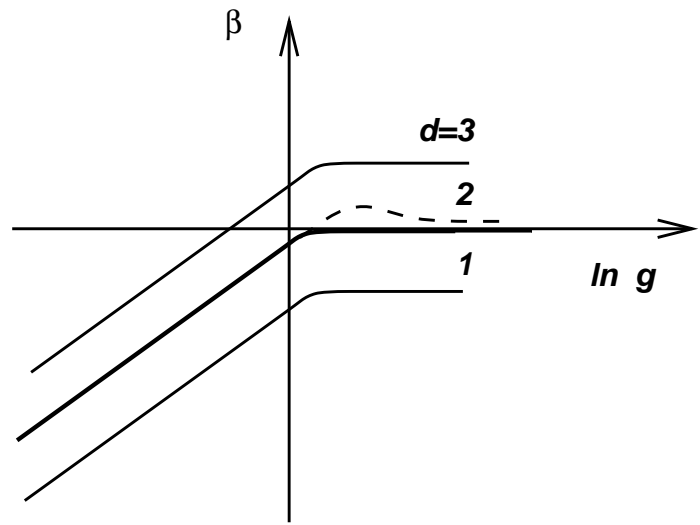
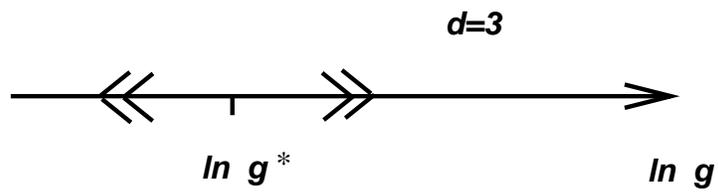
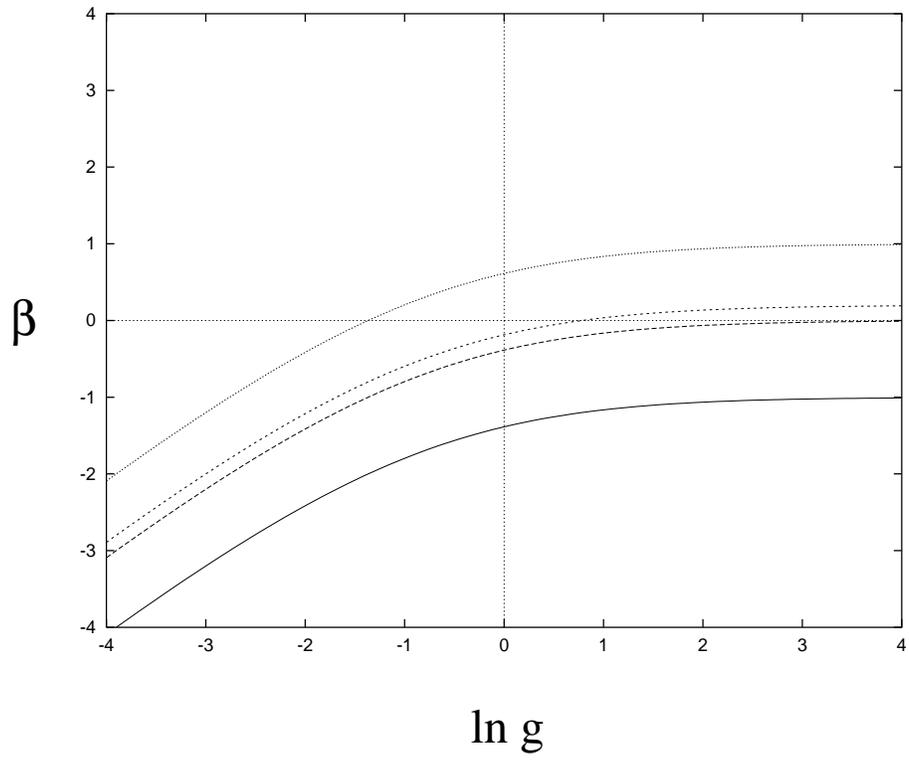
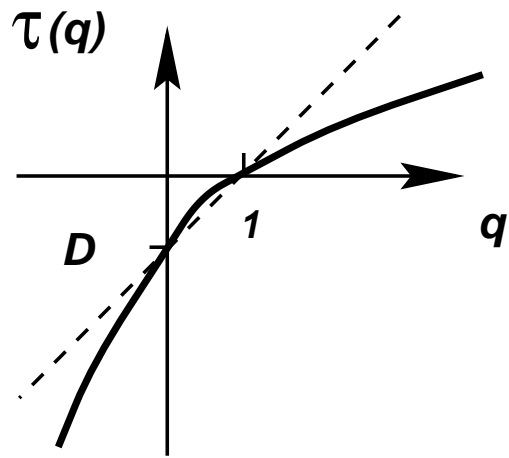
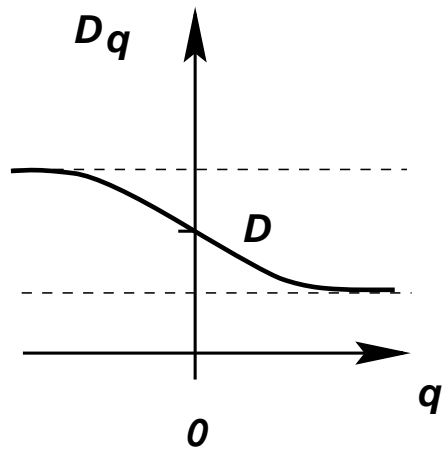
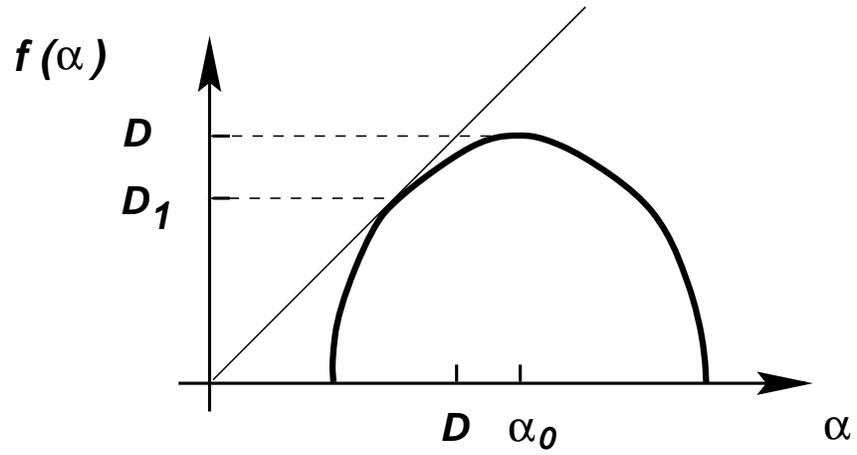
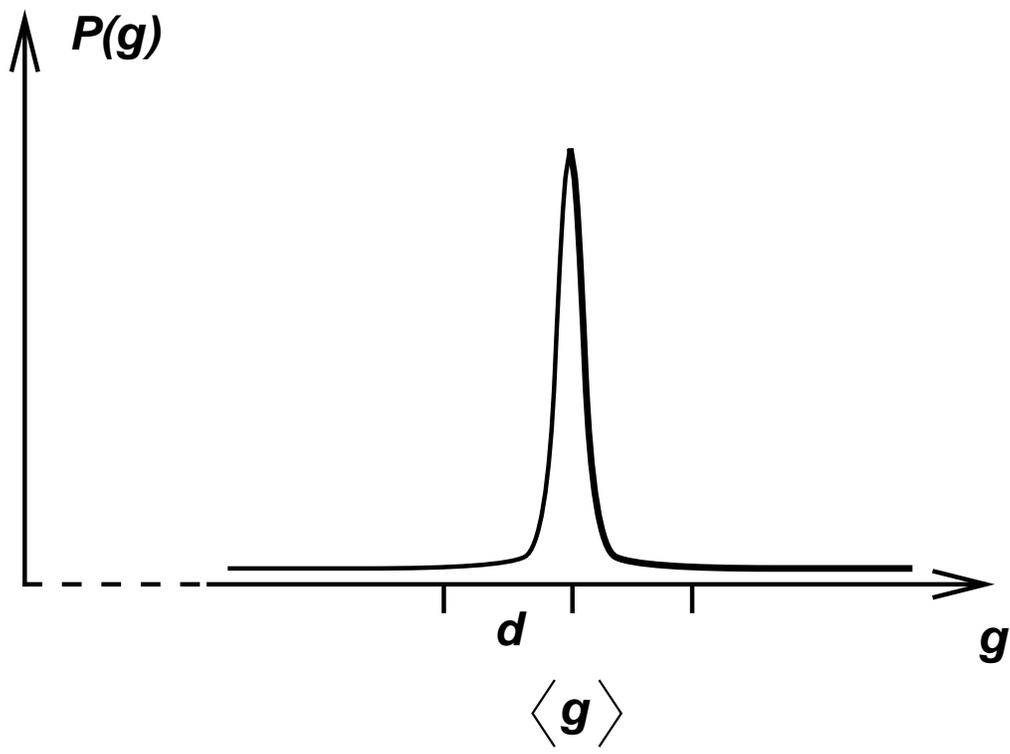
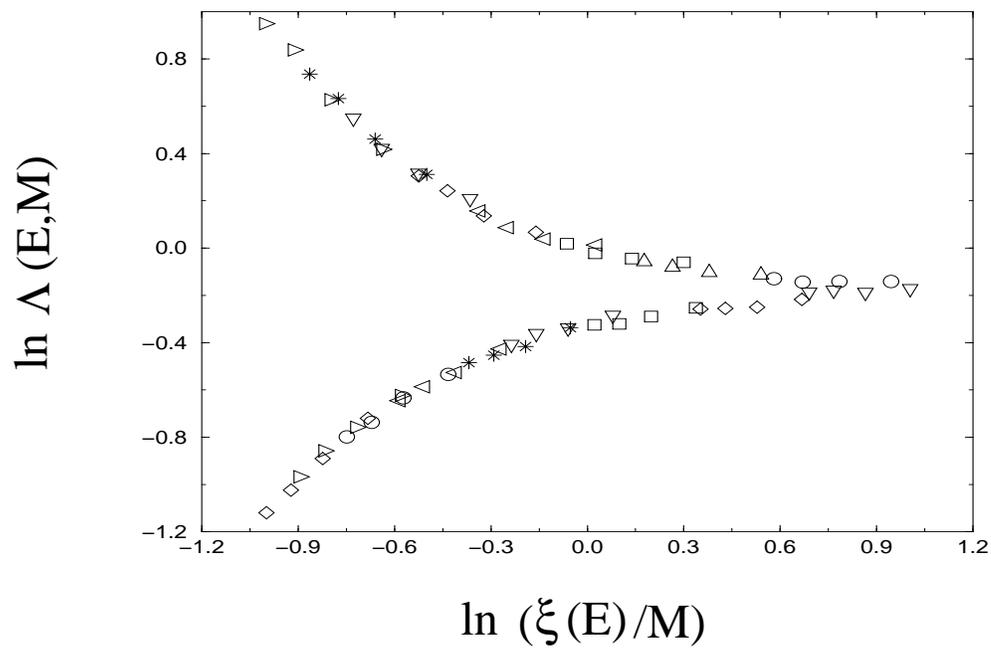
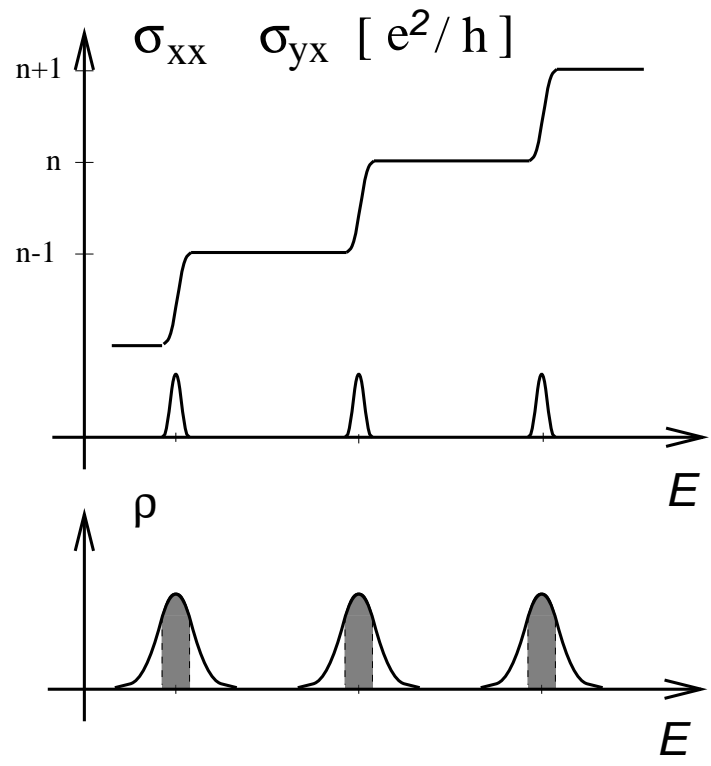
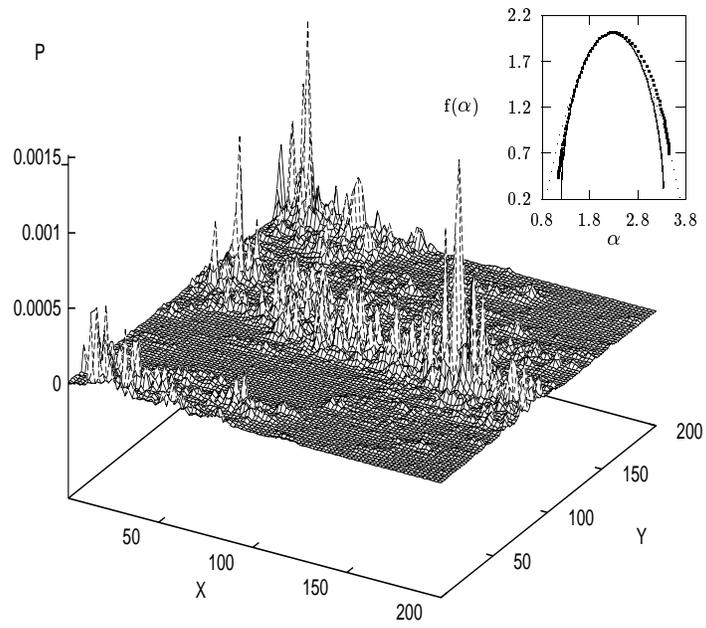
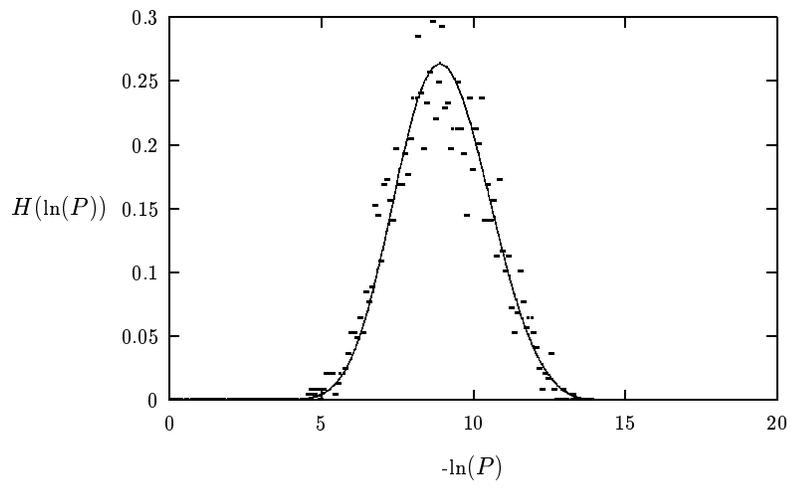
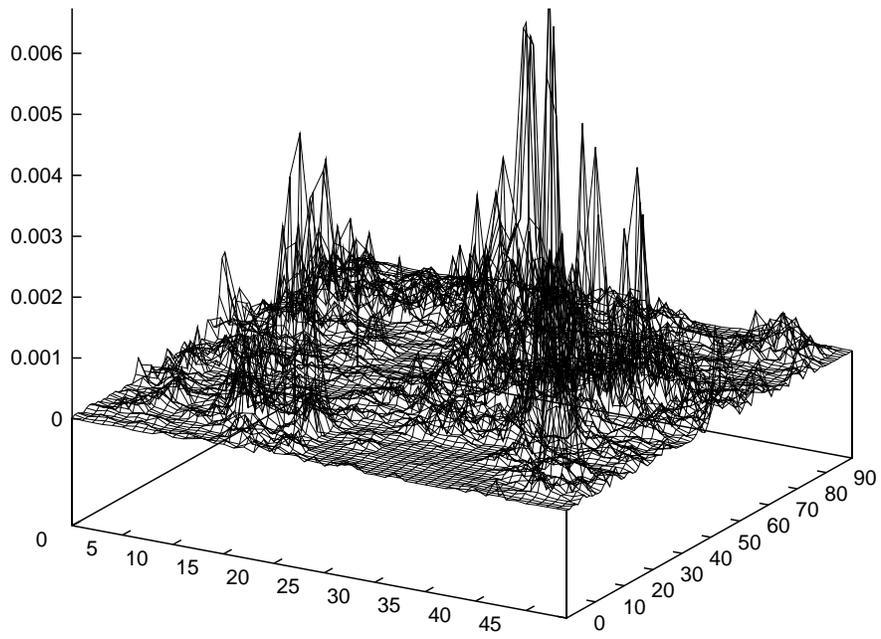
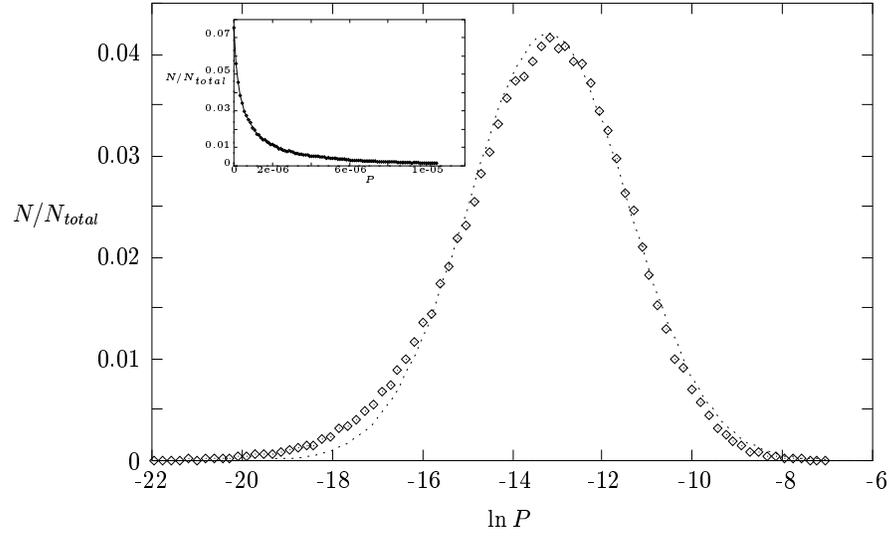
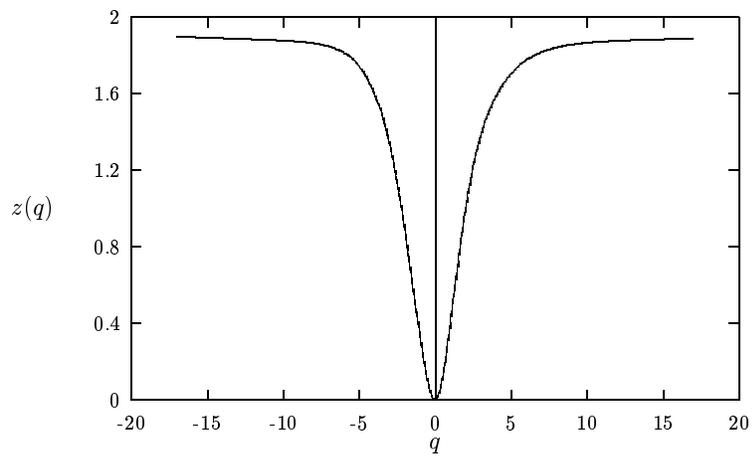
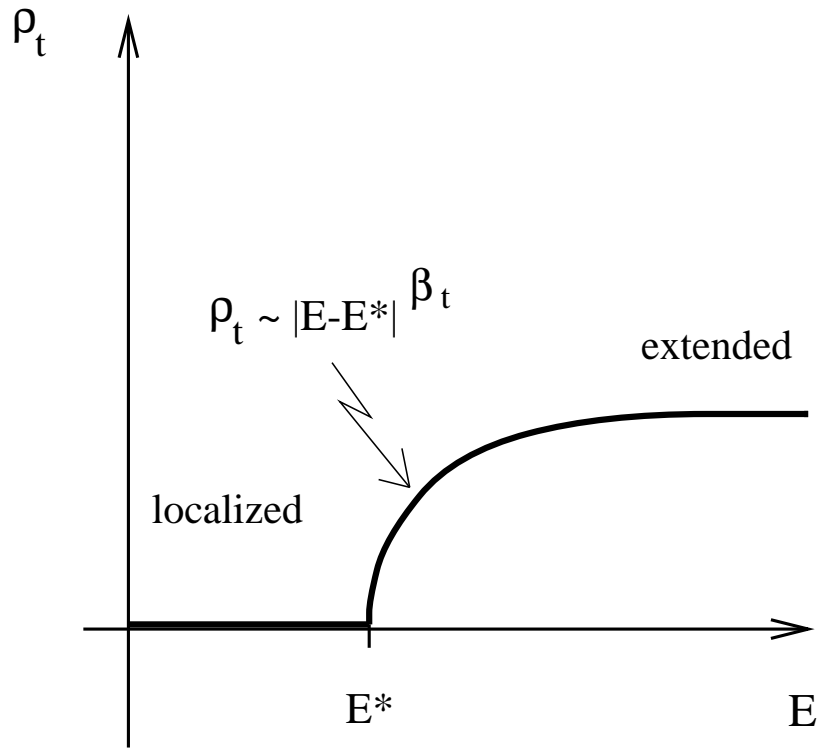
\end{document}